\documentclass{svjour3}

\usepackage{graphicx}

\usepackage{bigstrut}

\usepackage{multirow}

\newcommand{\bea}{\begin{eqnarray}}
\newcommand{\eea}{\end{eqnarray}}
\newcommand{\beq}{\begin{equation}}
\newcommand{\eeq}{\end{equation}}

\newcommand{\tref}[1]{(\ref{#1})}

\newcommand{\tnote}[1]{\textbf{(T)}\footnote{\textbf{(T)} #1}}
\newcommand{\nnote}[1]{} 
\newcommand{\bnote}[1]{} 

\newcommand{\czero}{c_0}
\newcommand{\crzero}{\langle c/r \rangle}
\newcommand{\cfindex}{c_\mathrm{f}}
\newcommand{\crindex}{c_\mathrm{r}}
\newcommand{\zfindex}{z_\mathrm{f}}
\newcommand{\zrindex}{z_\mathrm{r}}

\newcommand{\Jcal}{\mathcal{J}}
\newcommand{\Pcal}{\mathcal{P}}
\newcommand{\Scal}{\mathcal{S}}

\begin{document}

\title{Universality of Performance Indicators based on Citation
and Reference Counts\thanks{Revised version 18th February 2012, \texttt{arXiv:1110.3271}, \texttt{Imperial/TP/11/TSE/5}. Submitted to Scientometrics.}}
\author{T.S. Evans \and  N. Hopkins \and B.S. Kaube}
\institute{T.S. Evans \and N. Hopkins \and B.S. Kaube
            \at Department of Physics,
                Imperial College London, SW7 2AZ, U.K.}
\maketitle

\begin{abstract}
We find evidence for the universality of two relative bibliometric
indicators of the quality of individual scientific publications taken from different data sets.
One of these is a new index that considers both
citation and reference counts.
We demonstrate this universality for relatively well cited publications from
a single institute, grouped by year of publication and by faculty or by department.
We show similar behaviour in publications submitted to the arXiv e-print archive,
grouped by year of submission and by sub-archive. We also find that for reasonably well cited papers this distribution is
well fitted by a lognormal with a variance of around $\sigma^2=1.3$ which is consistent with the results of Radicchi, Fortunato, and Castellano \cite{Radicchi}.
Our work demonstrates that comparisons can be made between publications from different disciplines and publication dates, regardless of their citation count and without expensive access to the whole world-wide citation graph. Further, it shows that averages of the logarithm of such relative bibliometric indices deal with the issue of long tails and avoid the need for statistics based on lengthy ranking procedures.
\end{abstract}

\section{Introduction}

The use of relative bibliometric indicators to provide robust measures has been discussed in several contexts \cite{SB86,V86,MDV95,V97,AT04,Radicchi,RML07,CR09,BD09a,WELVR10,RC11,ACOR11,WER11,EF11}.  Radicchi et al.\ \cite{Radicchi} (hereafter referred to as RFC) found a
universal distribution for one such relative measure of the number of citations each paper received.
The universality found by RFC was demonstrated
across a wide range of scientific disciplines using the commercial
Thomson Reuters's Web of Science (WoS) database \cite{WoS} to
derive the citation counts. The indicator used by RFC applied to single
publications was $\cfindex=c/\czero$, where $c$ is the number of citations for a given
paper and $\czero$ is the average number of citations for all
papers published in the same field and in the same year as the
paper being considered\footnote{This is similar to the crown indicator \cite{MDV95} but applied to a single publication, see \cite{LBMO11} for other references on this.}. RFC \cite{Radicchi} used
Thomson Reuters's Journal of Citation Reports, which allocates one or more fields to each journal, to assign fields to each paper. This index $\cfindex$ gives a measure of the significance of a given paper
which can be used compare papers from a wide range of disciplines and
published at different times.  The big drawback is that it requires access to a global dataset of publications to calculate the average $\czero$.

In this paper we extend the work of RFC in three
ways. First, we work with a different subset of papers, either those published by authors of one institute, and later those put on the electronic preprint repository, arXiv. Secondly, we assign the research field of a paper in different ways, via the political divisions of the institute, using either faculty or departments, and for arXiv we use its predefined subdivisions.  Finally, we consider alternative indicators of a paper's performance, involving the number of references in its bibliography as well as the number of citations of that paper.  By showing that in all cases a lognormal distribution is a reasonable model for the data, we have demonstrated that these useful indices can be applied on a large number of smaller datasets.  As such data may already be available for other reasons, our results will lead to a reduction in the costs of research assessment, be this for academic research or for administrative reasons.

We will start in section \ref{sind} with the case of the papers from a single institute and use this example  to define the indicators we shall consider.  We then comment in section \ref{sresults} on the properties of our data from a single institute and the results for the indicators for the data from the institute.  In section \ref{sarXivdata} we repeat the analysis for data from arXiv.  We then discuss our results in terms of simple statistical models in section \ref{sinterpetation} and finish with some conclusions in section \ref{sconclusions}.  An extensive list of tables and additional plots are given in the supplementary material.

\section{Definition of Indicators}\label{sind}

We will define the indicators used in terms of our first example, the papers from a single institute.
The first index we use is defined in terms of two sets of papers:-
\begin{itemize}

\item[$\Pcal$ ---] Complete WoS data, including uncited items and those without references, published in 2010 or before.

\item[$\Scal$ ---] Any WoS item approved by staff of one faculty in a single
calendar year, or from one department in a three year interval, respectively, with at least one citation and one reference.

\end{itemize}
We assume that for any paper in the set $\Scal$ we know \emph{all} the citations coming from any paper in $\Pcal$.
Then we define the relative bibliometric indicator $\cfindex(s,\Scal,\Pcal)$ (later often abbreviated to $\cfindex$) \cite{Radicchi} to be
\beq
 \cfindex(s,\Scal,\Pcal) = \frac{c(s,\Pcal)}{\czero(\Scal,\Pcal)} \, , \;s \in \Scal \, ,
 \qquad
 \czero(\Scal,\Pcal) = \frac{1}{|\Scal|} \sum_{s' \in \Scal} c(s',\Pcal) \, .
 \label{cfdef}
\eeq
Here $s$ is a paper drawn from the set\footnote{Usually $\Scal$ is a subset of $\Pcal$, $\Scal \subseteq \Pcal$, but this is not strictly necessary.} $\Scal$, and
$c(s,\Pcal)$ is the number of citations to paper $s$ from the
set of papers $\Pcal$. Both here and in \cite{Radicchi} $\Pcal$ was the
whole Thomson Reuters database taken at some point in time. We differ over our choice of set $\Scal$ as in \cite{Radicchi}  this was chosen to be the subset of papers (excluding some other types of publication) published
in one year and in one field, as defined by the Thomson Reuters's
Journal of Citation Reports.  In our case $\Scal$ is either the set
of papers published in one year from one faculty or those published in three years
from one department, each faculty containing several departments
of related fields.

This index is successful because several
factors which might be expected to change the citations $c(s,\Pcal)$
of individual papers $s$ will be mirrored in the behaviour of the
average. For instance if we change the length of time papers have
had to gather citations, changing $\Pcal$, our first guess might be that this effect
would cancel in the ratio $\cfindex$. Likewise, the numbers of
citations change with the field but we might hope that this effect
cancels out in taking the ratio. The results of
\cite{Radicchi} show that for their definitions of $\Scal$ and
$\Pcal$ the statistical distribution of this ratio is
independent of the field and publication year used to choose
the subset $\Scal$. It is therefore not unreasonable for us to
hope that by looking at the same ratio but for a different set
of papers $\Scal$, we would see the same universality.

Our use of faculties and departments of an institute to define
academic field is a cruder way to split up the set of all papers
$\Pcal$.  For instance there are eight physics classifications
in the Thomson Reuters's Journal of Citation
Reports while we have but one physics department.  However the
greatest differences in RFC \cite{Radicchi} occur on broader
classifications, with the differences between citation
behaviour of medical, physical science and engineering fields.
In this sense we hope that our broader classification will
still be sufficient to show the universality of RFC
\cite{Radicchi}.  In this context we also note the work of
Rafols and Leydesdorff \cite{RL09} who showed that four
different classifications including the Journal of Citation
Reports had considerable differences but nevertheless they drew
similar conclusions about the statistical properties of sets of
papers whichever classification was used. One might hope that a
department is a dynamic entity responding to shifts and changes
organically and thereby it may well provide a good emergent
definition of a field. Basing the analysis on the political
structures of faculties and departments is a simple and workable
definition and the data required is likely to be already
available at many institutions. This may provide a simpler, cheaper and more practical method to analyse citation data.

Our final variation on RFC \cite{Radicchi} is to look at other
indicators involving the number of references from paper $s$ in
$\Pcal$ to other papers in the database, $r(s,\Pcal)$, a quantity
readily calculable from the usual databases. A
comparison of two fields with different average reference
counts per paper would also be expected to show corresponding
variation in citation counts. This suggests that the quantity
$c(s,\Pcal)/r(s,\Pcal)$ could be a useful measure.
However, it is clear that $r(s,\Pcal)$ can not be a good proxy for
$\czero(\Scal,\Pcal)$ as the former is fixed for each paper
while the latter grows in time. The solution is to use the
same trick as with the $\cfindex$ index \tref{cfdef} and to consider
$c(s,\Pcal)/r(s,\Pcal)$ for paper $s$ divided by its average.  We will use the
short hand notation $\crindex$ to denote this, where
\beq
 \crindex
 = \frac{c(s,\Pcal)}{r(s,\Pcal)}\frac{1}{\crzero(\Scal,\Pcal)} \, , \;\; s \in \Scal,
 \qquad
 \crzero(\Scal,\Pcal) = \frac{1}{|\Scal|}
 \sum_{s' \in \Scal} \frac{c(s',\Pcal)}{r(s',\Pcal)}
 \label{crdef}
\eeq
One advantage of such an indicator is that it will naturally
penalise review articles, which tend to have a large number of references and citations that can distort other indices.

Using the number of references to normalise citation counts is not a new idea.  For instance it has been used in the context of measuring the impact of a journal by Yanovsky \cite{Y81} and more recently by Nicolaisen and Frandsen \cite{NF08}.  Basically the total number of citations in a journal over a given period were divided by the total number of references in the journal.  We are not aware of the use of reference normalisations being used on a per article basis as we do but the principle is the same.  The refinements suggested by Nicolaisen and Frandsen in terms of limited windows in time for references \cite{NF08} could also be applied to our metric \tref{crdef}.  Our approach does suggest a different journal measure from those of \cite{Y81,NF08} by averaging our individual paper ratios \tref{crdef} for all papers published in a given period, as $\langle c \rangle/\langle r \rangle \neq \langle c/r \rangle$.  In fact we will give an explicit

\section{Results for a Single Institute}\label{sresults}


Our data set $\Pcal$, consists of all approved
publications authored by at least one current permanent staff member\footnote{This is the usual situation but some exceptions exist.} of
the institution providing our data and with at least one
citation, at least one item in the bibliography and a definite year of publication. They are necessarily in WoS \cite{WoS} which provides the number of citations, number of references and the year of publication. Publications were classified by Thomson Reuters as articles (78.8\%), proceedings (8.1\%), reviews (5.4\%), editorial material (2.5\%), letters (2.3\%), notes (1.4\%) and meeting abstracts (1.1\%) with a small number of other types of publication (0.2\%) (see table \ref{tpubtypes} in the supplementary material). Approval is through a web based interface in which staff confirm that they
authored a given publication.  This ensures that the assignment of
authors to their current faculty and department will be almost
perfect\footnote{While almost all papers are validated, the
status of a few papers is unclear but they are not included in
our set.  If staff have changed fields since the publication of
a paper, it is possible that some assignments will be
incorrect. We presume this is effect is small and worse for
older papers.}. It is an important feature of this data that name and address disambiguation problems are completely avoided.
The number of references is the length of the
bibliography even if not all elements in that bibliography are
included in WoS. For instance, a reference to a book will be
counted in $r$ but the citations from that book will not, since
books are not part of WoS.
We only include papers with positive citation counts, positive
reference counts and known publication year, of which there were
78267 (74\%)\footnote{There were 12089 (13\%)
papers which appear to have zero citations and a positive number of
references. For simplicity we did not do so in our study as their logarithm is infinite.
See the conclusions in section \ref{sconclusions} for further discussion of zero and low cited papers and how and where we could include them in our analysis.
The remaining papers have
a variety of signals that the entry is unreliable, e.g.\ no
publication year, zero references. We have also excluded this
remaining 13\%.}.

The papers were grouped into various sets $\Scal$, either papers published in the same year with at least one staff author from a particular faculty, or papers published in a three year interval with at least one staff author from a particular department.  These choices were made to get a reasonable number of papers in our sets $\Scal$ to ensure statistically significant results could be obtained.
If papers were written by multiple authors who are
part of different departments or faculties, the paper was
counted once for each relevant department or faculty. Hence
the category definitions are not mutually exclusive.

\begin{figure}[htbp]
\begin{center}
\includegraphics[width=7cm]{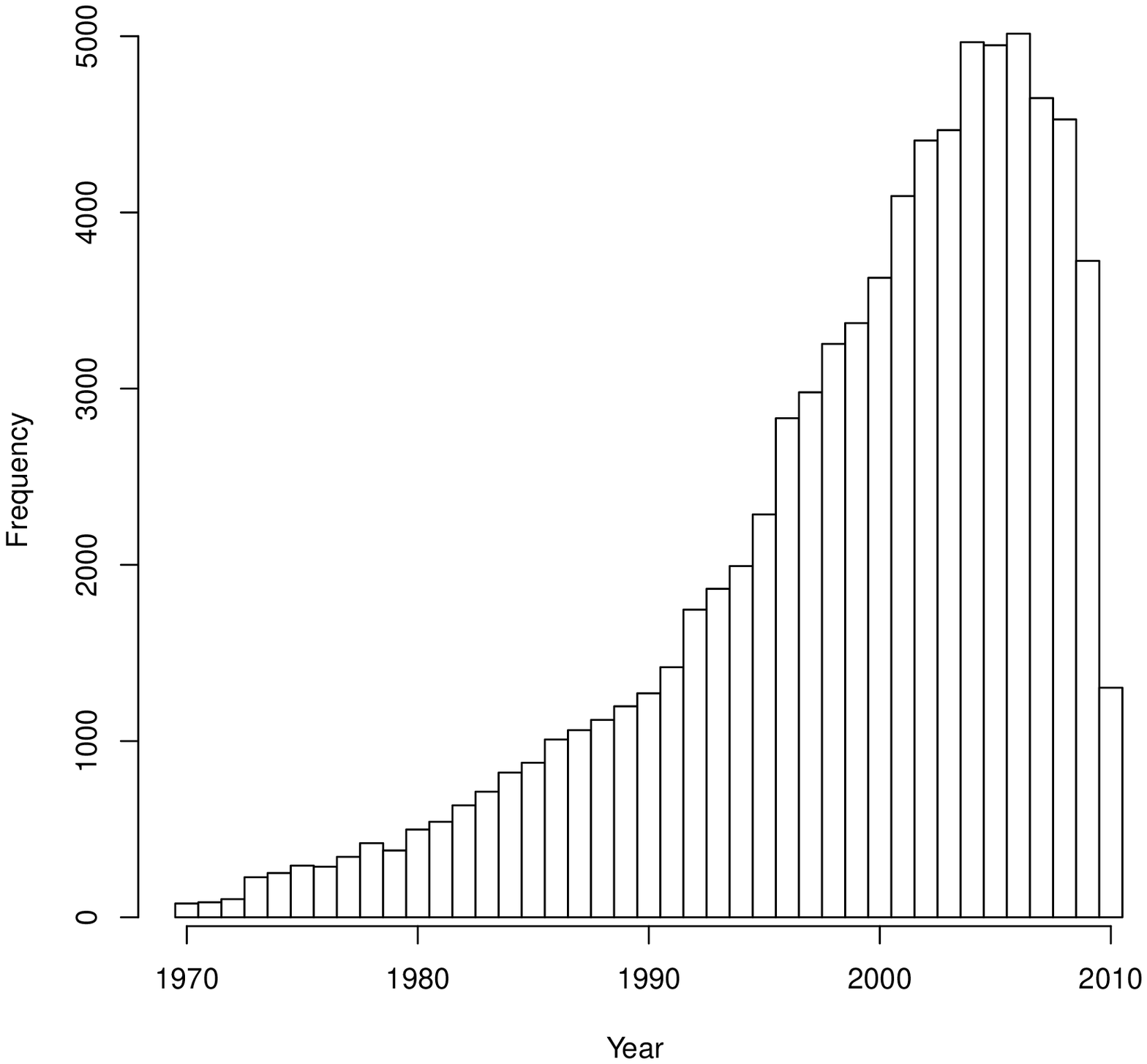}
\includegraphics[width=7cm]{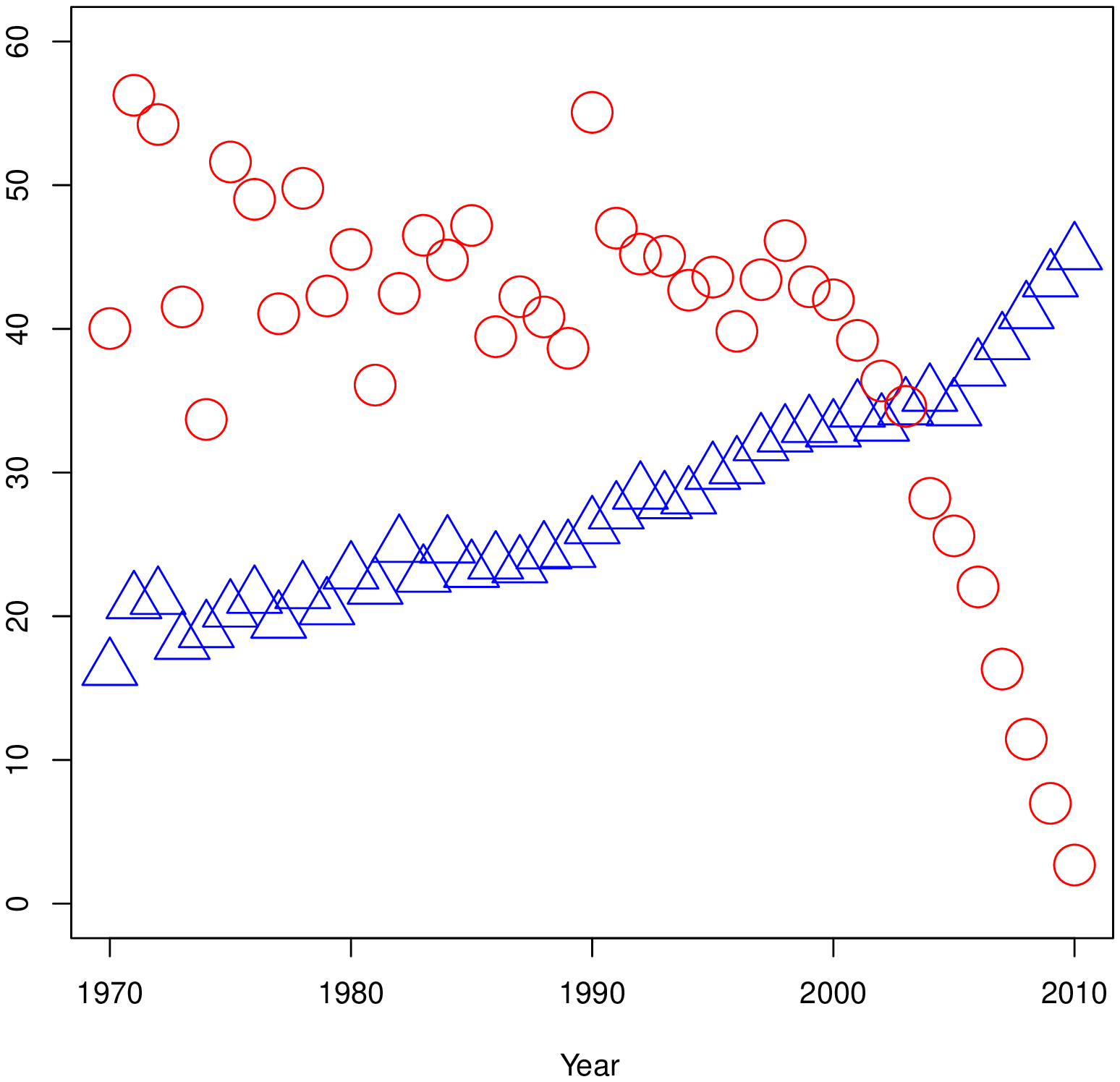}
\end{center}
\caption{On the left is a histogram of the number of papers published each year
with at least one author from the institute and with both a
positive citation count and a positive number of references,
$c,r>0$. On the right the average number of references $\langle r \rangle$ (blue triangles)
and the average number of citations $c_0=\langle c \rangle$ (red circles)
for publications with $c,r>0$ published in each year.}
\label{fnpapers}
\label{fncrav}
\end{figure}

The distribution of publications in our dataset $\Pcal$ is shown in Figure
\ref{fnpapers}. The data tails off markedly after 2008 and before the year
1996. This is due to local factors influencing the collection
of this data.
The behaviour of the citations and references is familiar from
elsewhere e.g.\ \cite{BMG04}.\tnote{Other references
here?} Given these variations in the data, our focus will be on
the data for 1997-2007.

\subsection{The $\cfindex$ measure for faculties}
\label{sCrownMeasure}

RFC \cite{Radicchi} showed that the relative bibliometric index,
$\cfindex$ \tref{crdef}, for individual papers published in a single
year and in a single field as defined by the Thomson Reuter categories, followed a
universal form which was well approximated by a lognormal
distribution with probability density
\begin{equation}
F(c_{\mathrm{f}}; \mu, \sigma^2) = \frac{1}{\sigma c_{\mathrm{f}} \sqrt{2 \pi}}
\exp \left\{ \frac{-[\log(c_{\mathrm{f}}) - \mu]^{2}}{2\sigma^{2}} \right\} \, .
\label{eLognormal}
\end{equation}
Since $\langle c_{\mathrm{f}} \rangle=1$ this leads to the constraint $\sigma^{2} = -2\mu$. If we use this and the normalisation constraint, we perform a one-parameter fit of the pdf of the data to\footnote{To be more precise we put our data for $c_f$ into bins with lower and upper boundaries $C(b)$ and $C(b+1)=r.c(b)$ where $r$ is a constant.  The smallest and largest value always fall in the middle of the first and last bins respectively.  The number of bins was chosen by hand to ensure a reasonable number of non-zero data points. We compare the actual count in each bin against the number expected to lie in that bin $\int_{C(b)}^{C(b+1)} F(c_{\mathrm{f}}; \mu=-\sigma^{2}/2, \sigma^2)$.  The points shown on plots correspond to value for a single bin, using the midpoint of the bins to locate the points horizontally.  Same approach used for other lognormal fits performed here.} $F(c_{\mathrm{f}}; \mu=-\sigma^{2}/2, \sigma^2)$. This was the approach used by RFC who found $\sigma^2$ to lie between 1.0 and 1.8
for the scientific fields considered with an average value of 1.3 \cite{Radicchi}.

\begin{figure*}
\begin{center}
\includegraphics[width=7cm]{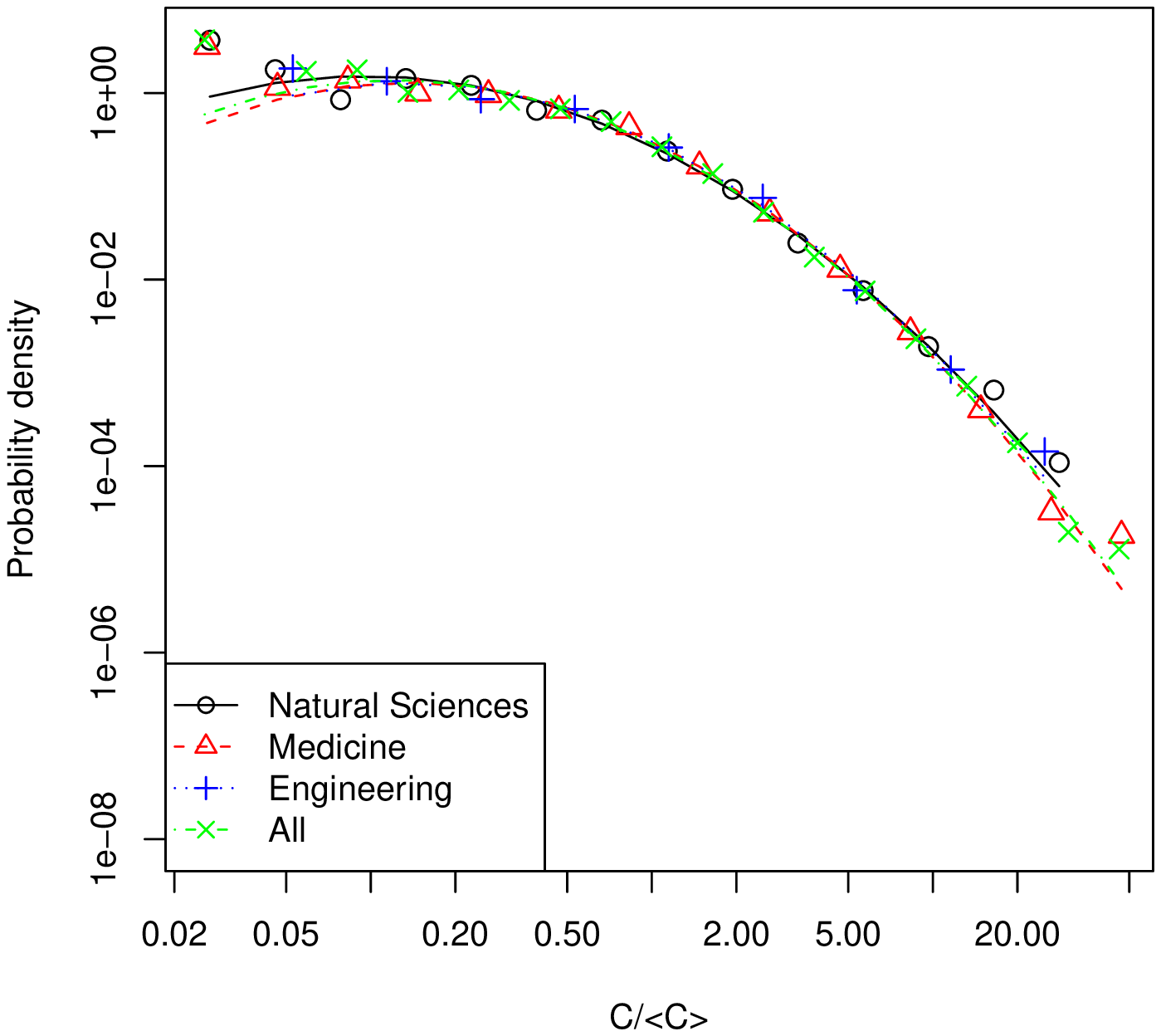}
\includegraphics[width=7cm]{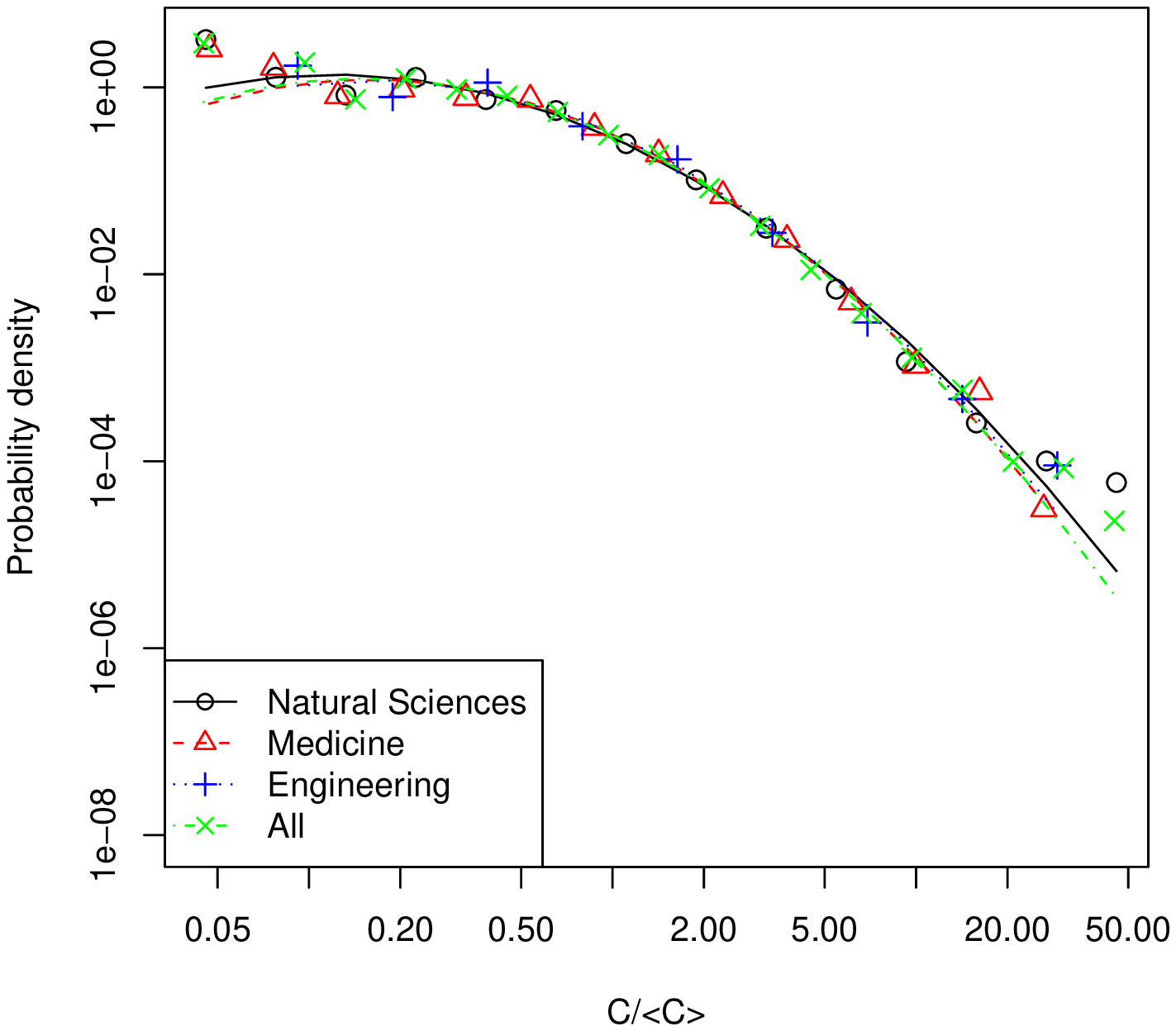}
\end{center}
\caption{The symbols show the distribution of $c_{\mathrm{f}}$ for
faculty data for all papers published in the year 2001 (left) or
in 2006 (right). The lines are the best fits to lognormal with one
free parameter. The values of $\sigma^{2}$ for Natural Sciences
(black solid line and circles), Medicine (red triangles and dashed
line) and Engineering (blue crosses and dotted line) respectively
were $1.49\pm0.10$, $1.34\pm0.06$, and $1.25\pm0.09$ for 2001, and
$1.38\pm0.08$, $1.19\pm0.08$, and $1.21\pm0.19$ for 2006.}
\label{fCnormdist}
\end{figure*}

Using the three faculties of Science (Medicine, Natural Sciences
and Engineering) and a single year of publication to define our research disciplines, our subsets $\Scal$ of
papers $\Pcal$, we found that we had between 389 and 4501 papers in each subset $\Scal$ (see table \ref{tRadtab1para} in the supplementary material) which proved sufficient to perform our analysis.

The data from our single institution
produces curves for $c_{\mathrm{f}}$ shown for a couple of typical
years in Figure \ref{fCnormdist}. These distributions are very
similar in shape to those found by RFC and we also
found that a lognormal with a single free parameter, $\sigma^{2}$,
was a good fit to the data for $c_{\mathrm{f}}$ from each faculty in any one
year. As in \cite{Radicchi} the small $c_{\mathrm{f}}$ head of the distribution
and extreme tail seem to fit the least well. For the large $c_{\mathrm{f}}$
values this may be attributed to statistical errors caused by
having fewer heavily cited publications while the lower $c_{\mathrm{f}}$
suggest a systematic deviation from the lognormal distribution\footnote{Lognormal can only be an approximation to true behaviour for low $\cfindex$ as it does not include uncited publications}.
A $\chi^2$ goodness of fit test applied to the
single parameter distribution resulted in $\chi^2$ values per degree of freedom ranging
from 2.91 to 38.4 with a mean value of 15.1. See table \ref{tabChi2fac} in the supplementary material for
$\chi^2$ values for each number of bins used in grouping the data.

The predominant source of discrepancy here, also visible in \cite{Radicchi}, was caused by publications with very
low citation counts, i.e.\ roughly those with less than 10\% of the
mean citation counts for a given faculty. The number of papers with low citation counts can be an order of magnitude higher than suggested by the lognormal curves.  With large numbers of such items, this is not a problem of low statistics.
We suggest that the dominant processes leading to citation of an item with an ultimately low citation count are different from the processes prevailing at higher citation counts.  We found that the meeting abstracts in particular were numerous yet had far lower citation counts (most were already removed since we studied only papers with a non-zero number of citations). Thus one explanation for the change in behaviour at low citation count is that it is due to the way different types of publication are cited coupled with the fact that the relative proportions of different types of item is different between low cited items and medium/high cited items.   This would not explain the same low citation issue seen in \cite{Radicchi} as they limit their data to articles and letters. Alternatively, or perhaps in addition, a larger proportion of citations may be self-citations for low cited articles and self-citation processes are likely to be different. Finally errors in data collection may lead to several records associated with one publication, and often all but one of these will have just one or two citations \cite{B77}.  Again this will cause most distortion for low cited publications.

To deal with the low citation issue\footnote{In \cite{Radicchi} papers with zero citations are excluded but otherwise all articles and reviews (as classified by WoS) are included in their analysis. Lundberg \cite{L07} uses $\ln(c+1)$ to avoid problems with zero citation count.},
we only fitted the lognormal to data above a minimum cutoff of $c_{\mathrm{f}}>0.1$. The value of 0.1 reflected a compromise between goodness of
fit and including as much data as possible, with 88\% of
publications in our data set used in the fits. The resulting $\chi^2$
values per degree of freedom were between 1.47 and 24.4 with an average of 3.98.
\begin{figure}[htbp]
\begin{center}
\includegraphics[width=7cm]{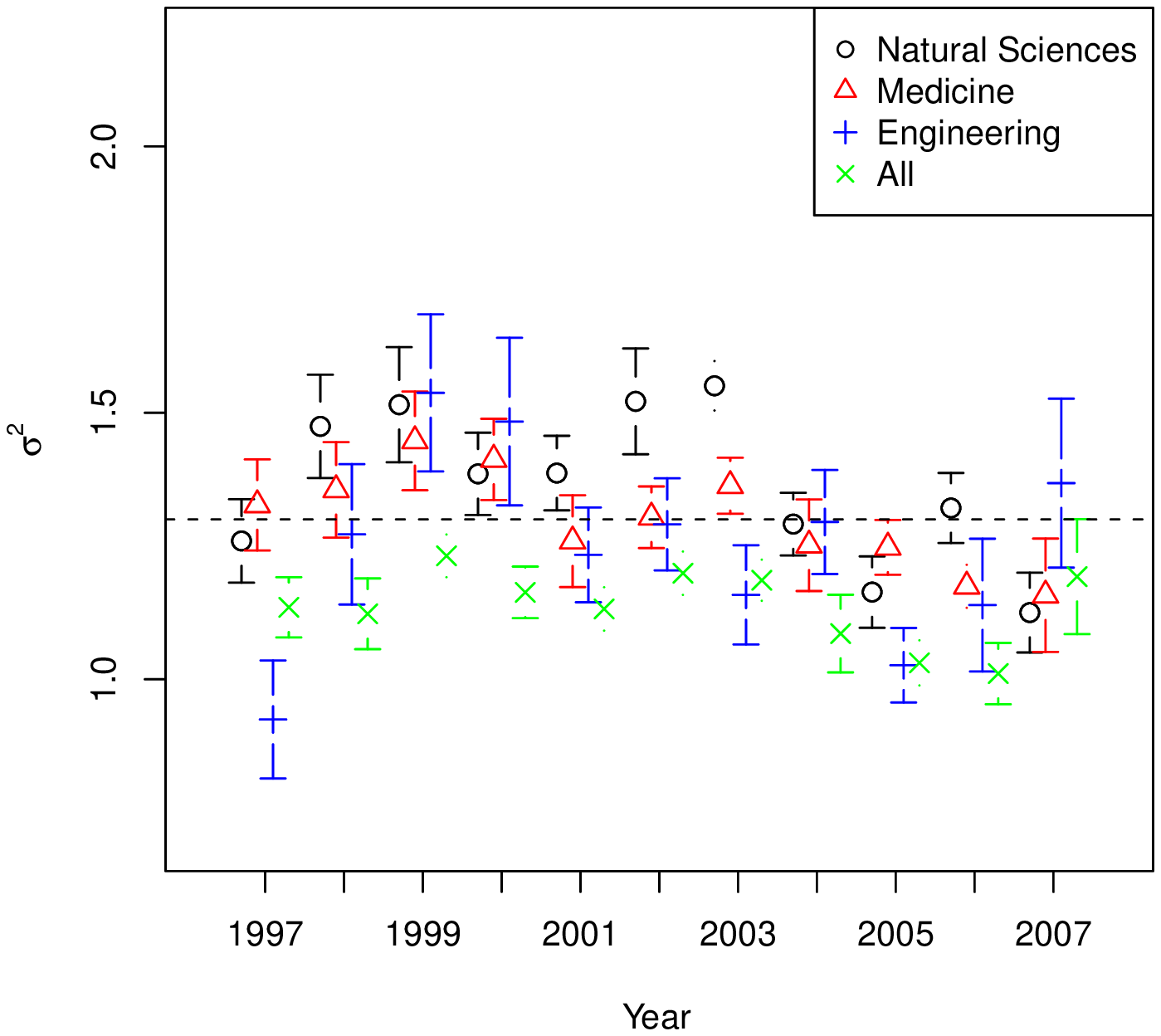}
\includegraphics[width=7cm]{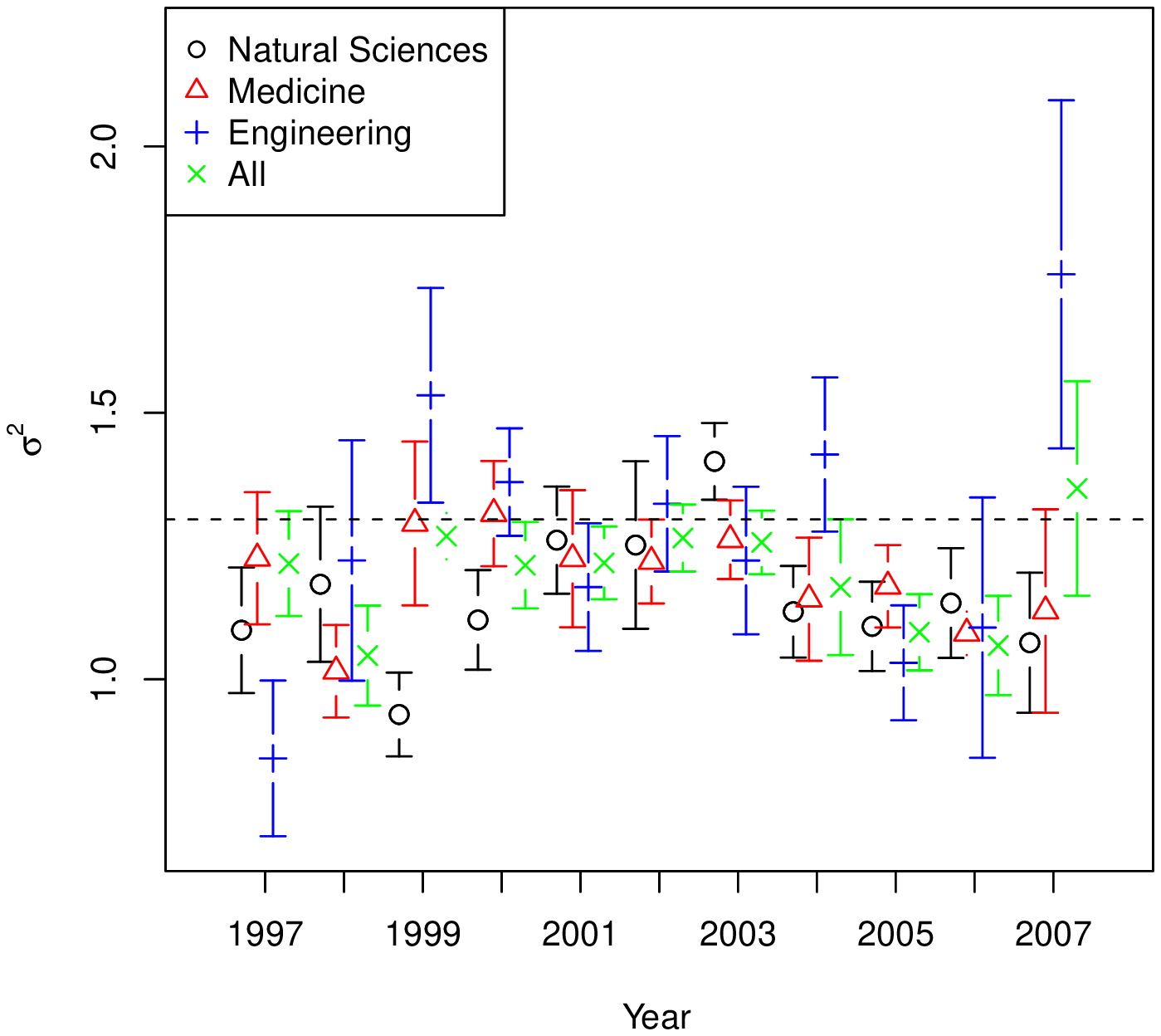}
\end{center}
\caption{A plot of $\sigma^2$ against year resulting from a one
(left) or three (right) parameter fit of a lognormal to the
$\cfindex$ measure.  Done for papers published in a single year from
each science faculty separately with Natural Sciences (black
circles), Medicine (red triangles) and Engineering (blue crosses).
The dashed line indicates the universal value 1.3 suggested by
RFC while the arithmetic average of all our results gives $1.44
\pm 0.13$ from the one parameter fit.  The data labelled All (green crosses) was found by taking the $\cfindex$ for each paper, using the $\czero$ value appropriate to the faculty and year of publication, and fitting a single lognormal to the whole dataset.} \label{fsigmaC1and3}
\end{figure}

For the years 1997 - 2007, the values of $\sigma^2$ are shown in the left
hand plot of Figure \ref{fsigmaC1and3}. We found this to range from $0.92\pm 0.11$ (Engineering in 1997)
to $1.56\pm 0.06$ (Natural Sciences in 1999). The average values for
$\sigma^2$ across all these years for each faculty were
$1.36\pm0.09$, $1.30\pm0.08$ and $1.25\pm0.13$ for Natural
Sciences, Medicine and Engineering respectively. A simple arithmetic average gives $1.3\pm0.1$.\bnote{I calculate this global average to be $1.14\pm0.06$ - surprisingly lower than all the other values.}
The coincidence of the results across all three faculties is
striking, especially as we have found that the average citation counts
for the three faculties is quite different, matching what has been
seen in other studies including \cite{Radicchi} with Medicine being
higher than Natural Sciences and Engineering having the lowest
citation average. Likewise the disciplines are ranked in the same
way in terms of the number of papers produced, Engineering has half
the number of papers as Natural Science and a third the number of
Medicine in each year.

Thus despite using a much broader definition of scientific field
with a much narrower selection of papers, those from one
institute, we find the same type of universality as RFC. Notwithstanding the differences in the subset $\Pcal$ being used in the two studies the universal values for $\sigma^2$, $1.3(1)$ for us, $1.3$ in
\cite{Radicchi} are in encouraging agreement. Alternatively we can create a weighted average by fitting a lognormal to the $\cfindex$ values for all papers published in a single year, using the $\czero$ value appropriate to the faculty and year of publication.  This gives points labelled `All' in Figure \ref{fsigmaC1and3} with values of $\sigma^2$ a little lower, around $1.2$ though still statistically consistent with our other values.

\begin{figure}[htbp]
\begin{center}
\includegraphics[width=7cm]{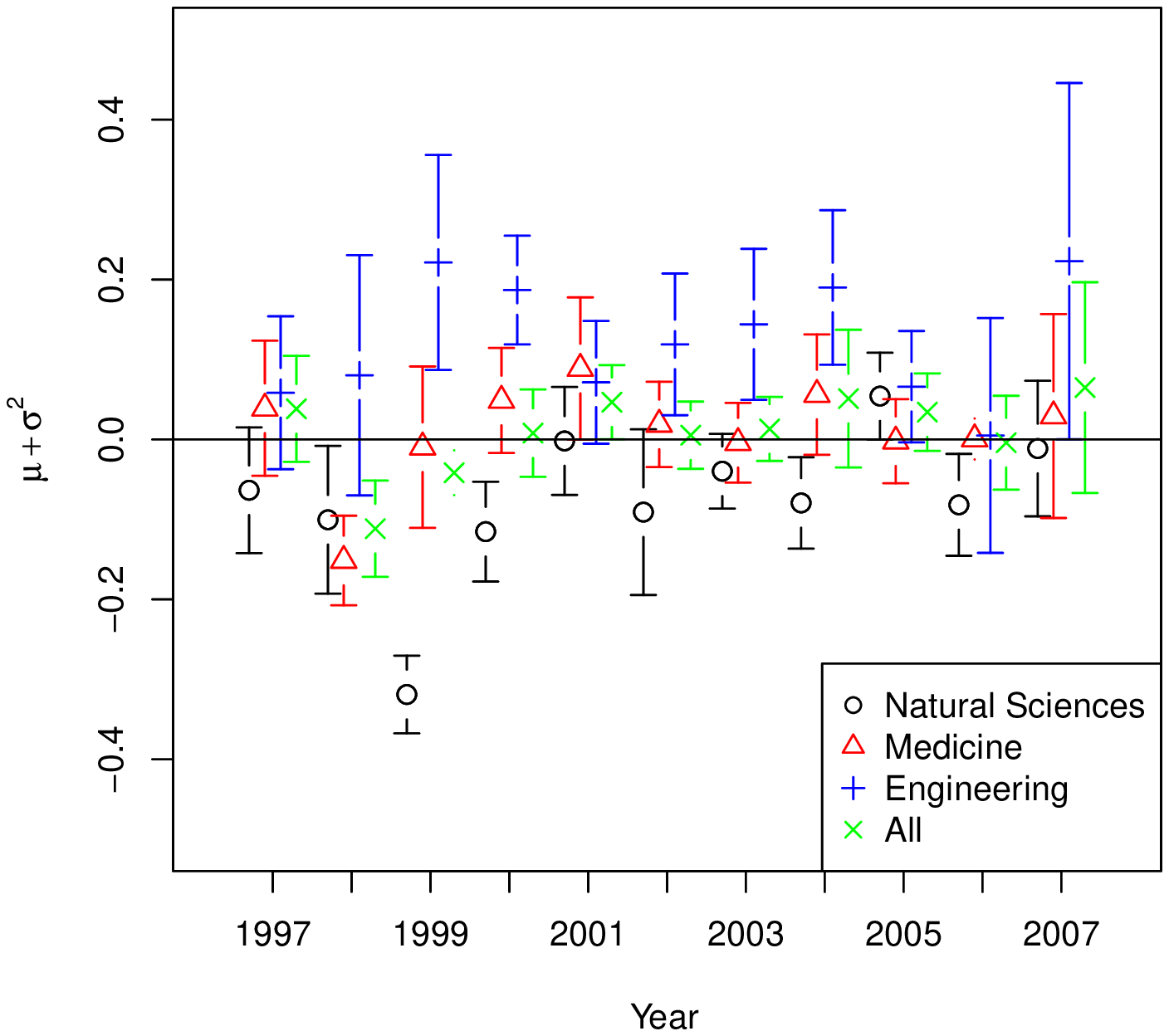}
\includegraphics[width=7cm]{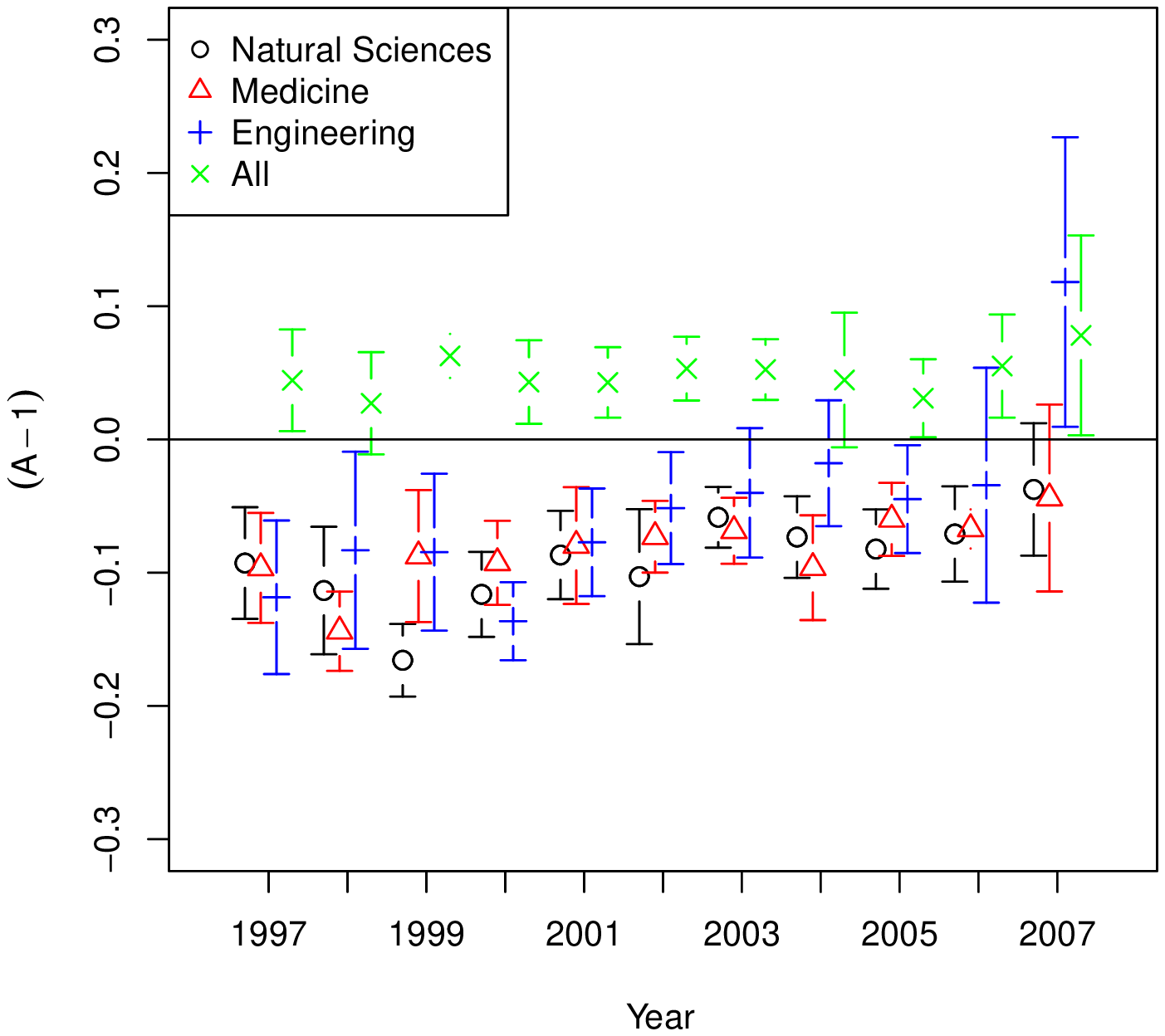}
\end{center}
\caption{A plot of $(\mu+\sigma^2/2)$ (left) and $(A-1)$ (right)
against year obtained by fitting a lognormal to the
$\cfindex$ measure for which zero is expected for both quantities.
For papers published in a single year
from each science faculty separately with
Natural Sciences (black circles), Medicine (red triangles) and
Engineering (blue crosses).}
\label{fdmudAC}
\end{figure}

As a check on our fitting, we also fitted our data to
$A \cdot F(\cfindex;\mu,\sigma^2)$, a lognormal with three independent
parameters, $\sigma^2$, $\mu$ and the overall normalisation $A$.
The values of $\sigma^2$ we obtain are equivalent statistically to
the values from our one parameter fit\footnote{The arithmetic averages for
Natural Sciences, Medicine and Engineering are respectively $1.15
\pm 0.11$, $1.19 \pm 0.12$, and $1.27 \pm 0.20$ giving an overall
average of $1.21 \pm 0.14$.}. Since $\langle \cfindex\rangle=1$ by
definition, the value of $(\mu + \frac{\sigma^2}{2})$
should be zero if the data for $\cfindex$ fits a lognormal
distribution. The normalisation $A$ should be unity by construction.
Figure \ref{fdmudAC} shows a plot of $(\mu + \frac{\sigma^2}{2})$
and $(A-1)$ against year for our data using the faculties to
define $\Pcal$ and our research disciplines. These values are consistent with
zero, confirming that the lognormal is a good fit.

\subsection{The $\crindex$ measure for faculties}

\begin{figure*}
\begin{center}
\includegraphics[width=7cm]{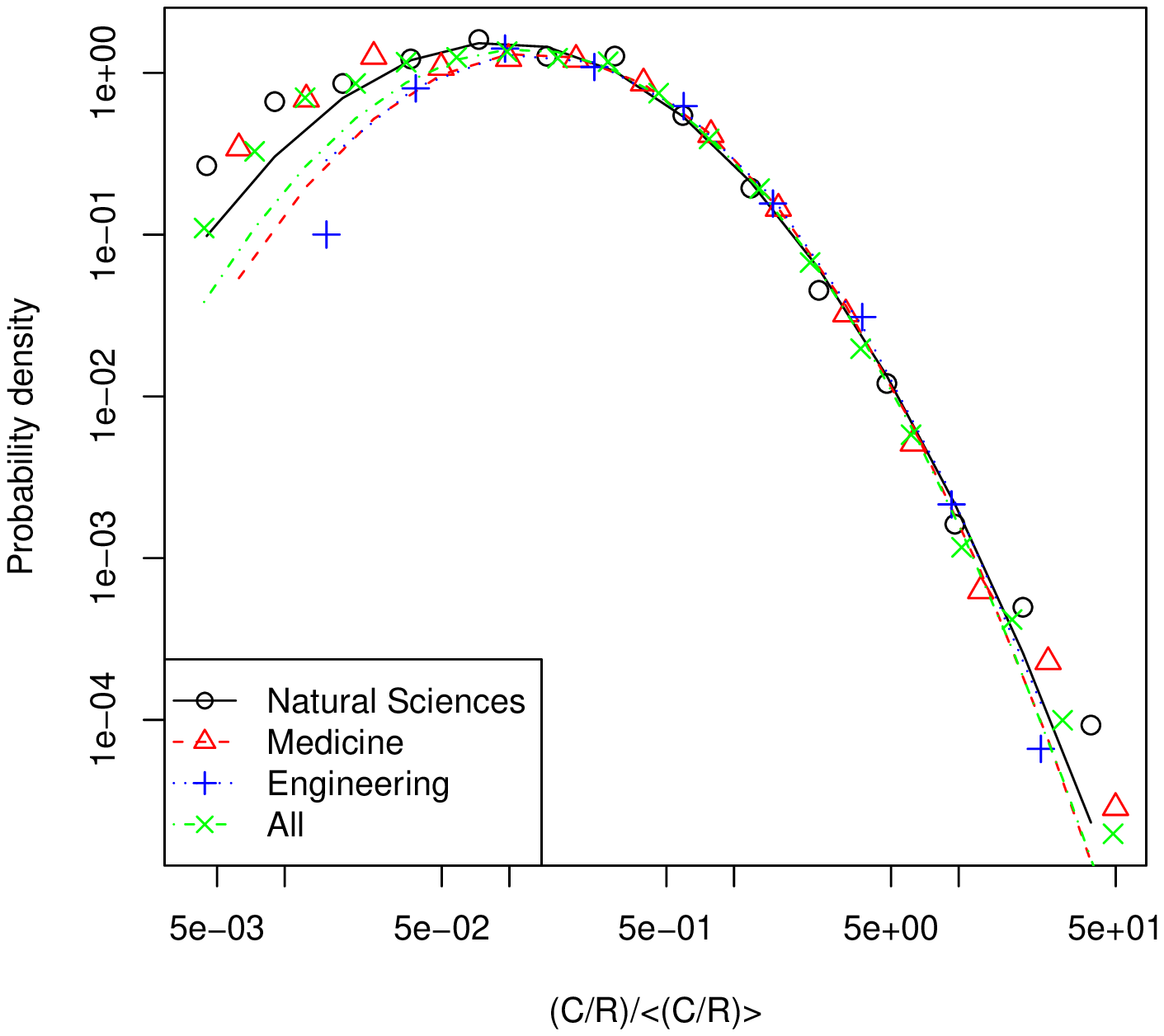}
\includegraphics[width=7cm]{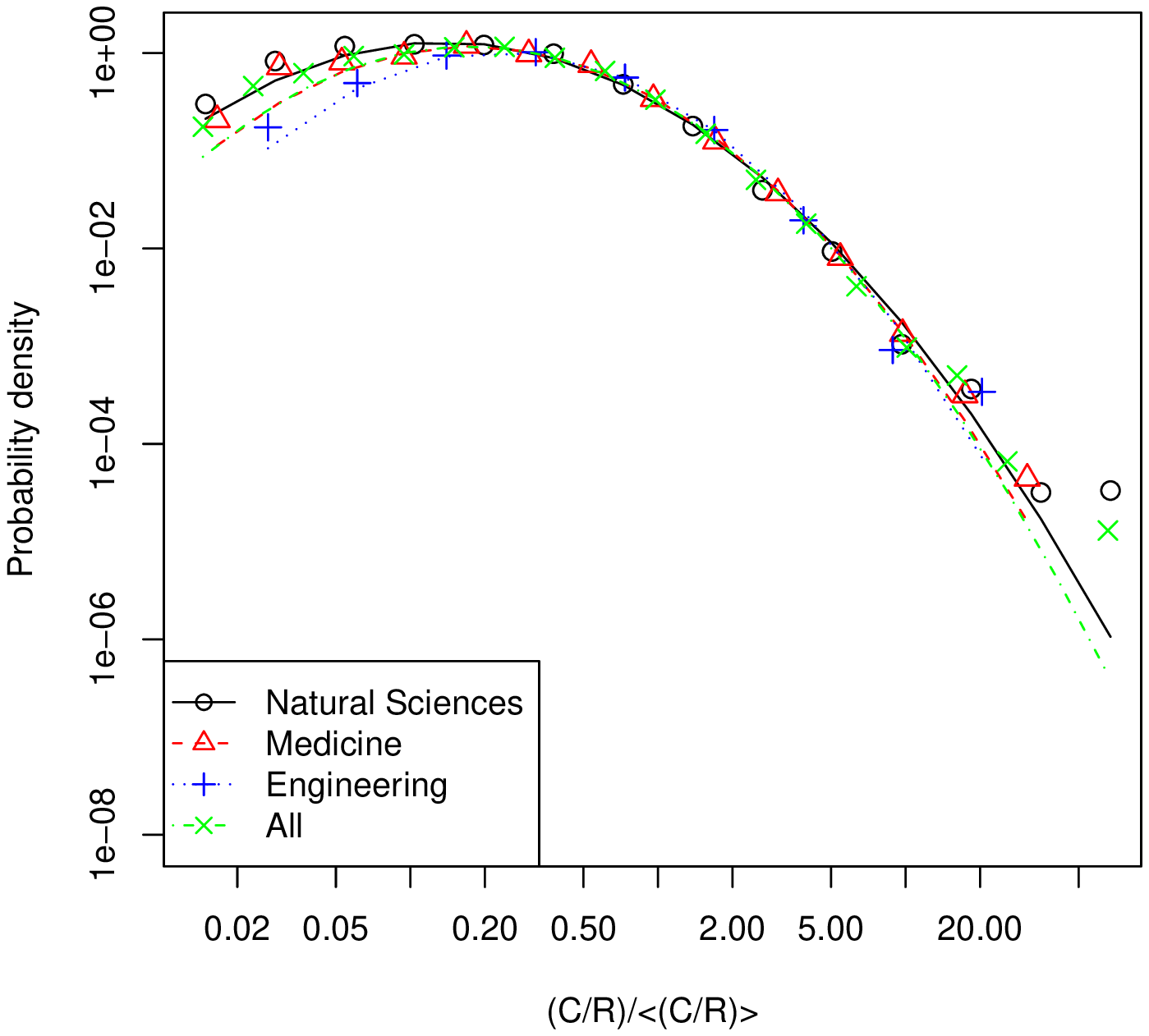}
\end{center}
\caption{The symbols show the distribution of $c_{\mathrm{r}}$ for
the papers published in 2001 (left) or 2006 (right) from each
science faculty. The lines are the best fits to a lognormal with
one free parameter. The values of $\sigma^{2}$ for Natural
Sciences (black solid line and circles), Medicine (red triangles
and dashed line) and Engineering (blue crosses and dotted line),
respectively were $1.65\pm0.10$, $1.37\pm0.05$, and $1.40\pm0.06$ for 2001, and $1.33\pm0.06$ ,
$1.17\pm0.04$ , and $0.98\pm0.02$  for 2006.} \label{fCRnormdist}
\end{figure*}

We also calculated our adjusted measure of $c_{\mathrm{r}}$ \tref{crdef} for
papers published in one year from one faculty, the same dataset $\Scal$
used in Figure \ref{fCRnormdist}. Again a lognormal of the form \tref{eLognormal} provided a good fit with
one or three free parameters; examples are shown in Figure
\ref{fCRnormdist}.
One difference is that with $c_{\mathrm{r}}$ we get a considerable number of
points to the left of the peak whereas with $c_{\mathrm{f}}$ in both
\cite{Radicchi} and Figure \ref{fCnormdist} only the peak of the lognormal parabola and points to its right
are seen.

The values of $\sigma^2$ obtained by fitting $c_{\mathrm{r}}$ to the
different subsets of papers $\Pcal$ are shown in Figure
\ref{fsigmaC1and3}, for both one and three parameter fits. There was no
marked improvement in goodness of fit when a cutoff was imposed, so all
publications were included in the fit resulting in an average $\chi^2$ of
5.31 per degree of freedom for the one parameter fit. The goodness of fit
data for each bin size computed are given in table \ref{tabChi2fac} in the supplementary material.
\begin{figure}[htbp]
\begin{center}
\includegraphics[width=7cm]{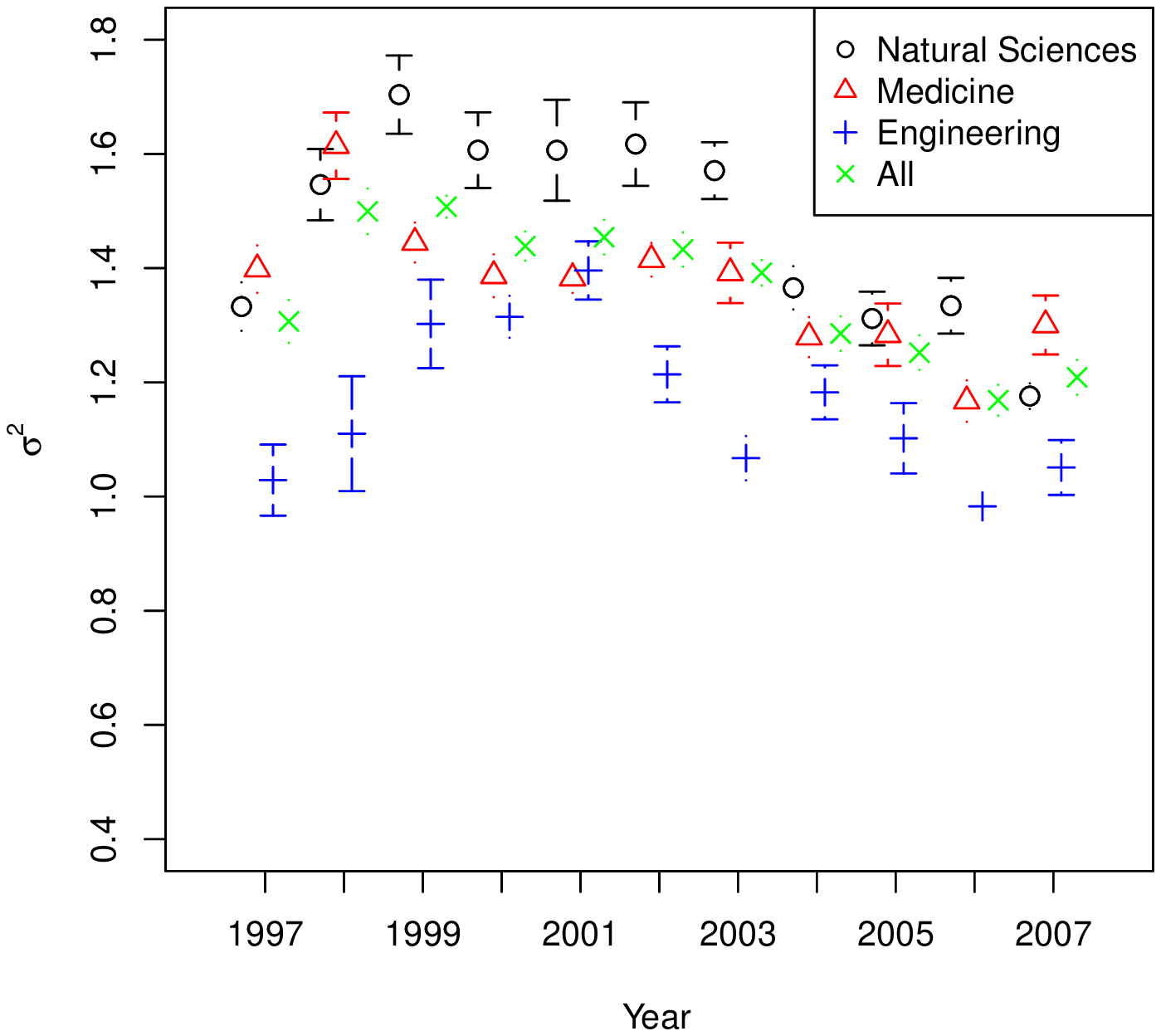}
\includegraphics[width=7cm]{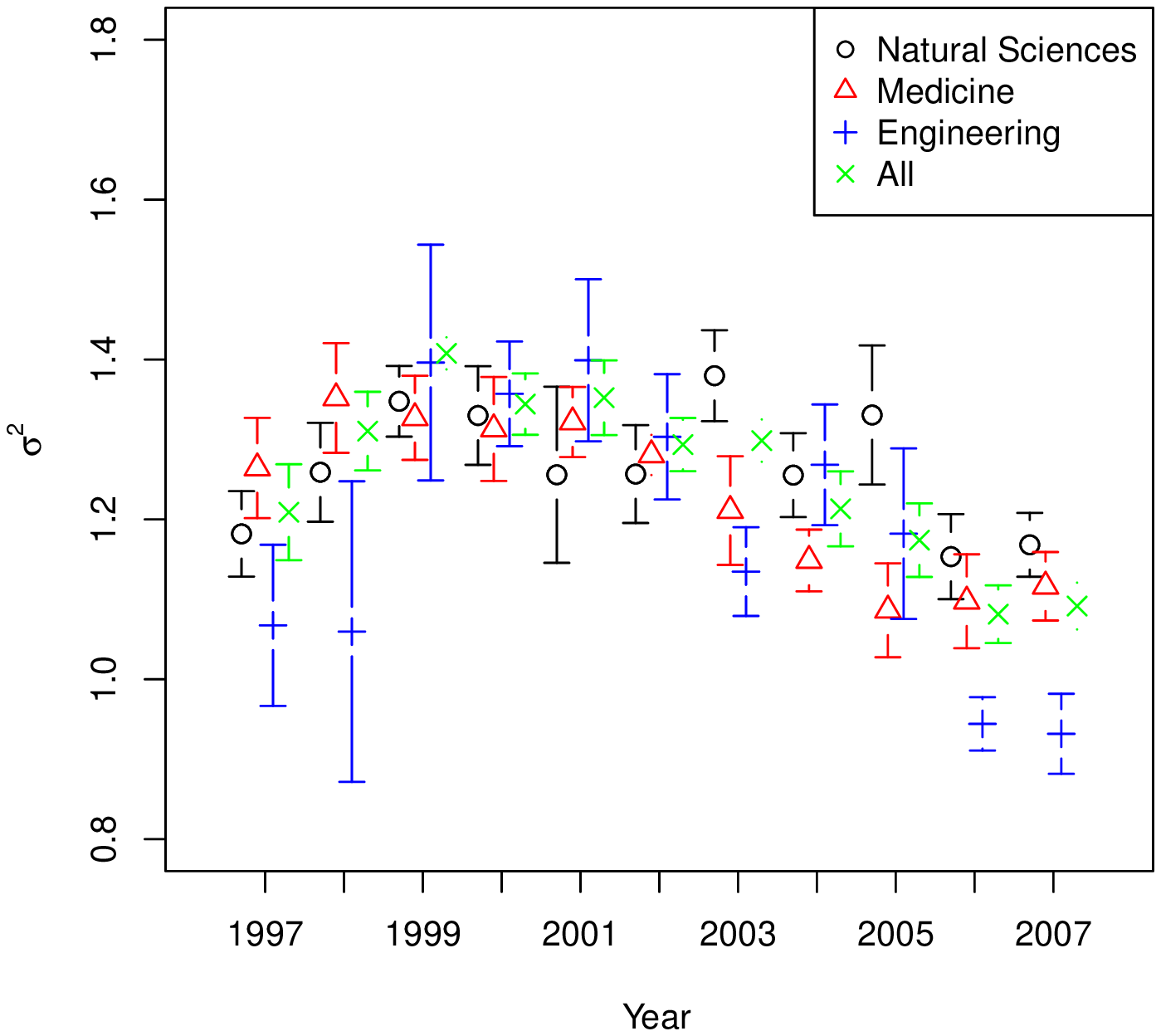}
\end{center}
\caption{A plot of $\sigma^2$ against year resulting from a one (left) or
three (right) parameter fit of a lognormal to the $\crindex$ measure.
Error bars are for one standard deviation.
The papers used for each point are published in a single year
from one science faculty:
Natural Sciences (black circles), Medicine (red triangles) or
Engineering (blue crosses). }
\label{fsigmaCR1and3}
\end{figure}
Considering the results for the one parameter fit first, we find
that the average over all years for the $\sigma^2$ of Natural
Sciences, Medicine and Engineering are respectively $1.47\pm
0.07$, $1.37 \pm 0.05$, and $1.16\pm 0.06$. The results suggest a
universal value for $\sigma^2$ of $1.33\pm0.06$.

For the one parameter fit, the Natural Sciences values for
$\sigma^2$ are either similar to or higher than those for
papers from the Medicine faculty. Both are invariably higher
than the Engineering faculty $\sigma^2$ results. In most years some of these values of $\sigma^2$ are three or more standard deviations apart.

\begin{figure}[htbp]
\begin{center}
\includegraphics[width=7cm]{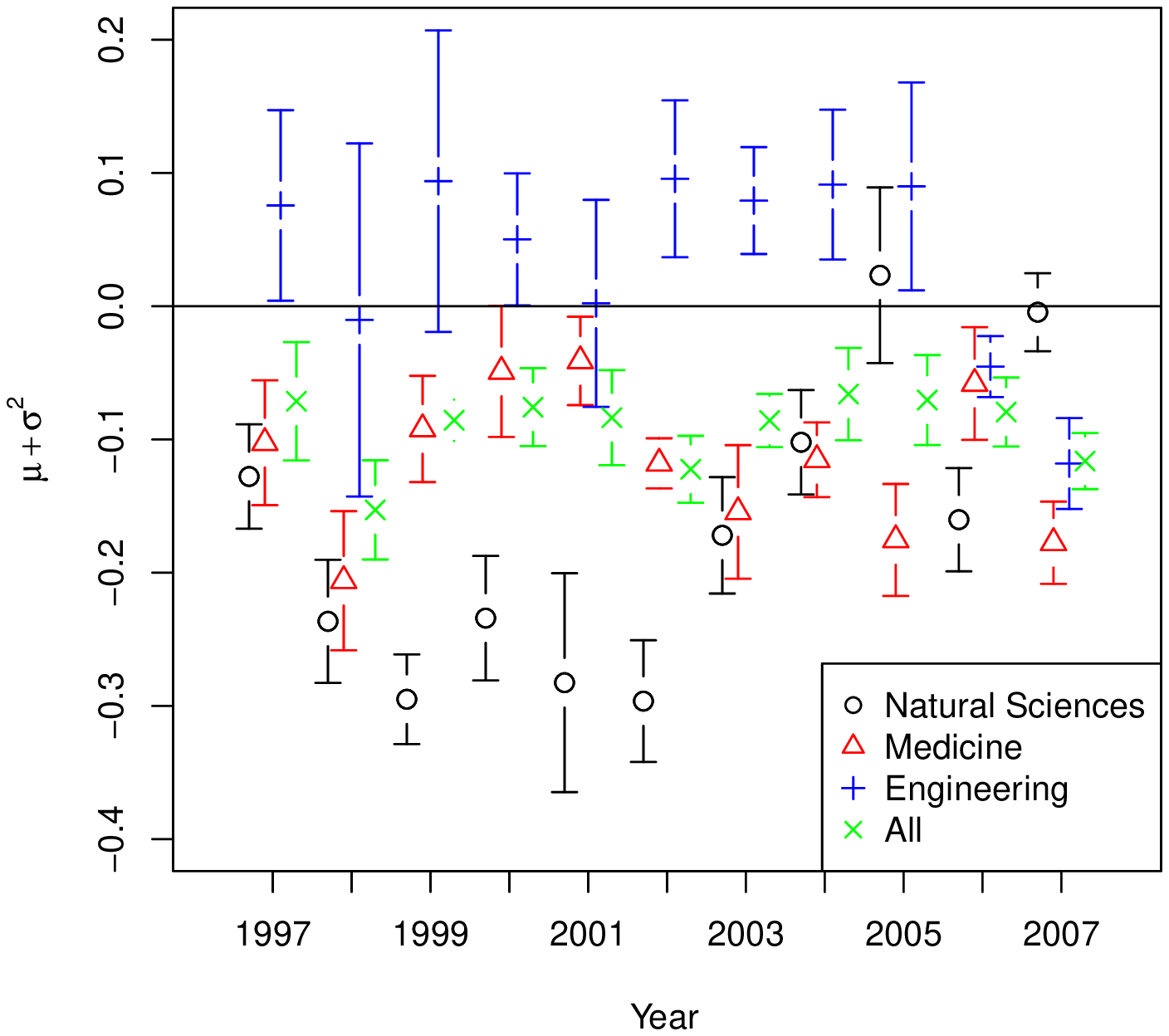}
\includegraphics[width=7cm]{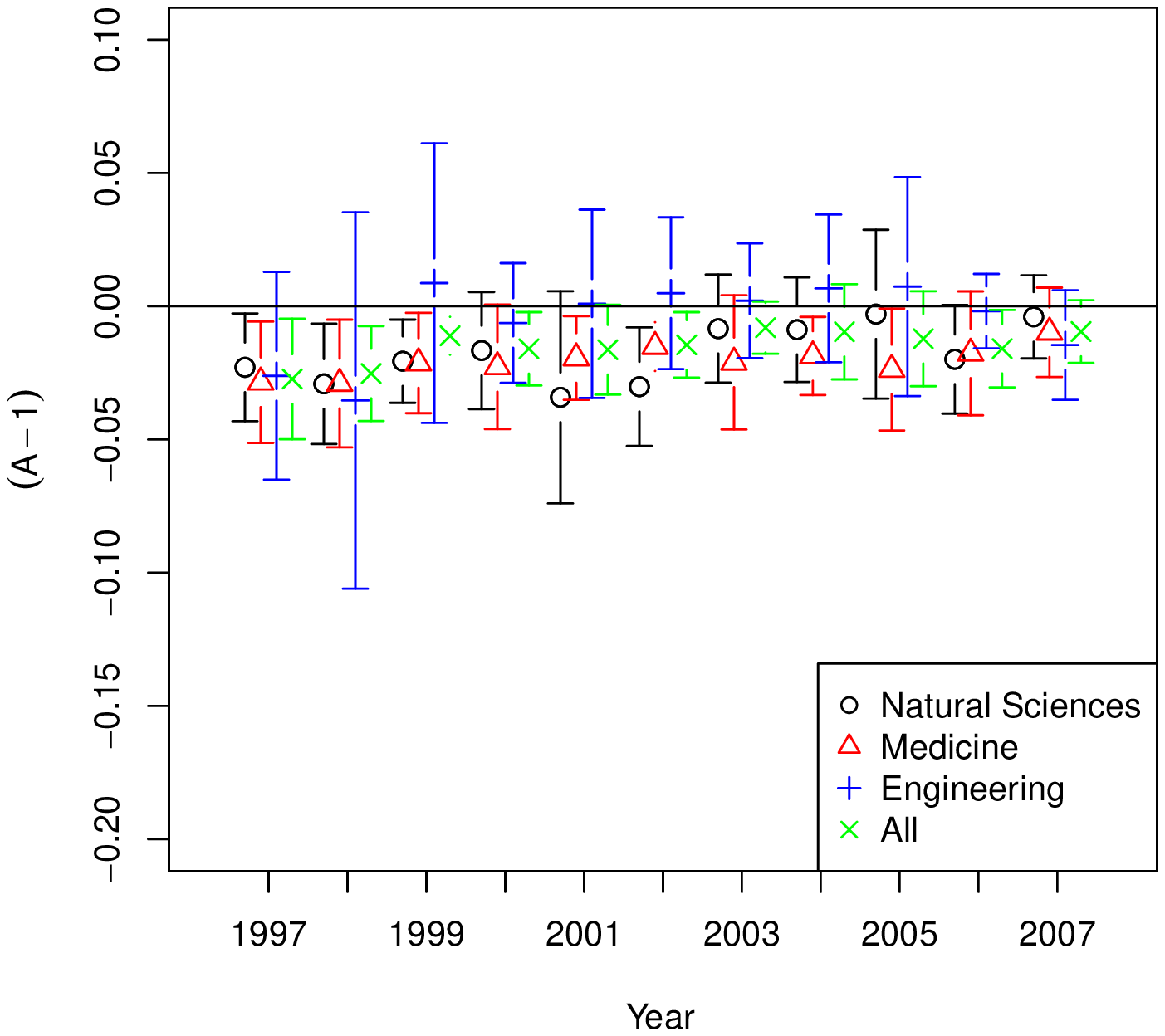}
\end{center}
\caption{A plot of $(\mu+\sigma^2/2)$ (left) and $(A-1)$ (right)
against year obtained by fitting a lognormal to the
$\cfindex$ measure for which zero is expected for both quantities.
For papers published in a single year
from each science faculty separately with
Natural Sciences (black circles), Medicine (red triangles) and
Engineering (blue crosses).}
\label{fdmudACR}
\end{figure}
On checking $\crindex$ data with a three parameter fit, the values of
$\sigma^2$ are now found to be consistent at each year\footnote{The
averages for Natural Sciences, Medicine and Engineering are
respectively $1.27\pm 0.07$, $1.23 \pm 0.06$ and $1.19 \pm 0.10$
with the global average of $1.23 \pm0.08$.}. The normalisation is
also consistent with unity. The problem is now seen in the value of
$(\mu+\sigma^2/2)$ (see Figure \ref{fdmudACR}) which is now more
than three standard deviations away from zero for Medicine and/or
Natural Science in many years.  Thus while the $c_{\mathrm{r}}$ appears to have a universal distribution,
it is not best described by a lognormal form.

\tnote{It can be seen that the residual error per degree of freedom for the one parameter
fit is very similar to the error on the three parameter fit (see tables \ref{tOurtab1para} and
\ref{tOurtab3para} in the supplementary material).
It was also noted that the residual error per degree of freedom
was generally smaller for the $c_{\mathrm{r}}$ measure than
the $\cfindex$ measure. Therefore it can be concluded that
the $c_{\mathrm{r}}$ measure is more accurately described by a
lognormal distribution than the $c_{\mathrm{f}}$ indicator.}

\subsection{Comparison of $c_{\mathrm{f}}$ and $c_{\mathrm{r}}$ for faculties}\label{scomparison}

Since both the measures $\cfindex$ \tref{cfdef} and $\crindex$ \tref{crdef} lie on universal distributions, it is interesting to compare them.  We may factor out the statistically insignificant variations in $\sigma$ by working with
\beq
 \zfindex(s,\Scal,\Pcal) = \frac{\ln(\cfindex(s,\Scal,\Pcal))-\mu_f(\Scal,\Pcal)}{\sigma_f(\Scal,\Pcal)} \, ,
 \;\;
 \zrindex(s,\Scal,\Pcal) = \frac{\ln(\crindex(s,\Scal,\Pcal))-\mu_r(\Scal,\Pcal)}{\sigma_r(\Scal,\Pcal)} \, ,
 \qquad
 \;s \in \Scal  \, ,
 \label{zdef}
\eeq
where we will use abbreviations $\zfindex$ and $\zrindex$ when unambiguous.
Here $\mu_f(\Scal,\Pcal)$ and $\sigma_f(\Scal,\Pcal)$ are the mean and standard
deviation parameters obtained from fitting a lognormal curve to $\cfindex>0.1$ data as described above,
with equivalents for the $\crindex>0.1$ data. It is important to note that it is sensible to work with these indices $\zfindex$ and $\zrindex$ \tref{zdef} since they are defined in terms of the logarithms of the normalised indices,  $\ln(\cfindex)$ and $\ln(\crindex)$, where there is an approximate normal distribution.

\begin{figure}[htbp]
\begin{center}
\includegraphics[width=7cm]{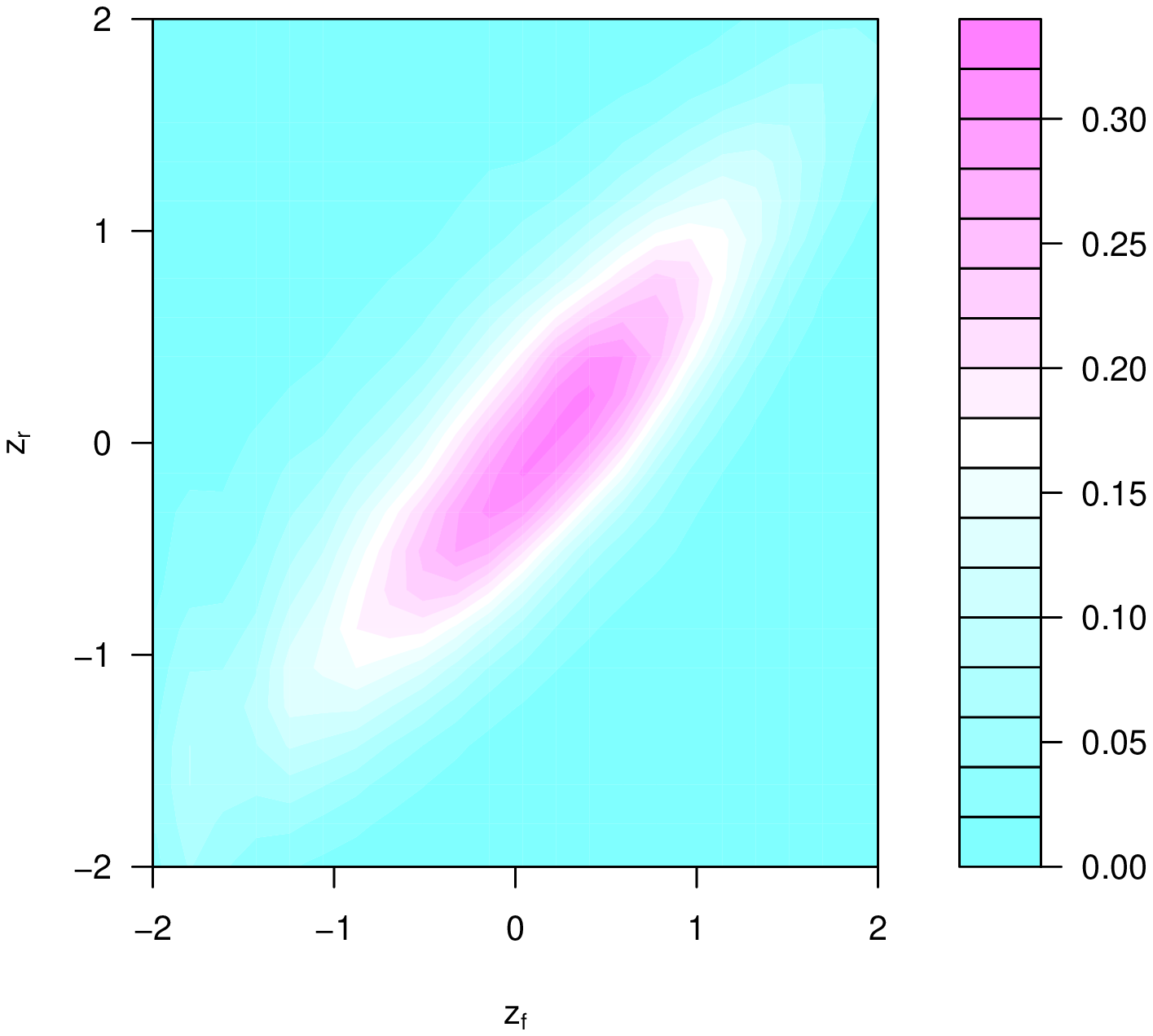}
\includegraphics[width=7cm]{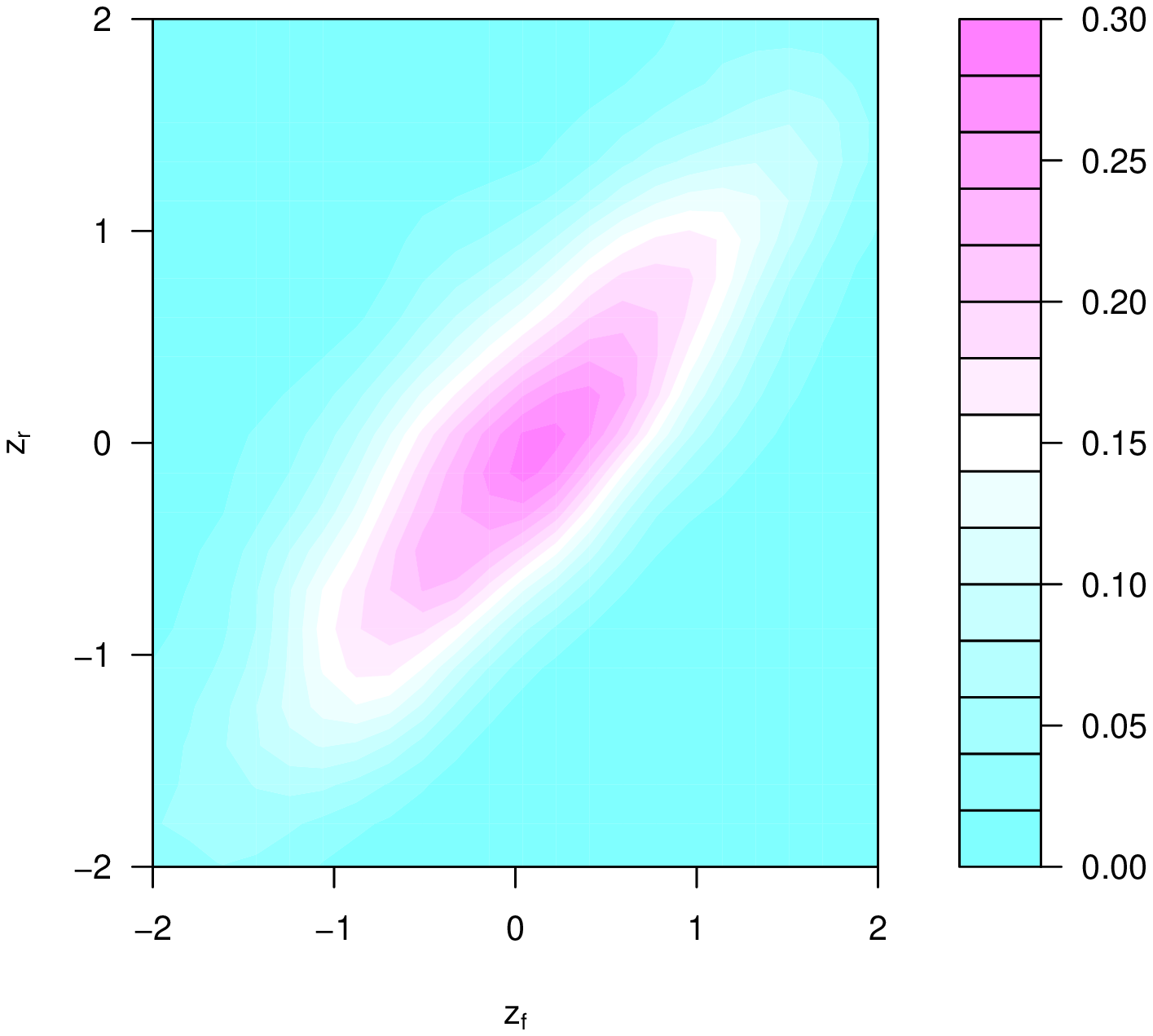}
\end{center}
\caption{Density plot of $\zfindex$ vs.\ $\zrindex$ of \tref{zdef} for all items (left)
and review articles only (right).}
\label{fZDensity}
\end{figure}
\begin{figure}[htbp]
\begin{center}
\includegraphics[width=7cm]{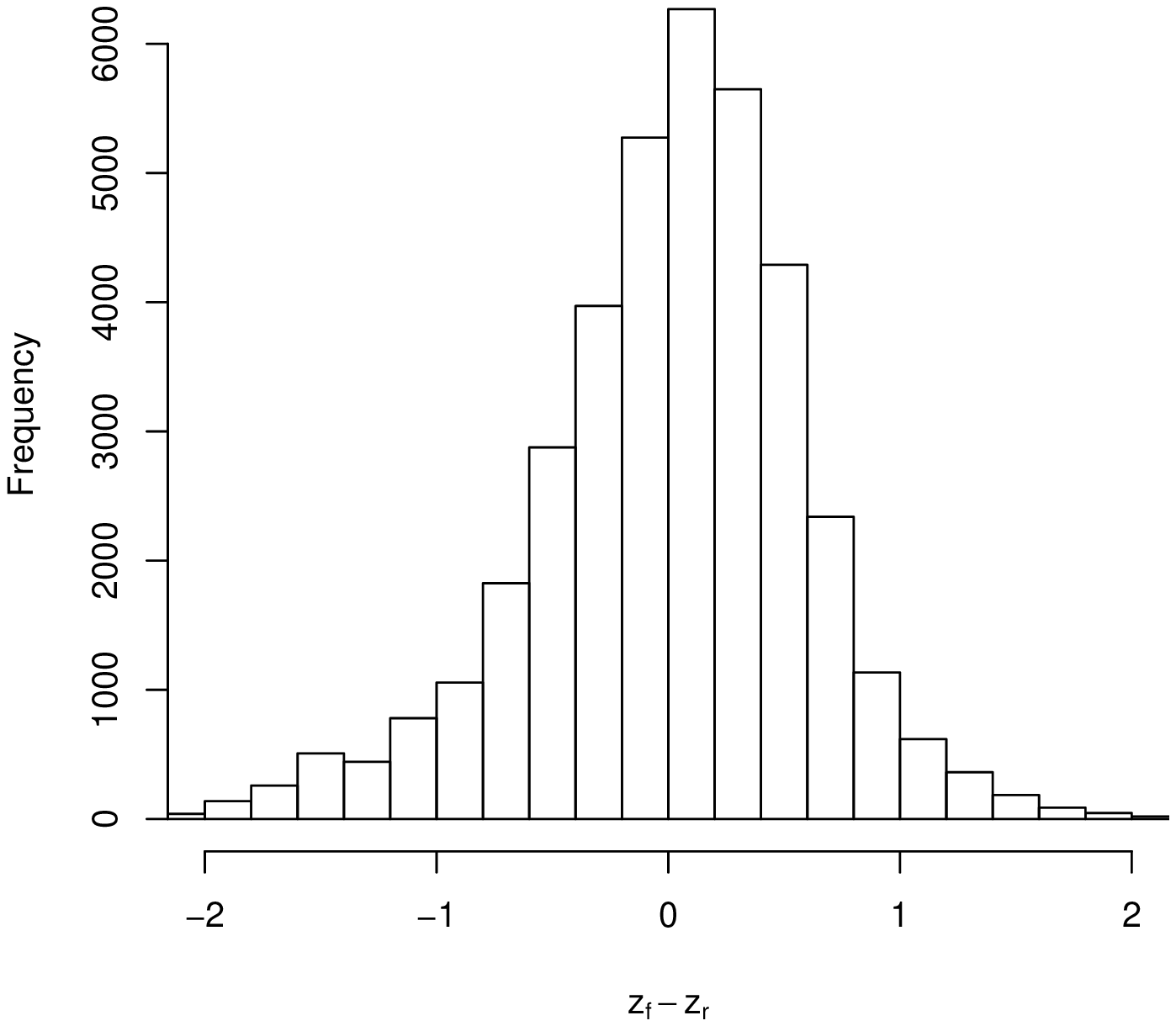}
\includegraphics[width=7cm]{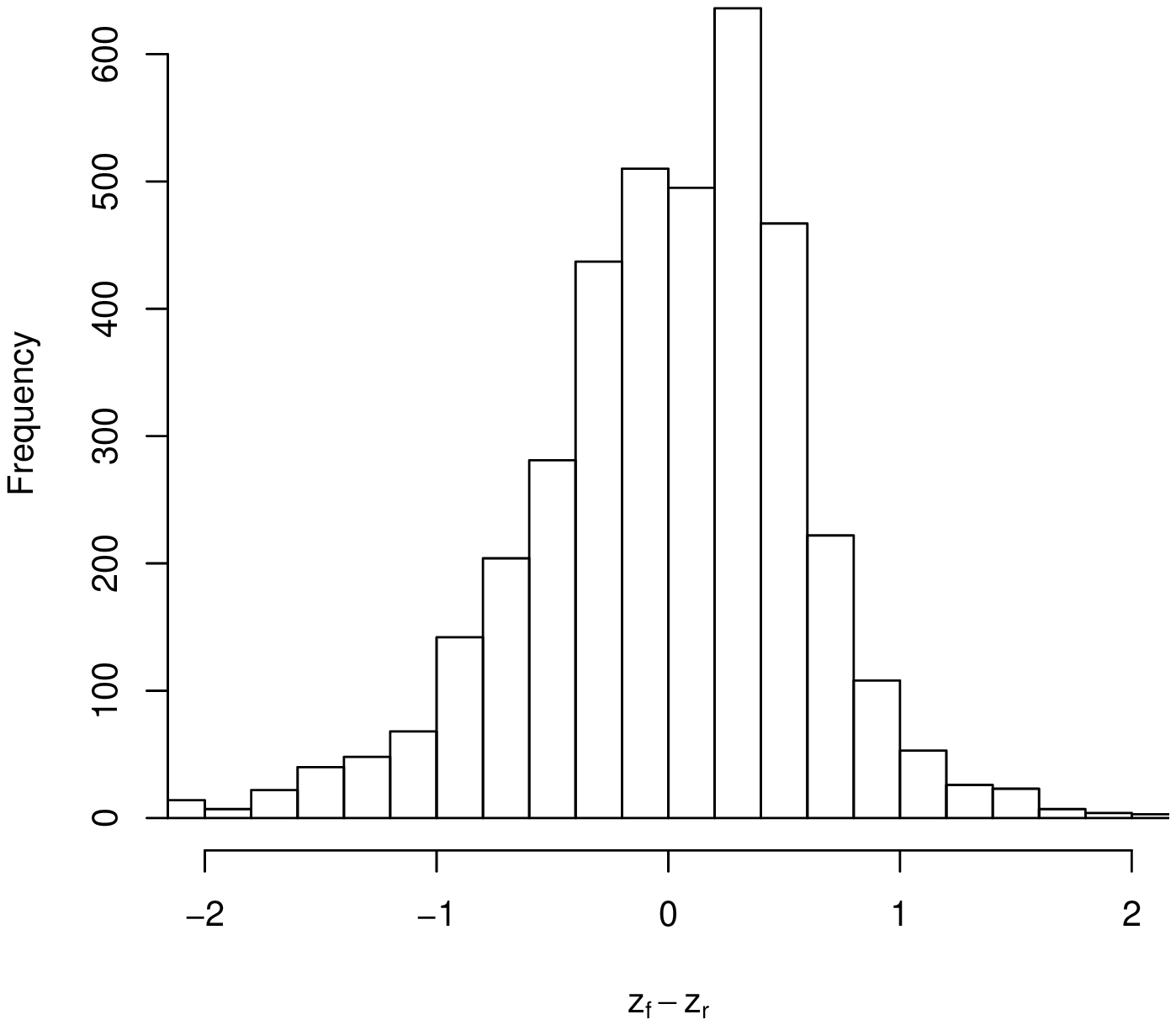}
\end{center}
\caption{Histograms of $\zfindex$ vs.\ $\zrindex$ of \tref{zdef} for all items (left)
and review articles only (right).}
\label{fZHist}
\end{figure}

The comparison of $\zfindex$ and $\zrindex$ in Figures \ref{fZDensity} and \ref{fZHist} shows that for the vast majority of the data, the difference between $\zfindex$ and $\zrindex$ is less than one.  If we restrict ourselves to just review papers, as defined by WoS, we expect a larger difference since reviews have a higher than average number of references.  While there is now some difference between $\zfindex$ and $\zrindex$ it is still less than one. As can be seen in Figures \ref{fZDensity} and \ref{fZHist} (see also Figure \ref{fZCorrel} in supplementary material) there does not appear to be any significant difference between the two measures.

\subsection{Departments}

The data set for the institute was also analysed using the
departments to define the research discipline of a paper and our
subset $\Pcal$. As some departments were found not to publish
enough papers per year to draw statistically significant conclusions,
it was instead decided to focus on the two most prolific departments
from each of the faculties, taking papers published in three consecutive years rather than in one single year.
This produced subsets $\Scal$ of between 209 and 1643 publications (see table \ref{tRadDepttab1para} in supplementary material).
The single parameter lognormal distribution produced a reasonable fit when all publications were
included with $\chi^2$ values per degree of freedom ranging from
2.10 to 63.8 with an average value of 17.7. If we repeat the fit but only on publications with a reasonable number of citations, that for $c_{\mathrm{f}} > 0.1$, the
goodness of fit was greatly improved with $\chi^2$ per degree
of freedom subsequently ranging from 1.06 to 55.8 with a mean of
6.98 for the $c_{\mathrm{f}}$ measure.
\begin{figure}[htbp]
\begin{center}
\includegraphics[width=7cm]{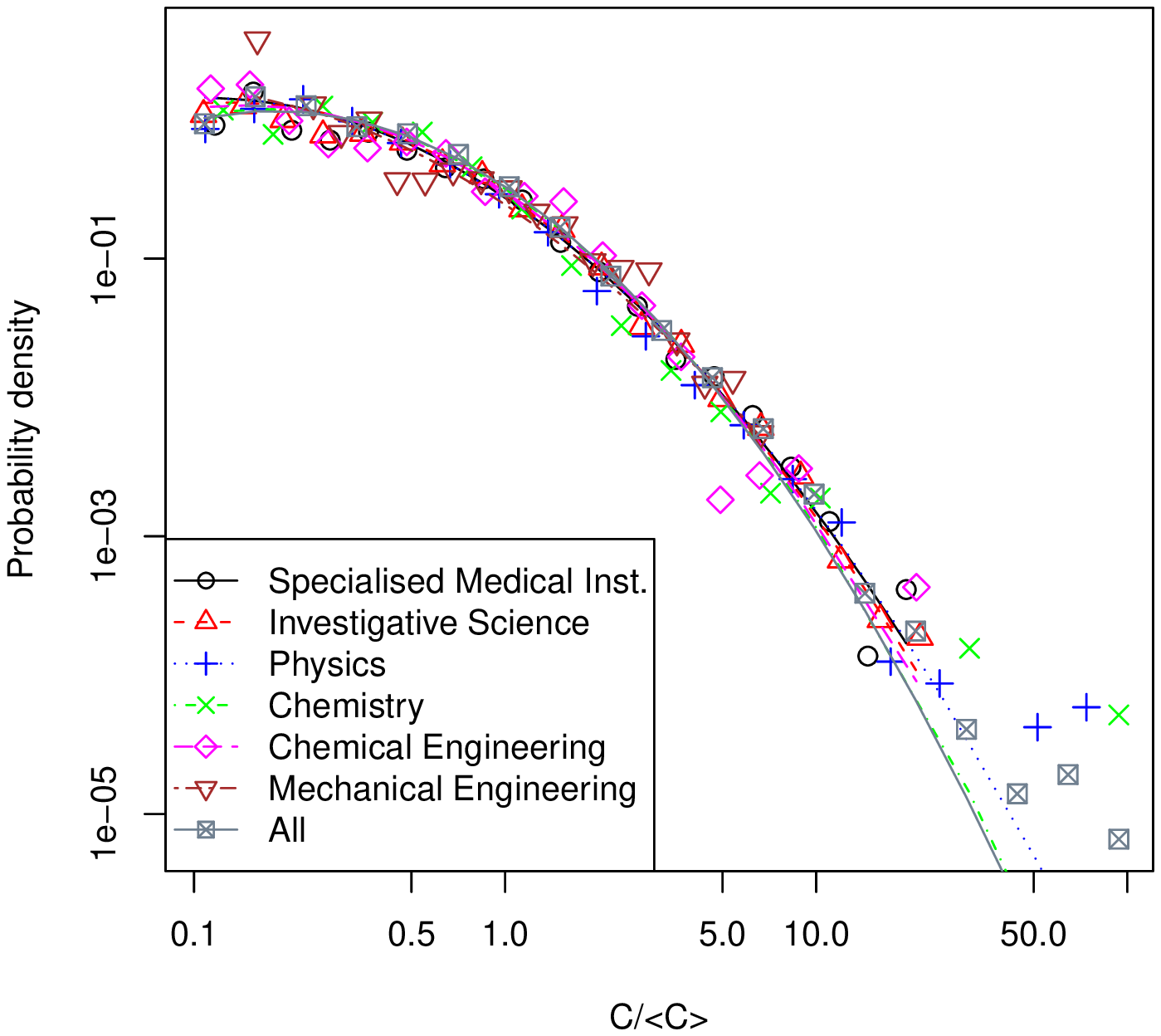}
\includegraphics[width=7cm]{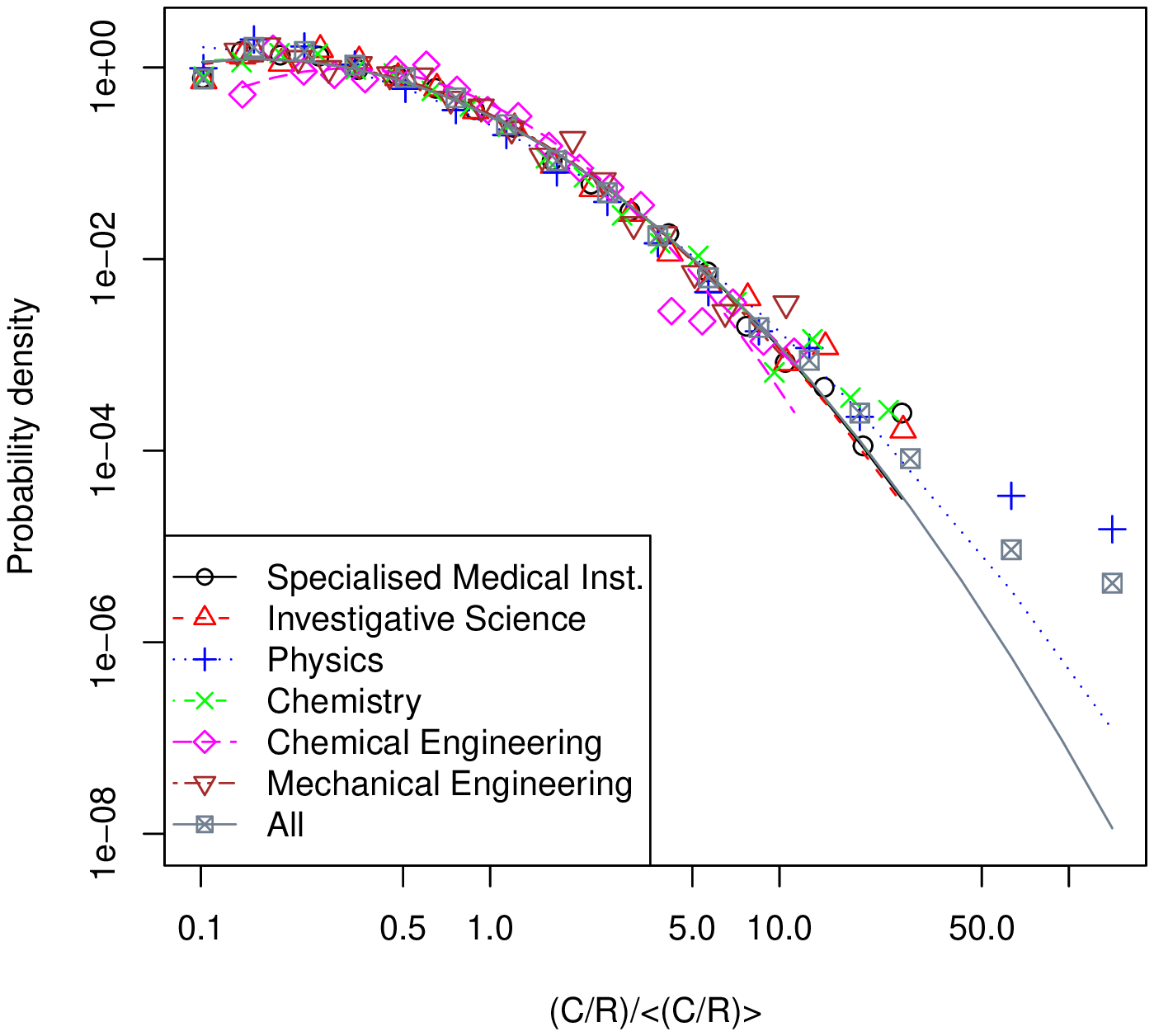}
\end{center}
\caption{The symbols show the distribution of $c_{\mathrm{f}}$ (left)
$c_{\mathrm{r}}$ (right) for department data for all papers with
$c_{\mathrm{f}} > 0.1$ published between years 1999--2001.
The lines are the best fits to lognormal with one free parameter.}
\label{cdeptnorm}
\end{figure}

\begin{figure}[htbp]
\begin{center}
\includegraphics[width=7cm]{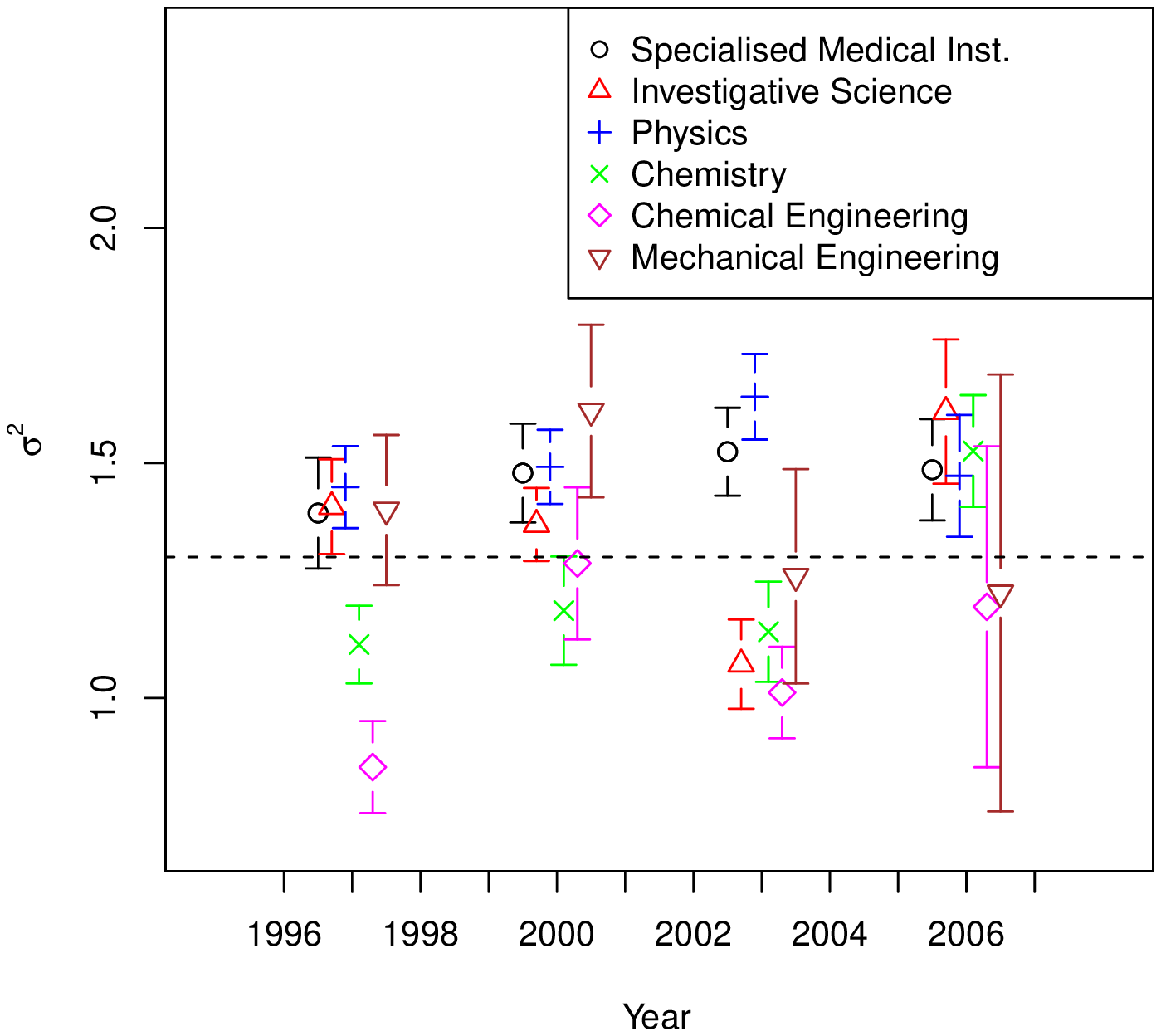}
\includegraphics[width=7cm]{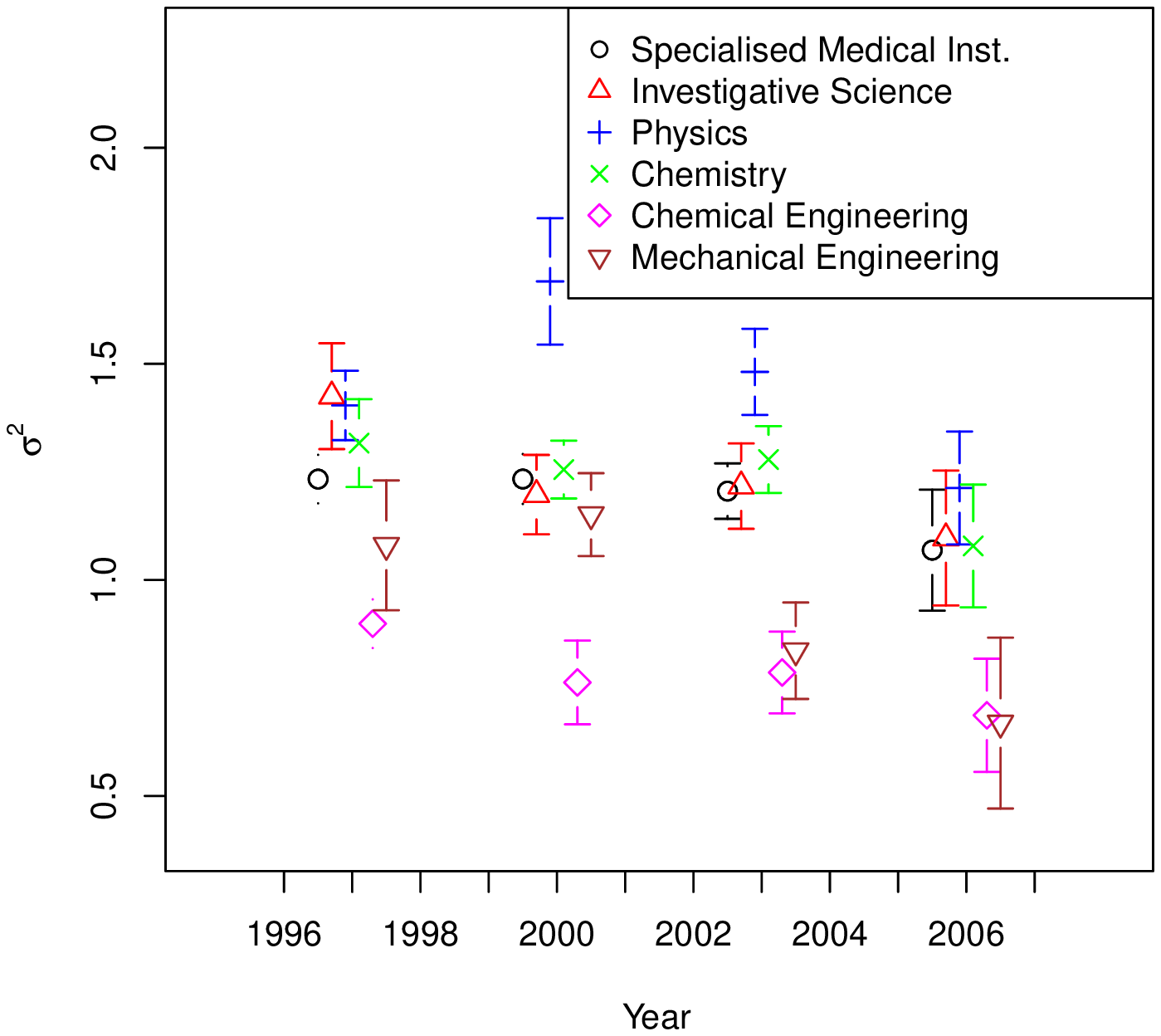}
\end{center}
\caption{A plot of $\sigma^2$ against year resulting from a one
 parameter fit of a lognormal to the $c_{\mathrm{f}}$ (left)
$c_{\mathrm{r}}$ (right) measure. Error bars correspond to one standard deviation.
The papers used for each point correspond to publications with $\cfindex > 0.1$ binned into three year intervals
for the two most prolific departments of each faculty.}
\label{cdeptsig2}
\end{figure}

When the data was fitted with a single parameter lognormal
we found the value of $\sigma^{2}$ varied between 0.9 and 1.7,
with a typical value around 1.3 (see Figure \ref{cdeptsig2} in supplementary material).
This compares against the universal value for $\sigma^{2}$ of 1.3 suggested in \cite{Radicchi}.
Using a three parameter fit
to check the fit it was found that $(\mu+\sigma^2/2)$ took values
between -0.4 and 0.2 for large departments publishing around 500
papers per year. Smaller departments, publishing only 30 or
so papers per year, showed a much bigger range for $(\mu+\sigma^2/2)$ of around -1 to 4,
indicative of insufficient data.

\bnote{Global average is $1.13\pm0.07$}

Repeating the analysis with the $\crindex$ measure yielded comparable results
with consistent variations between fields. Application of the same $\crindex >0.1$ cutoff
improved the $\chi^2$ statistic per degree of freedom from ranging between 0.52 and 9910 with a mean of 415 to within 0.76 and 20.1 around an average value of 3.92. The single parameter logarithmic fit had $\sigma^{2}$ falling between
0.8 and 1.7. The more recent years (2006--2007) showed greater deviations in the three
parameter fit as these publications had less time to accumulate citations relative to
the number of references. \bnote{Global average is $1.17\pm0.09$}

\section{arXiv Data}\label{sarXivdata}

The analysis here so far and in \cite{Radicchi} has used global data from WoS as the set $\Pcal$ and so as the source of all citation counts.  To see if universality applies when other data sets are used we have used the arXiv e-print archive.  We used citations from papers in eight sub-archives between the years 1991 and 2006.  We then analysed the four larger sub-archives each corresponding to different subject areas within physics. To be precise the sets $\Pcal$ and $\Scal$ used in the definitions of $\cfindex$ \tref{cfdef} and $\crindex$ \tref{crdef} are now:-
\begin{itemize}

\item[$\Pcal$ ---] All items in the eight sub-archives (astro-ph, gr-qc, hep-ex, hep-lat, hep-ph, hep-th, nucl-ex and nucl-th) of the arXiv preprint archive with an initial deposit date between 1991 and 2006 inclusive.

\item[$\Scal$ ---] All items belonging to one sub-archive (astro-ph,hep-ph, hep-th or gr-qc)
published in a single calendar year (any one between 1997 and 2004) with at least one reference to and at least one  citation from an item in $\Pcal$.

\end{itemize}

\begin{figure}[htbp]
\begin{center}
\includegraphics[width=7cm]{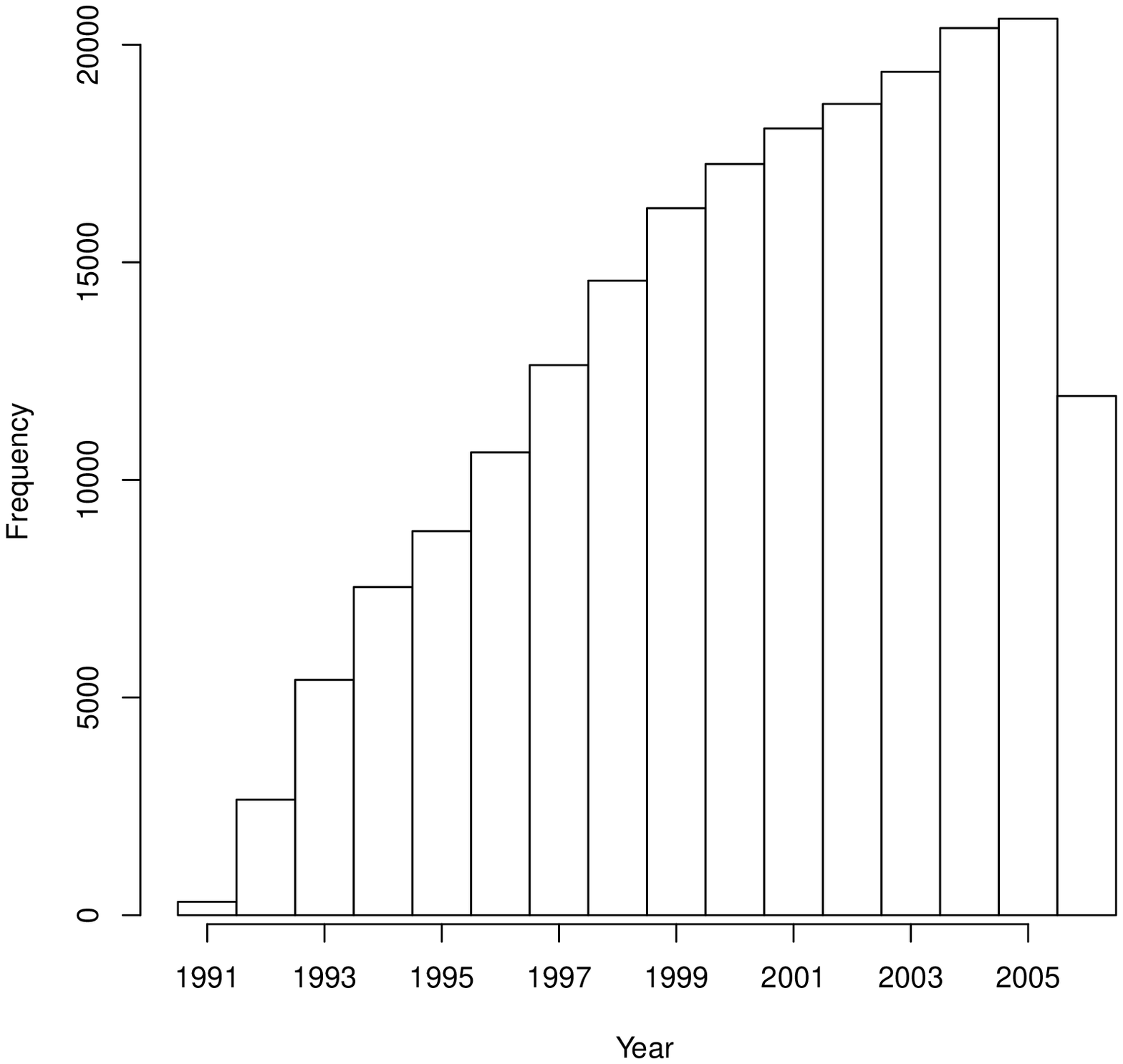}
\includegraphics[width=7cm]{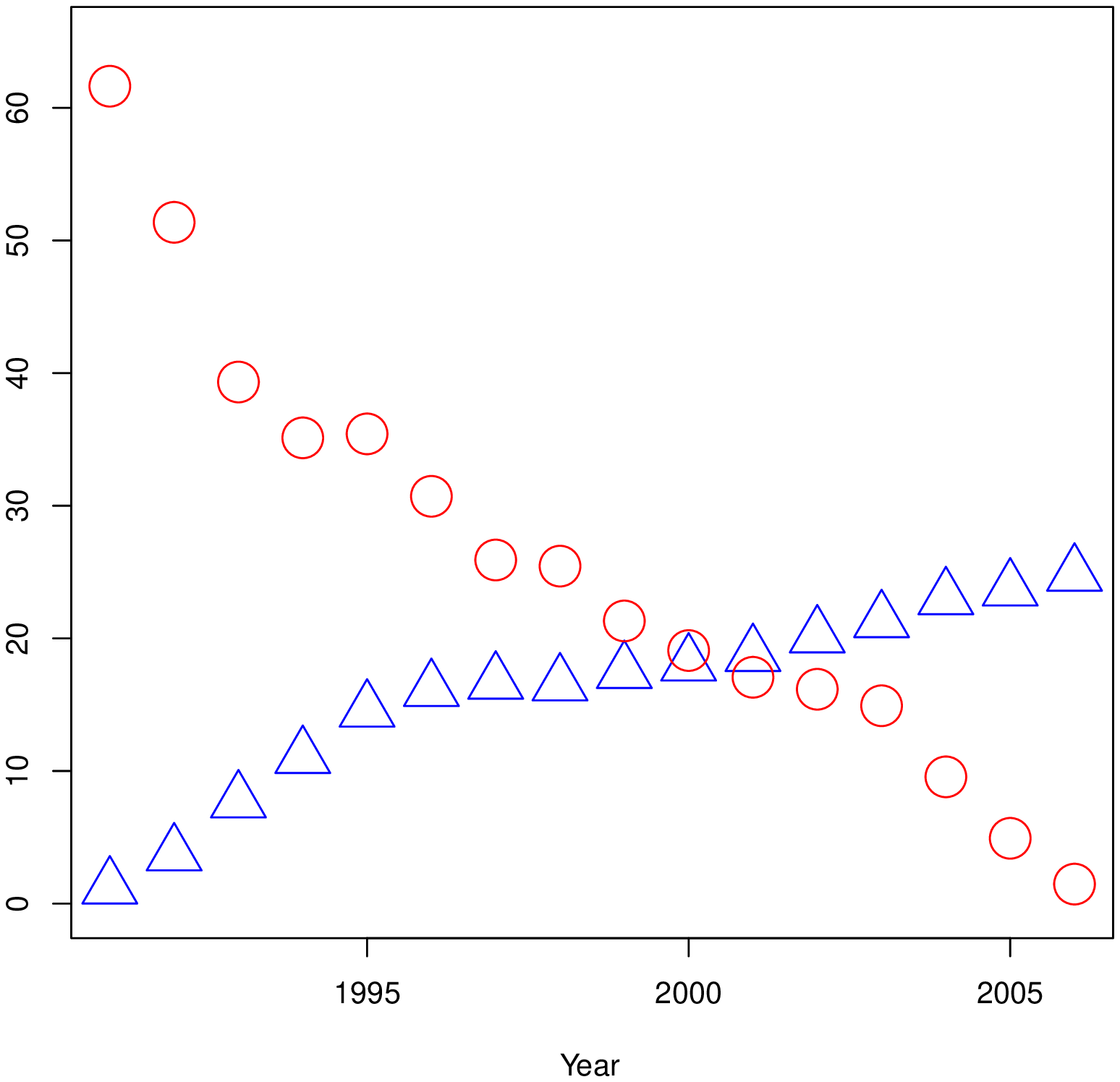}
\end{center}
\caption{On the left, the number of publications in our arXiv data.  On the right the average number of citations (red circles) and references (blue triangles) for publications initially deposited in a given year.}
\label{farXivinfo}
\end{figure}

\begin{figure}[htbp]
\begin{center}
\includegraphics[width=7cm]{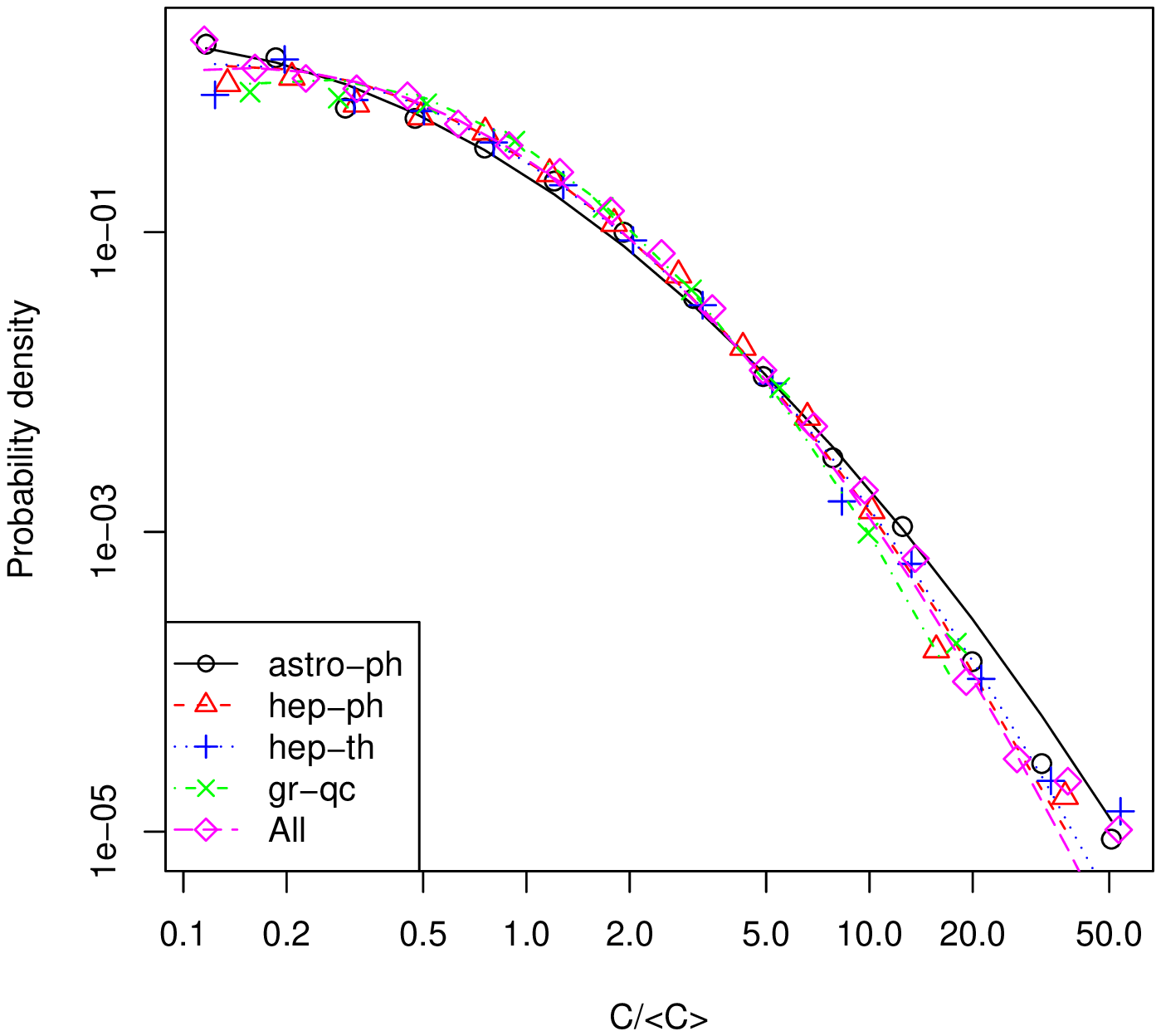}
\includegraphics[width=7cm]{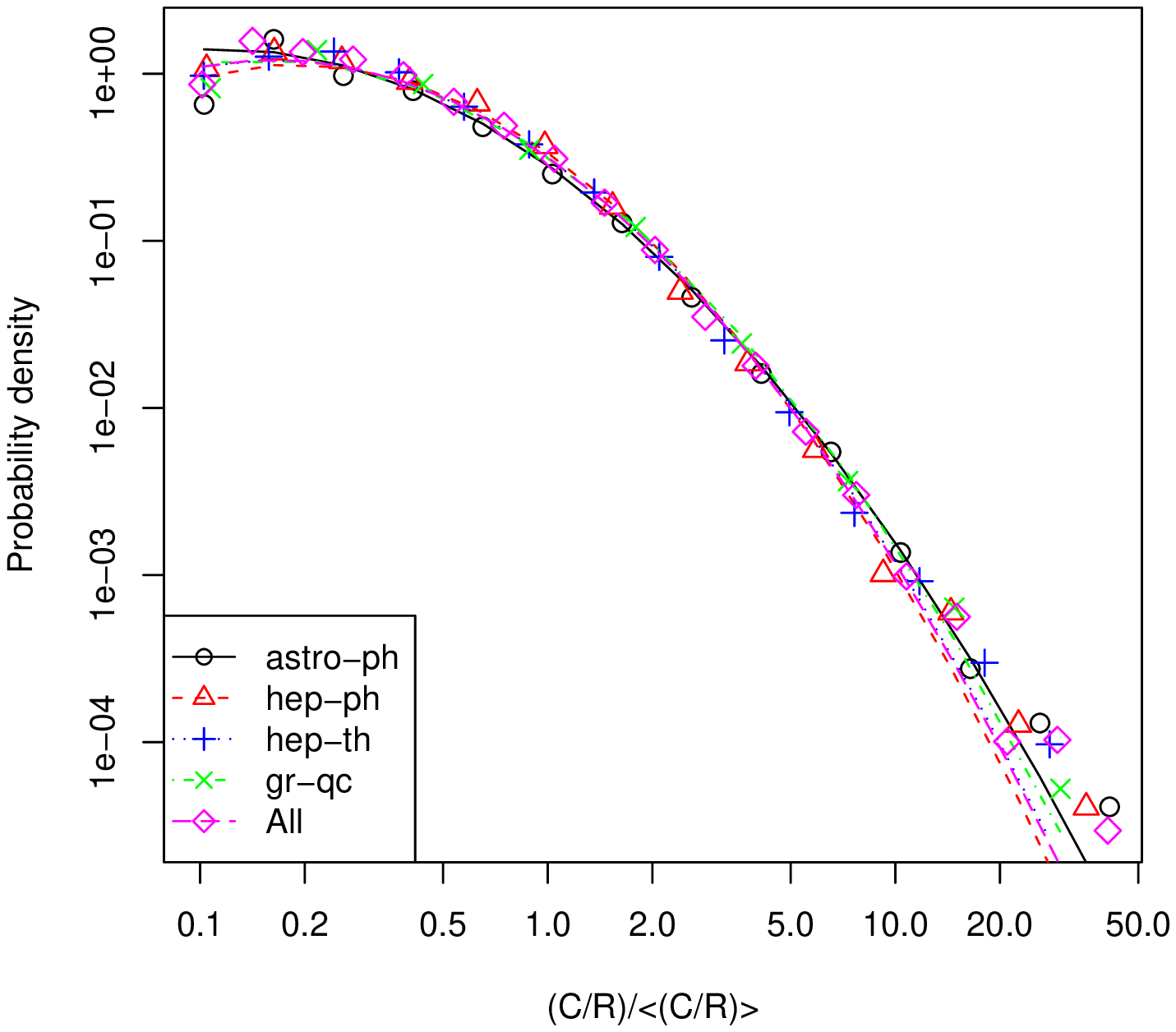}
\end{center}
\caption{The symbols show the distribution of $c_{\mathrm{f}}$
(left) and $c_{\mathrm{r}}$ (right) for arXiv data for publications
of four major sub-archives with $c_{\mathrm{f,r}} > 0.1$
published between in  2002. The lines are the best fits to lognormal with one
free parameter.}
\label{carXivnorm}
\end{figure}

\begin{figure}[htbp]
\begin{center}
\includegraphics[width=7cm]{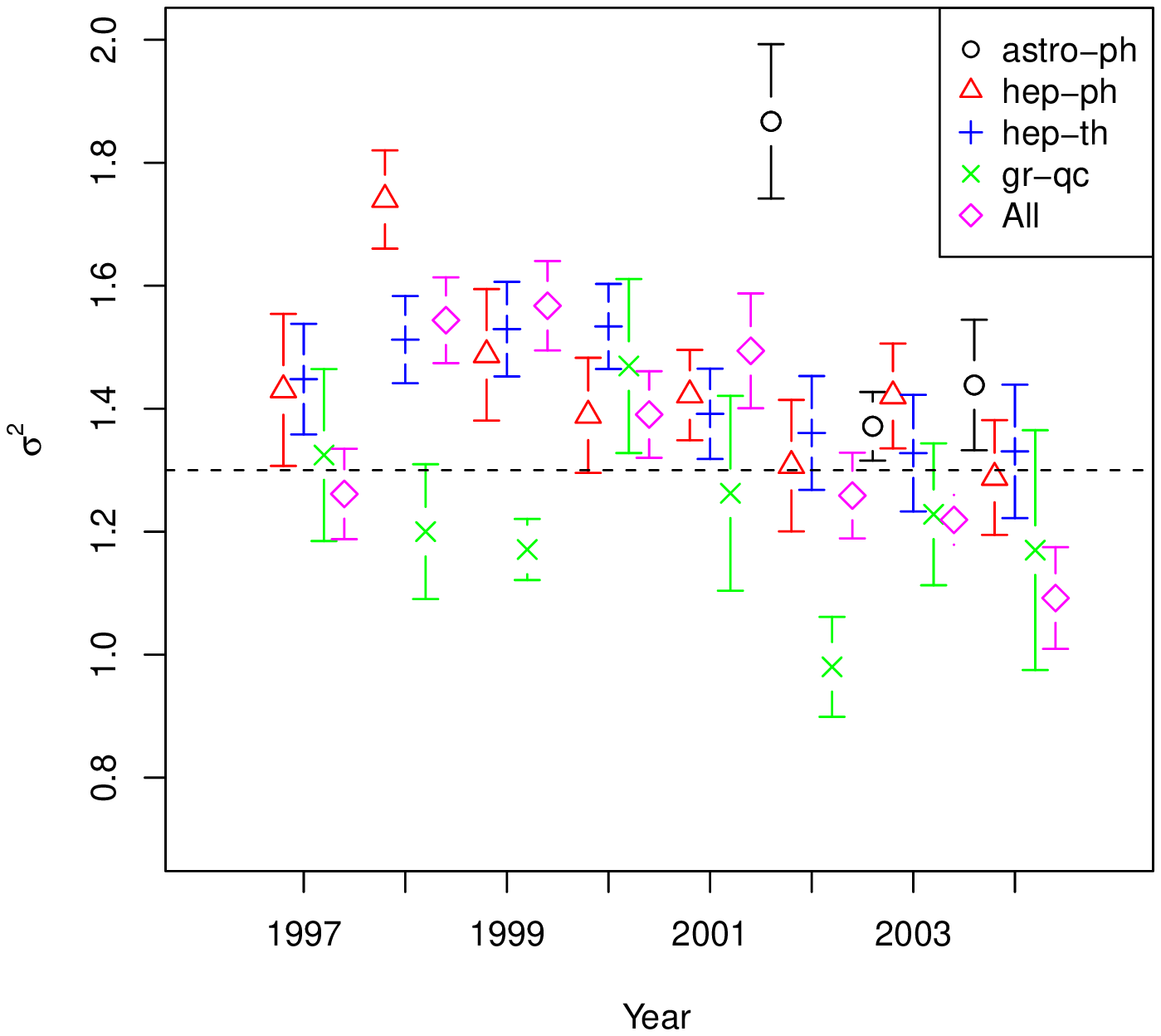}
\includegraphics[width=7cm]{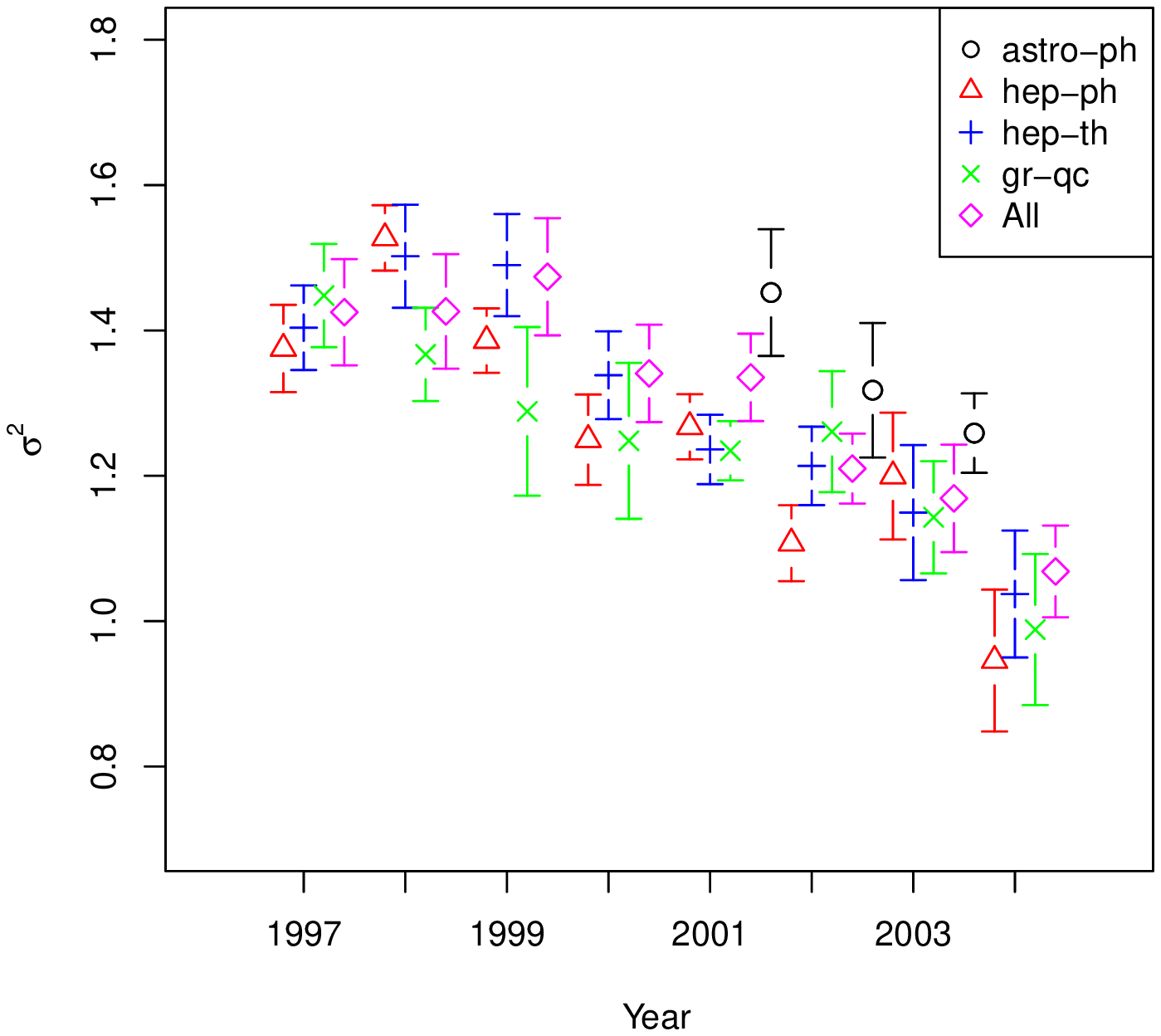}
\end{center}
\caption{A plot of $\sigma^2$ against year resulting from a one (left) or
three (right) parameter fit of a lognormal to the $\cfindex$ measure.
Error bars are for one standard deviation. Not shown on left plot are markers
corresponding to astro-ph 1997, astro-ph 1998, astro-ph 1999, astro-ph 2000 and astro-ph 2001
with values $3.86\pm23$,$3.26\pm28$, $2.92\pm37$, $2.75\pm16$ and, $2.47\pm26$ respectively.
Omitted from the right plot are markers corresponding to astro-ph 1997,
astro-ph 1998, astro-ph 1999 and astro-ph 2002
with values $2.73\pm23$,$2.69\pm16$ , $2.62\pm22$ and, $2.55\pm11$ respectively.}
\label{cfarXivsig2}
\end{figure}

Employing the same $c_{\mathrm{f}}>0.1$ cutoff to the one parameter lognormal fit,
the $\chi^2$ per degree of freedom was reduced from ranging from 3.92
to 59.6 with a mean of 30.8 to between 1.49 and 87.0 around an average of
8.98 whilst retaining 84\% of publications. This fit resulted in $\sigma^2$
values ranging from $2.73\pm0.23$ for astro-ph in 1997 to $0.97\pm0.09$ for gr-qc
in 2002.
The averages for each sub-archive were astro-ph $2.49\pm0.20$, hep-ph
$1.44\pm0.11$, hep-th  $1.43\pm0.10$ and gr-qc  $1.23\pm0.14$ resulting in an
overall average of $1.35\pm0.08$. These values are notably higher than those
for the faculties and departments considered. This is in part due to some of the astro-ph
data distorting the global average. A confirmation of the lognormal
distribution fit was provided by a three parameter lognormal fit. Nearly
all the $\sigma^2$ values are consistent with the constraint $\sigma^{2} = -2\mu$,
see Figure \ref{cfarXivsig2}.

Using a corresponding reference count, the $c_{\mathrm{r}}$ measure was evaluated
for each publication. As imposing a cutoff on $c_{\mathrm{r}}$ did not improve the
goodness of fit, it was decided to use values from all publications. A single
parameter lognormal fit resulted in a $\chi^2$ per degree of freedom value ranging from 0.50 to 10.3
with an average value of 4.39. Imposing a minimum $c_{\mathrm{r}}$ cutoff did not result in
any improvement in goodness of fit. The value of $\sigma^2$ was found to vary between
$2.49\pm0.06$ for astro-ph in 1997 and $1.23\pm0.06$ for gr-qc in 2004.  The
resulting average $\sigma^2$ values were found to be $1.75\pm0.22$, $1.43\pm0.09$,
$1.34\pm0.08$ and $1.35\pm11$ for astro-ph, hep-ph, hep-th and gr-qc. The overall
average value was $1.68\pm0.04$.

The astro-ph data appears be less consistent with the other sub-archives.
This is in part caused by a much longer distribution tail with more publications with very high citation counts ($>50c_0$)
which are not typically seen for the other sub archives. The three parameter fit
confirms that hep-ph, hep-th and gr-qc are well approximated by the lognormal
distribution, with the constraints on the normalisation and mean preserved. So one explanation is that the processes involved in citing older astro-ph publications is different from those behind other physics sub-archives and indeed different from all other papers described here and in \cite{Radicchi}.  Alternatively, the citations in astro-ph are described by the same process and there is some unknown problems with the older astro-ph data.

\section{Interpretation}\label{sinterpetation}

So far no detailed model has been proposed which adequately explains the
origin of the universality seen here and in \cite{Radicchi}.

The Price model of citations \cite{Price76} and its variations invariably
result in power law behaviour for the whole population of papers.
This fails to account for the low citation count part of actual citation
distributions.  However, we only study citations of papers over one or three years and for single fields
and we found power laws to be visibly worse fits than a lognormal to the large citation part of our data.
From the analytical results of Dorogovtsev et al.\ \cite{DMS00} we derived citation distributions within the Price model for papers published within some short interval. These degree distributions depend on the number of citations and some configurable
initial attractiveness. Only around the peak of the distribution can an approximate
lognormal distribution be fitted but this is at far too high a value with too narrow a width. This is because all early publications have had longer to accrue citations so that almost all pick up a substantial number of citations. In reality the majority of publications pick up few citations however old they are.

One potential treatment of this problem is to introduce some artificial ageing of publications to
reduce the rate at which older publications are cited. Wang et al.\ \cite{WYY09} modified the standard attachment
kernel by including an exponential damping factor $\propto \exp(-\lambda t)$. This, however, results in
an exponential tail to the citation distribution for papers published in one year which falls off too fast for the fat-tailed distributions we see.

\bnote{Do we mention the Van Raan approach? This is already mentioned in the RFC paper.}

Lognormal distributions are typically the hallmark of multiplicative growth processes.  So consider a simple stochastic process in which the citations of each publication at time $t$, $c_i(t)$, are assumed to evolve independently at each time step according to $c_i(t+1) \rightarrow c_i(t) \xi_i(t)$.
Here $\xi_i(t)$ is chosen from a suitable
probability distribution function with mean $1+\lambda (c_{i}(t))^{\beta}$, where
$\lambda$ is the citation growth rate (which varies with field)  and $\beta$ a configurable parameter.
Making a reasonable assumption that scientific knowledge propagates on
the time scale of months and years\bnote{Reference needed} and that a
typical publication has a citation accruing lifetime of around
10 years,\bnote{Need reference for this.} iterating the map for 10--100 time
steps would appear appropriate.
Initialising each publication with a uniform citation count, the model was
iterated over 25 discrete time steps and the emergent distribution analysed
as in section \ref{sCrownMeasure}.
By dividing through by the mean citation count, the scale determining growth
factor $\lambda$ is effectively cancelled out. The resulting distribution
for one million papers was found to be reasonably well described
by a lognormal for a wide range of parameters.  However these had variances $\sigma^2$ which were much too small for a range of $\beta$ values around zero. This can be changed by choosing the initial value to be some measure of intrinsic fitness, $c_i(0)=q_i$.  We can adjust the distribution of the paper fitness parameters $q_i$ to obtain better results but this would require some a priori justification.

In any case such a model has an intrinsic problem in that its variance changes with time. For the case $\beta=0$ the central limit theorem tells us that
the variance should scale as $\sigma^2\sim t^{-1}$ where $t$ denotes the number of elapsed time steps.
This would be manifested in a systematic temporal variation in the $\sigma^2$ parameter and we simply do not see this feature in our results, see Figures \ref{fsigmaC1and3} and \ref{cdeptsig2}. Under the assumption of
the simple multiplicative growth process one would expect a factor of $4$ between
the variances of 1997 and 2007 for any given faculty in Figure \ref{fsigmaC1and3} which is just not observed.  Even if time $t$ is better measured in terms of the number of citations accrued (since the rate at which citations are accrued dies off with time after a few years) there is no suggestion in the data of any systematic decrease in variance over time. The data for arXiv in Figure \ref{cfarXivsig2} suggests a possible variation but it is an increase in variance with time, not a reduction. This invalidates the assumption that each multiplicative increase is independent of the last
suggesting the system is governed by strong temporal correlations.

The simplest model which has no change over time in the variance of a resulting lognormal distribution is just
$c_i(t) = q_i g(t) \prod_{t} \xi_i(t)$ with $g(t)$ defining the growth in the mean citation, $q_i$ a measure of the intrinsic quality of a paper and  $\xi_i(t)$ a random variable drawn from a suitable distributions.  The distribution for $\xi$ has to give $\ln(\xi)$ a mean of zero and a finite variance.  Then the variance in citation counts $c_i(t)$ coming from the noise $\xi$ will die off as $1/\sqrt{t}$.  So provided the variation at initial times coming from $\xi_i$ is small enough, the noise will be unimportant at any time. This explains the universality of the citation distributions of $\cfindex$ over time which we have seen.  The differences in the citations of each paper are controlled only by the intrinsic quality $q_i$ along with the growth in the average number of citations.  To explain the universality over research field means that only $g(t)$ can depend on field, the distribution of $q_i$ can not.  The reason for this is that it is only when we look at the ratio $\cfindex$ does the field dependent growth factor $g(t)$ cancel.  That then leaves $\cfindex$ as a universal measure across time and field, as it is controlled only by the intrinsic quality $q_i$.

The distribution of the $\cfindex$ still has to be explained in terms of the distribution of the intrinsic qualities of a paper.  The lognormal form of the curve we have seen for reasonably well cited papers (roughly for $\cfindex >0.1$) leads us to conjecture that the quality of a paper is made up of a \emph{product} of factors, $q_i = \prod_a q_i^a$ where each factor $q_i^a$ is the effect of issue labelled $a$.  Issues may include \tnote{Others?  Check for citations and look for others?} the quality of publishing journal \cite{LG09}, prestige of home institutions, faculties or departments \cite{H71}, differences between subdisciplines, and even a measure of the true quality of the work in the publication.  Whatever the nature of these distributions over different effects, the central limit theorem will ensure only a few are needed to lead to the lognormal being a good description of normalised citation indices such as $\cfindex$ \tref{cfdef}.

Of course such a model can only capture the general behaviour of citations for a reasonable number of publications, but it does suggest that the universality seen here and in \cite{Radicchi} means that other effects are smaller.  As mentioned before the low citation results may fit a universal distribution but they are not well described by a log normal.  One problem is that the lognormal form describes a continuous variable so mapping this onto the discrete values taken by $\cfindex$ is most problematic for low citation count.  Alternatively we have suggested that other processes such as self-citation, the increased fraction of different types of publication (such as meeting abstracts) and data errors \cite{B77} may be important only for low citation count behaviour in data.

Our results and those of \cite{Radicchi} give a lognormal with variance of around $\sigma^2 \approx 1.3$.  This is comparable to the variances typically measured in a wide range empirical lognormal distributions \cite{LSA01}.  However our simple model above gives no insight as to why the value is not O(10) or O(0.1). As such is it best used as a framework for discussion.

\section{Conclusions}\label{sconclusions}

We have shown that citation measures taken relative to averages, in particular $\cfindex$ \tref{cfdef} and $\crindex$ \tref{crdef}, appear to conform to a universal behaviour independent of the source of our data.  The lognormal form is a good description of this form for all publications, except for those with low citation count (say $\cfindex<0.1$). We have shown this for papers from a single institute with the citations coming from Web of Science (WoS) and divisions made by the political structure of the institute, either by department or by faculty, as well as by year. We saw the same universal form in data taken from the e-print archive arXiv where now the source of citations is not WoS but arXiv itself.  The earlier work of Radicchi, Fortunato and Castellano (RFC) \cite{Radicchi} found the same universality in $c_{\mathrm{f}}$ for the whole WoS data but where publications were grouped by year and by field, there defined by the Journal of Citation Report of Thomson Reuters. Thus we have shown that useful comparisons of publications across diverse scientific fields and times can be made on subsets of papers, defined in a variety of ways.  This greatly extends the practical applications of the results of \cite{Radicchi}.  It also means that evaluation of publications across different disciplines and time can be achieved from many data sets, and this choice will lead to lower costs for such evaluations.

One area that deserves further investigation is to look at emergent definitions of research field. The definition of field in our work has been done through top-down methods: the faculty or department of authors and the arXiv classifications here, the Thomson Reuters Journal of Citation Reports in \cite{Radicchi}.  The alternative is to define fields of research from the relationships between papers themselves, using network clustering (community detection) methods \cite{F09}. Such bottom-up methods gave similar results on a broad statistical scale in \cite{RL09} but it would be interesting to try such emergent definitions of field them in this context. In particular using modern overlapping community detection methods such as \cite{EL09,EL10,E10,ABL10,G11} allow papers to be in more than one category and provide a better definition of field.

One example of a practical application of our results is that it can be used to cut costs of research assessment.  For instance the Research Excellence Framework (REF) run by Higher Education Funding Council for England (HEFCE) will assess the quality of research in UK higher education institutions.  For the 2014 exercise, it is proposed that staff submit up to four publications for assessment. An expert opinion is to be sought on each publication and for some fields (sub-panels in the language of REF) the experts will be provided with some citation information.  This is a citation count for each paper along with as yet unspecified ``discipline-specific contextual information about citation rates for each year of the assessment period to inform, if appropriate, the interpretation of citation data'' \cite{HEFCE2012}.  More sophisticated measures, including the normalisation of citation counts using world citation averages for different fields, were highlighted in a report commissioned by HEFCE \cite{RML07} but are not to be provided.  However these measures are not to be provided presumably on the grounds of cost as access to world-wide data sets $\Scal$ are then needed.  On the other hand, our work suggests that for the REF we could define a similar measure $\cfindex$ \tref{cfdef} but now in terms of the average values found from all those submitting.  That is we define the averages in \tref{cfdef} in terms of subset $\Scal$ of \textit{all} papers authored by the staff, in a given year and in a given field.  For organisational purposes, e.g.\ to select appropriate expert referees, the REF has defined its fields of research so these could be used much as we have used faculties or departments as a convenient definition of research field.  Since four papers are already required for the REF, extending its requirements to all papers published by each contributor does not require major changes or additional cost in the data collection since most institutions collect data on all published papers for a variety of reasons.  As the data from additional papers are only used to find averages, the extra processing required is minimal.  The drawback of this approach is that the measures $\cfindex$ or $\crindex$ would be relative to a UK standard in this case.  If one field of UK research was weaker than another, this would not be apparent in the normalised measures based on UK counts.  Still our normalised indices $\cfindex$ or $\crindex$ would be considerably better than raw citation counts, are cheap to calculate and allow simple comparisons between Institutes within each field which is key goal of the REF. In any case, should data on the global position of each field be available separately, for instance some were given in HEFCE's own report \cite{RML07} or may be part of the unspecified ``contextual information'' provided \cite{HEFCE2012}, a correction for global difference between research fields in the UK could be made if that was deemed important.

By dividing citation counts by references and scaling by the average of this quantity, it was hoped to capture more of the variation in citation patterns between research fields. The $\crindex$ measure appears reasonably
well described by the lognormal distribution.  However this measure seems to be largely correlated with $c_f$, even for review articles which one might expect to have unusually large numbers of references.  So it appears that $\crindex$ is most useful in identifying the occasional publication with unusual characteristics.  As $\crindex$ is trivial to calculate alongside $\cfindex$, it is also a useful check on any calculation.

Though we have focused on using  $\cfindex$ \tref{cfdef} and $\crindex$ \tref{crdef}
for individual papers, there is no reason why these could not be used as the basis
for the analysis of individuals \cite{S92}, groups of researchers
\cite{BWL08}, an institution \cite{M05a}, or a journal \cite{Y81,D93,L07,NF08}.
There has been some debate about the best way to combine measures
for individual papers into a measure for a group of papers, centred
round the crown indicator \cite{MDV95}, see \cite{LBMO11} for one
view and other references on this topic.  One of the criticisms \cite{H71,S92,BM11,LO11}
focuses on the long-tailed nature of the distribution of citations, even
for those in a single year and a field, a problem in many other ways too \cite{AGJ08,LB06}.
The long-tail suggests that simple arithmetic averages of citations measures (normalised or not)
are inappropriate.  By way of comparison, for all its other faults, the h-index \cite{H05}
is specifically designed to take such long-tails into account.  However our approach
suggests this is unnecessary.  Our results and those in \cite{Radicchi} show that
the logarithm of our normalised citation measure is well approximated by a normal
distribution, for which there is no long tail.  The idea of using logarithms to overcome the long tail has appeared elsewhere \cite{L07} but was used in a different way. Thus the issue of long tails can
be dealt with simply by taking averages of the \emph{logarithm} of our normalised
citation indices.  For instance our $\zfindex$ and $\zrindex$ indices of \tref{zdef}
are working in terms of $\ln(\cfindex)$ and $\ln(\crindex)$, and use the mean and
average of the distribution of the space of the logarithm of the normalised
citation indices.

To illustrate what we mean consider the example of journals.  Suppose we consider a set of papers published in one journal, $\Jcal$.  For simplicity assume that each paper, $j \in \Jcal$, is considered to be in a single subset $\Scal(j)$, i.e.\ from a unique field\footnote{Should papers be assigned to more than one field we would suggest weighting contributions from one paper to each field.  So a paper assigned to two fields would be treated a two separate `half-papers'.}.  Then for that paper $\cfindex(j,\Scal(j),\Pcal) = c(j,\Pcal) / \czero(\Scal(j),\Pcal)$. By studying the data on papers in that field $S(j)$ we can fix a mean $\mu_f(\Scal(j),\Pcal)$ and standard deviation $\sigma_f(\Scal(j),\Pcal)$ from the distribution of $\ln(\cfindex)$ (ignoring low cited papers when doing this fit as we suggest). Then each paper $j$ published in that year is assigned a score, $\zfindex(j,\Scal(j),\Pcal) = [\ln(\cfindex(j,\Scal(j),\Pcal))-\mu_f(\Scal(j),\Pcal) ] / \sigma_f(\Scal(j),\Pcal)$ of \tref{zdef}.  Papers with zero citation do not affect the fitting of the $\ln(c_f)$ distributions but would give $\zfindex=-\infty$, a problem also encountered in \cite{L07} when using logarithms of citations.  A simple trick we would suggest to deal with this would be to treat zero cited papers as having a quarter of a citation.  The motivation is that we envisage associating the discrete valued citation count of zero with a bin of a continuous variable running from zero to one half. A citation value of quarter is the midpoint of this bin\footnote{Alternatively a more precise measure is that papers with zero citation are assigned an effective count of $c_\mathrm{eff}$ where
$0.5F(c_\mathrm{eff}/c_0;\mu_f,\sigma_f) = \int_0^{0.5} dc' \; F(c'/c_0;\mu_f,\sigma_f)$ and $F$ is the lognormal distribution of \tref{eLognormal}. However with typical values of $\sigma^2=1.3$, $\mu=-\sigma^2/2$ and $c_0=10$ we find $c_\mathrm{eff}\approx 0.248$ which is a tiny error. In this case zero cited papers would score $\zfindex =-2.67$.}. An obvious measure of a journal in one year would then be the arithmetic average of the $\zfindex$ values of all published papers.  That is our journal index would be $\zfindex(\Jcal) = |\Jcal|^{-1} \sum_{j \in \Jcal} \zfindex(j,\Scal(j),\Pcal)$. It makes sense to take the arithmetic average of values as the $\ln(\cfindex)$ distribution is not fat-tailed.  For such measures, the set of papers in a journal need not be from a single field, thus  providing a practical method of comparing multi-disciplinary journals with specialised ones.

If the journal has papers from only one subset $\Scal$ (so $\Scal=\Scal(j) \; \forall \; j \in \Jcal$), so from one field and for one window in time used to select the subsets $\Scal$, our measure $\zfindex(\Jcal)$ of journal $\Jcal$ is then simply \bea
 \zfindex(\Jcal) &=& \frac{\ln[\cfindex(\Jcal,\Pcal)]-\mu_f(\Scal,\Pcal) ) }{ \sigma_f(\Scal,\Pcal)}
 \\
 \ln[\cfindex(\Jcal,\Pcal)] &=& \frac{1}{|\Jcal|} \ln\left[\prod_{j \in \Jcal}  c(j,\Pcal) \right]  - \ln[ \czero(\Scal,\Pcal)] \; .
\eea
This form highlights another feature of our approach.  Our use of the logarithm of citation count as a measure of an individual paper means that when we look at collections of papers and take arithmetic averages of our index values, the result contains \emph{geometric} means of the raw citation counts rather than the much criticised \cite{LBMO11} arithmetic mean of citation counts, i.e.\ we are exploiting $\sum_j \ln[   c(j,\Pcal) ] = \ln [\prod_{j \in \Jcal}  c(j ,\Pcal) ]$.  By way of comparison, Lundberg \cite{L07} works with the $\ln[   \sum_j  c(j,\Pcal) ]$ which is a very different quantity as it still involves an arithmetic mean of a values taken from a fat-tailed distribution.

We could apply exactly the same approach to assign an index to a journal but based on our index $\crindex$ of \tref{crdef}.  This is the same context in which normalisation by reference counts has been suggested before \cite{Y81,NF08}. However we note that in these earlier approaches the arithmetic average citation count was divided by the arithmetic reference count for a journal in a given time window.  We, however, would be considering something more like a geometric mean of the ratio $c/r$ for each paper.  So while the basic motivation is the same, the statistic produced will be quite different.

In a similar way if a collection of papers is from one field but covers a large time scale, e.g.\ an individual's publication record, this will also correspond to papers drawn from several different subsets $\Scal(j)$ but our measures ensure papers of different ages are weighted appropriately in the measures.

Not all authors reach similar conclusions to us.  Albarr\'{a}n et al.\ \cite{ACOR11} and Waltman et al.\ \cite{WER11} are much less optimistic about the universality the distributions of $\cfindex$ \tref{cfdef}, in contrast with \cite{Radicchi,CR09,BD09a,RC11} and our results.  One area where there are differences is in the treatment of zero cited papers which form a significant proportion of all papers \cite{S92}.  The uncited paper appears in three ways in our analysis: (a) through the definitions of average citations $c_0(\Scal,\Pcal)$ of \tref{cfdef}, (b) through the normalisation of data used to fitting probability distributions, and (c) in fitting zero citation counts to a distribution.

If we were to include uncited papers, point (a) would increase our values of $c_0$ but this can be absorbed into a shift in $\mu$ for our lognormal distribution.  We estimate\footnote{Across our data for the Institute we have 84\% of papers with $c,r>0$ used in our analysis with another 13\% of papers with $c=0$ but $r>0$.  If we use this to estimate the effect of zero cited papers, it suggests that including them would increase $\ln(\cfindex)$ by about 0.14.} that this effect is equivalent to increasing $\mu$ by about 0.14 whereas $\mu$ has a typical value of around $-\sigma^2/2 \approx -0.65$. This is noticeable but not overwhelming as it is similar in size to the deviations we found of $\mu+\sigma^2/2$ from zero
when we use a three-parameter fit Fig.\ref{fdmudAC} (where $\mu+\sigma^2/2=0$ is not enforced). However while this may explain part of the variation in $\mu+\sigma^2/2$, its deviations from zero are not consistently of one sign and certainly possible corrections to $c_0$ do not seem to interfere significantly with our analysis.  If this was significant we would find that the relation $\mu+\sigma^2/2=0$ would not hold. However our three-parameter fits showed no serious problems with this relation.  Radicchi et al.\ \cite{Radicchi} also noted that this shift had no effect on their results.

The normalisation issue of (b) does not affect our fitting as the lognormal distributions we predict have around 2\% of papers with a citation of one half or less\footnote{With $F$ the lognormal distribution of \tref{eLognormal} we have $0.0198 \approx \int_0^{0.5/c_0} dc' \; F(c'/c_0;\mu_f,\sigma_f)$ for  typical values of $\sigma^2=1.3$, $\mu=-\sigma^2/2$ and $c_0=10$.}.  So leaving our uncited papers will have a large effect on our fits through the normalisation of the whole data. Fig.\ \tref{fdmudAC} confirms that when the normalisation is left a free parameter, noise in the fit is much larger than the effect of excluding uncited papers from the total normalisation.

For us point (c) is irrelevant as we exclude low cited and hence zero cited papers from the fit.

Overall then we feel that while including zero cited papers would be an improvement in our analysis, they are unlikely to alter our conclusions.

In fact we go further and emphasise that papers with low numbers of citations do \emph{not} appear to fit a `universal' lognormal model even if there is a universal distribution for such publications. One problem is that the relation between discrete valued citation counts and a continuous distribution such as the lognormal is difficult for low cited papers. We have also suggested that there are additional processes involved for zero and low cited publications such as an increase in the proportion of non-standard types of publication, the nature of self-citation processes and errors in the data \cite{B77}.  In general another factor is that errors in bibliographic records often lead to the creation of a distinct record that has only one or two citations\footnote{One of the advantages of our data is that it is validated by the authors.  However we used a feed from WoS to provide the citation count.  So if the author validated record is linked to a WoS record which is a rare variant of the actual article, we may still retain an aspect of this problem.}. Of course such processes will be more important for disciplines with low numbers of citations and we interpret this as consistent with the observation by Waltman et al.\ \cite{WER11} that the deviations they discussed were worse for fields with low numbers of citations while they improved when zero cited papers were excluded.  We note in particular that if the proportion of low cited papers is variable from year to year or from field to field, then our results suggest that such variations will upset an analysis based on a ranking or  percentile using the whole data.  In our approach using $\zfindex$ the proportion of low cited papers in each field has little effect as they are excluded from our fit.  However the ranking of a paper with a high $\zfindex$ will change depending on the variations in the number of low cited papers.

To summarise, our approach is as follows.  To compare papers from different fields and published at different times from a large set of papers, first split the papers into subsets ($\Scal$) using publication date and an available definition of field. Then, using the data for citations to each paper which probably come from a larger set $\Pcal$,  the data for the indices $\cfindex = c/c_0$ \tref{cfdef} and $\crindex = (c / r) / \crzero$
of \tref{crdef} are fitted to a lognormal but only using reasonably well cited papers. We suggest an operational definition that $\cfindex,\crindex>0.1$ for any reasonably well cited publication. The position of each publication on this curve, even those not used to do the fit, gives a measure that gives a meaningful comparison across disciplines and time.

There still remains much uncertainty and many apparent differences even in the recent literature.  Different data sets are used as many are not publicly available for analysis by other groups.   The different treatment of zero cited papers, different preferred forms for citation curves and different schemes for fitting data, means that direct comparison between our results and other recent papers such as \cite{Radicchi,CR09,BD09a,RC11,ACOR11,WER11,EF11} is difficult. Nevertheless, despite these differences, our work leads us to highlight some general ideas which may produce robust measures of the performance of a publication and of collections of publications. In particular by working with the logarithm of citation measures normalised by time and field, $\ln(\cfindex)$, produces distributions without a fat tail.  Even if these are not a normal distribution (as we suggest for reasonably well cited papers) the mean and standard deviation of the $\ln(\cfindex)$ distribution will be a good characterisation of the data.  With no fat tail it then also makes sense to use arithmetic averages of $\ln(\cfindex)$ when looking at collections of papers.  One clear signal that such an approach makes sense is that we see no systematic variation in our measured parameter $\sigma$ across time or field.

\begin{acknowledgements}
We would like to than L.Waltman, N.J.van Eck and A.F.J.van Raan for useful comments.  NH would like to thank the Nuffield Foundation for a Summer Student bursary.  BSK would like to thank the Imperial College London UROP scheme for a bursary.  We thank O.Kibaroglu and D.Hook for help in obtaining and interpreting the raw data, Thomson Reuters for allowing us to use the citation and reference counts for the data for the Institute, and P.Ginsparg for providing the data from arXiv.
\end{acknowledgements}


\newpage
\appendix
\setcounter{figure}{0}
\renewcommand{\thefigure}{A\arabic{figure}}
\setcounter{table}{0}
\renewcommand{\thetable}{A\arabic{table}}
\vspace*{0.5cm}\begin{center}{\Large Supplementary Material}\end{center}

\section*{Fitting Process}

Our subsets of papers $\Scal$ are either (i) papers published by authors in one faculty of our institute in one calendar year, (ii) papers published by authors in one department of our institute in three consecutive calendar years, or (iii) papers placed on sub-archive of arXiv in a calendar year.  We also place the restriction that papers in each $\Scal$ have positive numbers of both citations and references. The latter are likely not to be significant publications but zero cited papers form a significant proportion of all academic papers, see Tables \ref{tpubtypes} and \ref{tarXivNumbers}.

Once we have defined our set we fix the average number of citations for that set, $c_0(\Scal,\Pcal)$ following \tref{cfdef} and  $\crzero(\Scal,\Pcal)$ of \tref{crdef}.
With the average we can find the indices $\cfindex$ and $\crindex$ for each individual paper.  In the following we will refer to $\cfindex$ alone but we used an identical  procedure for our analysis of $\crindex$.

We define bins with boundaries $C(b)$ such that $C(b+1)=r.c(b)$ where $r$ is a constant.  The number of bins is given and then we choose values of $r$ and the bottom and top bin boundaries such that the smallest and largest values of $\cfindex$ in the data under consideration always fall in the middle of the first and last bins respectively.  The number of bins was chosen by hand to ensure a reasonable number of non-zero data points.

When fitting the data we compare the actual count in each bin against the number expected to lie in that bin $\int_{C(b)}^{C(b+1)} F(c_{\mathrm{f}}; \mu=-\sigma^{2}/2, \sigma^2)$.  The points shown on plots correspond to value we used for a single bin, using the midpoint of the bins to locate the points horizontally.  It makes more sense to use the geometric mean of the bin boundaries to represent the position of the bin but $\sqrt{c(b)c(b+1)}$ differs from the midpoint $(c(b)+c(b+1))/2$ only by a factor of $O((r-1)^2)$ which is negligible in terms of visualisation.

\newcommand{\tsecaption}[1]{\caption{\small #1}}

\section*{Single Institution}

\begin{table}[htbp]\small
\begin{tabular}{|r|c|c|c|c|c|c|}
\hline
 & \multicolumn{2}{c|}{\textbf{All items}} & \multicolumn{2}{c|}{$\mathbf{c,r >0}$}  & \multicolumn{2}{c|}{$\mathbf{c,r >0}$ \textbf{and date}} \\ \hline
\textbf{Type} & \textbf{Number} & \textbf{\%} & \textbf{Number} & \textbf{\%} & \textbf{Number} & \textbf{\%} \\ \hline
Poetry & 1 & 0.00\% &  &  &  &  \\ \hline
Bibliography & 1 & 0.00\% &  &  &  &  \\ \hline
Abstract of Published Item & 2 & 0.00\% &  &  &  &  \\ \hline
Software Review & 9 & 0.01\% & 2 & 0.00\% & 2 & 0.00\% \\ \hline
Item About an Individual & 15 & 0.01\% & 3 & 0.00\% & 3 & 0.00\% \\ \hline
Reprint & 16 & 0.02\% & 5 & 0.01\% & 5 & 0.01\% \\ \hline
Biographical-Item & 58 & 0.05\% & 9 & 0.01\% & 9 & 0.01\% \\ \hline
Book Review & 157 & 0.15\% & 10 & 0.01\% & 10 & 0.01\% \\ \hline
News Item & 64 & 0.06\% & 26 & 0.03\% & 26 & 0.03\% \\ \hline
Correction, Addition & 109 & 0.10\% & 34 & 0.04\% & 34 & 0.04\% \\ \hline
Discussion & 167 & 0.16\% & 54 & 0.07\%  & 54 & 0.07\% \\ \hline
Correction & 403 & 0.38\% & 124 & 0.16\% & 124 & 0.16\% \\ \hline
Meeting Abstract & 15222 & 14.39\% & 875 & 1.12\% & 863 & 1.10\% \\ \hline
Note & 1250 & 1.18\% & 1129 & 1.44\% & 1128 & 1.44\% \\ \hline
Letter & 3251 & 3.07\% & 1767 & 2.26\% & 1767 & 2.26\% \\ \hline
Editorial Material & 3553 & 3.36\% & 1936 & 2.47\% & 1936 & 2.48\% \\ \hline
Review & 4649 & 4.40\% & 4251 & 5.43\% & 4248 & 5.43\% \\ \hline
Proceedings Paper & 9211 & 8.71\% & 6355 & 8.12\% & 6340 & 8.11\% \\ \hline
Article & 67629 & 63.94\% & 61687 & 78.82\% & 61667 & 78.84\% \\ \hline
TOTAL & 105767 &  & 78267 &  & 78216 &  \\ \hline
\end{tabular}
\tsecaption{Different Types of Publication in Data for Single Institution. $c,r >0$ indicates that papers counted must have at least one references and one citation.  The last two columns the publications must also have a valid year of publication. Data stretches from 1970 to 2010 as shown in Figure \ref{fnpapers}.}
\label{tpubtypes}
\end{table}

\newpage
\subsection*{Faculties}

\begin{table}[htbp]\small
\begin{center}
\begin{tabular}{|l|l|l|l|l|l|l|l|}
\hline
Year & Faculty & $N_p$ & $c_0$ & $\sigma^{2}$ &  res.err./d.o.f.  & Bins & $\chi^2/\mathrm{d.o.f.}$ \\ \hline
\multirow{4}{*}{1997}
& Natural Sciences & 869 & 38.88 & 1.26(9) & 0.93 & 14 & 1.8 \\
& Medicine & 1357 & 49.07 & 1.33(10) & 1.54 & 14 & 3.8 \\
& Engineering & 389 & 23.50 & 0.92(11) & 1.30 & 9 & 1.7 \\
& All & 2615 & NA & 1.13(5) & 0.63 & 24 & 3.4 \\\hline
\multirow{4}{*}{1998}
& Natural Sciences & 902 & 42.92 & 1.47(12) & 1.01 & 14 & 2.7 \\
& Medicine & 1381 & 51.89 & 1.36(10) & 2.15 & 12 & 24.4 \\
& Engineering & 471 & 22.63 & 1.27(15) & 1.96 & 8 & 3.7 \\
& All & 2754 & NA & 1.13(5) & 0.82 & 24 & 65.9 \\\hline
\multirow{4}{*}{1999}
& Natural Sciences & 842 & 51.41 & 1.52(13) & 1.22 & 14 & 7.7 \\
& Medicine & 1529 & 44.41 & 1.45(11) & 1.78 & 14 & 2.8 \\
& Engineering & 478 & 22.59 & 1.54(18) & 1.56 & 9 & 1.8 \\
& All & 2849 & NA & 1.24(7) & 1.03 & 24 & 12.4 \\\hline
\multirow{4}{*}{2000}
& Natural Sciences & 921 & 39.91 & 1.39(9) & 1.08 & 13 & 5.6 \\
& Medicine & 1660 & 44.31 & 1.41(9) & 1.60 & 14 & 2.3 \\
& Engineering & 515 & 21.91 & 1.48(19) & 2.42 & 8 & 3.7 \\
& All & 3096 & NA & 1.19(5) & 0.78 & 24 & 7.5 \\\hline
\multirow{4}{*}{2001}
& Natural Sciences & 1084 & 38.70 & 1.39(8) & 0.96 & 14 & 1.9 \\
& Medicine & 1879 & 39.83 & 1.26(10) & 2.27 & 14 & 3.5 \\
& Engineering & 663 & 20.37 & 1.23(10) & 1.50 & 9 & 1.6 \\
& All & 3626 & NA & 1.13(5) & 0.93 & 24 & 3.8 \\\hline
\multirow{4}{*}{2002}
& Natural Sciences & 1136 & 37.95 & 1.52(12) & 1.56 & 13 & 3.7 \\
& Medicine & 2016 & 38.69 & 1.30(7) & 1.55 & 14 & 3.7 \\
& Engineering & 774 & 18.43 & 1.29(10) & 1.77 & 9 & 2.1 \\
& All & 3926 & NA & 1.25(5) & 0.98 & 24 & 5.3 \\\hline
\multirow{4}{*}{2003}
& Natural Sciences & 1147 & 36.13 & 1.55(6) & 0.64 & 14 & 1.9 \\
& Medicine & 2024 & 39.16 & 1.36(6) & 1.41 & 14 & 2.3 \\
& Engineering & 845 & 16.67 & 1.16(10) & 2.03 & 9 & 1.8 \\
& All & 4016 & NA & 1.20(6) & 1.19 & 24 & 9.0 \\\hline
\multirow{4}{*}{2004}
& Natural Sciences & 1342 & 28.90 & 1.29(7) & 1.06 & 14 & 2.0 \\
& Medicine & 2140 & 31.00 & 1.25(10) & 2.75 & 13 & 4.3 \\
& Engineering & 944 & 15.05 & 1.3(11) & 2.20 & 9 & 2.4 \\
& All & 4426 & NA & 1.09(5) & 1.27 & 24 & 6.3 \\\hline
\multirow{4}{*}{2005}
& Natural Sciences & 1377 & 23.00 & 1.16(7) & 1.25 & 14 & 1.7 \\
& Medicine & 2181 & 29.87 & 1.25(6) & 1.59 & 14 & 5.4 \\
& Engineering & 927 & 13.46 & 1.03(7) & 1.87 & 9 & 3.0 \\
& All & 4485 & NA & 1.01(6) & 1.61 & 24 &  9.6 \\\hline
\multirow{4}{*}{2006}
& Natural Sciences & 1242 & 22.63 & 1.32(8) & 1.04 & 14 & 3.7 \\
& Medicine & 2278 & 21.93 & 1.17(4) & 1.20 & 14 & 1.8 \\
& Engineering & 981 & 11.61 & 1.14(13) & 3.14 & 9 & 4.1 \\
& All & 4501 & NA & 1.07(6) & 1.49 & 24 & 7.7 \\\hline
\multirow{4}{*}{2007}
& Natural Sciences & 1254 & 16.02 & 1.13(8) & 1.24 & 14 & 3.9 \\
& Medicine & 2267 & 17.65 & 1.16(11) & 3.39 & 14 & 9.0 \\
& Engineering & 929 & 8.63 & 1.37(19) & 3.12 & 9 & 5.7 \\
& All & 4450 & NA & 1.13(14) & 3.24 & 24 & 35.9 \\\hline
\end{tabular}
\tsecaption{Faculty data from graphs generated using Radicchi measure
using  1 parameter fit of Equation (\ref{eLognormal}). Here, and in later tables, the column labelled res.err./d.o.f.\ is the just the sum of squares of residuals divided by the degree of freedom (number of bins minus number of parameters) squared.  This differs from $\chi^2/\mathrm{d.o.f.}$ as we weight the residuals by the expectation for that bin in finding $\chi^2$.}
\label{tRadtab1para}
\end{center}
\end{table}

\clearpage
\begin{table}[htbp]\small
\begin{center}
\begin{tabular}{|l|l|l|l|l|l|l|l|}
\hline
Year & Faculty & $N_p$ & $c_0$ & $\sigma^{2}$ & $\mu + \frac{\sigma^{2}}{2}$ & residual error/df  \\ \hline
\multirow{3}{*}{1997}
& Natural Sciences & 869 & 38.88 & 1.09(12) & -0.1(1) & 1.0 \\
& Medicine & 1357 & 49.07 & 1.23(14) & 0.0(1) & 1.5 \\
& Engineering & 389 & 23.50 & 0.85(13) & 0.1(1) & 1.4 \\\hline
\multirow{3}{*}{1998}
& Natural Sciences & 902 & 42.92 & 1.18(16) & -0.1(1) & 1.1 \\
& Medicine & 1381 & 51.89 & 1.02(9) & -0.2(1) & 1.6 \\
& Engineering & 471 & 22.63 & 1.22(25) & 0.1(2) & 2.6 \\\hline
\multirow{3}{*}{1999}
& Natural Sciences & 842 & 51.41 & 0.93(8) & -0.3(0) & 0.8 \\
& Medicine & 1529 & 44.41 & 1.29(17) & 0.0(1) & 2.0 \\
& Engineering & 478 & 22.59 & 1.53(25) & 0.2(1) & 1.5 \\\hline
\multirow{3}{*}{2000}
& Natural Sciences & 921 & 39.91 & 1.11(10) & -0.1(1) & 1.0 \\
& Medicine & 1660 & 44.31 & 1.31(11) & 0.0(1) & 1.4 \\
& Engineering & 515 & 21.91 & 1.37(12) & 0.2(1) & 1.2 \\\hline
\multirow{3}{*}{2001}
& Natural Sciences & 1084 & 38.70 & 1.26(11) & 0.0(1) & 0.9 \\
& Medicine & 1879 & 39.83 & 1.23(14) & 0.1(1) & 2.2 \\
& Engineering & 663 & 20.37 & 1.17(13) & 0.1(1) & 1.5 \\\hline
\multirow{3}{*}{2002}
& Natural Sciences & 1136 & 37.95 & 1.25(18) & -0.1(1) & 1.8 \\
& Medicine & 2016 & 38.69 & 1.22(9) & 0.0(1) & 1.4 \\
& Engineering & 774 & 18.43 & 1.33(15) & 0.1(1) & 1.9 \\\hline
\multirow{3}{*}{2003}
& Natural Sciences & 1147 & 36.13 & 1.41(9) & 0.0(0) & 0.7 \\
& Medicine & 2024 & 39.16 & 1.26(8) & 0.0(0) & 1.3 \\
& Engineering & 845 & 16.67 & 1.22(15) & 0.1(1) & 2.2 \\\hline
\multirow{3}{*}{2004}
& Natural Sciences & 1342 & 28.90 & 1.13(9) & -0.1(1) & 1.1 \\
& Medicine & 2140 & 31.00 & 1.15(12) & 0.1(1) & 2.6 \\
& Engineering & 944 & 15.05 & 1.42(17) & 0.2(1) & 2.3 \\\hline
\multirow{3}{*}{2005}
& Natural Sciences & 1377 & 23.00 & 1.10(9) & 0.1(1) & 1.1 \\
& Medicine & 2181 & 29.87 & 1.17(8) & 0.0(1) & 1.6 \\
& Engineering & 927 & 13.46 & 1.03(11) & 0.1(1) & 2.2 \\\hline
\multirow{3}{*}{2006}
& Natural Sciences & 1242 & 22.63 & 1.14(11) & -0.1(1) & 1.1 \\
& Medicine & 2278 & 21.93 & 1.09(4) & 0.0(0) & 0.8 \\
& Engineering & 981 & 11.61 & 1.10(26) & 0.0(1) & 4.7 \\\hline
\multirow{3}{*}{2007}
& Natural Sciences & 1254 & 16.02 & 1.07(14) & 0.0(1) & 1.6 \\
& Medicine & 2267 & 17.65 & 1.13(20) & 0.0(1) & 4.2 \\
& Engineering & 929 & 8.63 & 1.76(43) & 0.2(2) & 4.2 \\\hline
\end{tabular}
\tsecaption{Faculty data from graphs generated using Radicchi measure
using a 3 parameter fit $A \cdot F(\cfindex;\mu,\sigma^2)$.}
\label{tRadtab3para}
\end{center}
\end{table}

\clearpage
\begin{table}[htbp]\small
\begin{center}
\begin{tabular}{|l|l|l|l|l|l|l|l|}
\hline
Year & Faculty & $N_p$ & $\langle \crindex \rangle$ & $\sigma^{2}$ & res.err./d.o.f. & Bins & $\chi^2/\mathrm{d.o.f.}$ \\ \hline
\multirow{3}{*}{1997}
& Natural Sciences & 982 & 1.49 & 1.33(5) & 0.6 & 14 & 3.6 \\
& Medicine & 1561 & 1.89 & 1.40(5) & 1.0 & 14 & 7.2 \\
& Engineering & 443 & 1.06 & 1.03(6) & 0.8 & 9 & 4 \\\hline
\multirow{3}{*}{1998}
& Natural Sciences & 1064 & 1.64 & 1.55(8) & 0.9 & 14 & 2.6 \\
& Medicine & 1646 & 2.02 & 1.61(7) & 1.4 & 14 & 4.0 \\
& Engineering & 543 & 1.04 & 1.11(11) & 1.8 & 9 & 5.5 \\\hline
\multirow{3}{*}{1999}
& Natural Sciences & 1030 & 1.79 & 1.70(9) & 1.0 & 14 & 2.1 \\
& Medicine & 1770 & 1.62 & 1.44(4) & 0.9 & 14 & 2.9 \\
& Engineering & 569 & 0.92 & 1.30(9) & 1.2 & 9 & 1.0 \\\hline
\multirow{3}{*}{2000}
& Natural Sciences & 1062 & 1.66 & 1.61(8) & 1.3 & 12 & 35.4 \\
& Medicine & 1939 & 1.57 & 1.39(4) & 1.1 & 14 & 3.7 \\
& Engineering & 637 & 0.92 & 1.31(4) & 0.8 & 8 & 27.2 \\\hline
\multirow{3}{*}{2001}
& Natural Sciences & 1244 & 1.56 & 1.61(11) & 1.6 & 14 & 3.4 \\
& Medicine & 2103 & 1.44 & 1.38(3) & 0.9 & 14 & 7.5 \\
& Engineering & 760 & 0.92 & 1.40(6) & 1.1 & 9 & 0.8 \\\hline
\multirow{3}{*}{2002}
& Natural Sciences & 1321 & 1.41 & 1.62(9) & 1.4 & 14 & 2.5 \\
& Medicine & 2262 & 1.36 & 1.41(4) & 1.0 & 14 & 3.9 \\
& Engineering & 858 & 0.78 & 1.21(5) & 1.2 & 9 & 1.1 \\\hline
\multirow{3}{*}{2003}
& Natural Sciences & 1319 & 1.26 & 1.57(6) & 0.9 & 14 & 2.1 \\
& Medicine & 2272 & 1.33 & 1.39(6) & 2.1 & 13 & 3.9 \\
& Engineering & 921 & 0.69 & 1.07(4) & 1.1 & 9 & 1.4 \\\hline
\multirow{3}{*}{2004}
& Natural Sciences & 1497 & 0.99 & 1.37(4) & 0.8 & 14 & 1.6 \\
& Medicine & 2484 & 1.05 & 1.28(4) & 1.5 & 14 & 17 \\
& Engineering & 1045 & 0.66 & 1.18(5) & 1.5 & 9 & 0.8 \\\hline
\multirow{3}{*}{2005}
& Natural Sciences & 1573 & 0.85 & 1.31(5) & 1.2 & 13 & 2.4 \\
& Medicine & 2409 & 1.01 & 1.28(6) & 2.2 & 14 & 4.9 \\
& Engineering & 1018 & 0.55 & 1.10(6) & 1.8 & 9 & 1.5 \\\hline
\multirow{3}{*}{2006}
& Natural Sciences & 1426 & 0.74 & 1.33(6) & 1.1 & 14 & 6.6 \\
& Medicine & 2581 & 0.71 & 1.17(4) & 1.5 & 14 & 3.3 \\
& Engineering & 1100 & 0.45 & 0.98(2) & 0.9 & 9 & 3.2 \\\hline
\multirow{3}{*}{2007}
& Natural Sciences & 1359 & 0.49 & 1.18(2) & 0.5 & 14 & 0.8 \\
& Medicine & 2463 & 0.59 & 1.30(6) & 2.0 & 14 & 6.4 \\
& Engineering & 929 & 0.33 & 1.05(5) & 1.4 & 9 & 1.1 \\\hline
\end{tabular}
\tsecaption{Data from graphs generated using the $c_{\mathrm{r}}$
measure using 1 parameter fit, see Equation (\ref{eLognormal}).} \label{tOurtab1para}
\end{center}
\end{table}

\clearpage
\begin{table}[htbp]\small
\begin{center}
\begin{tabular}{|l|l|l|l|l|l|l|}
\hline
Year & Faculty & $N_p$ & $\langle c_{\mathrm{r}} \rangle$ & $\sigma^{2}$ & $\mu + \frac{\sigma^{2}}{2}$ & residual error/df \\ \hline
\multirow{3}{*}{1997}
& Natural Sciences & 982 & 1.49 & 1.0(1) & -0.2(0) & 0.5 \\
& Medicine & 1561 & 1.89 & 1.1(1) & -0.1(1) & 1.3 \\
& Engineering & 443 & 1.06 & 1(0) & 0.1(0) & 0.4 \\\hline
\multirow{3}{*}{1998}
& Natural Sciences & 1064 & 1.64 & 1.0(1) & -0.3(0) & 0.8 \\
& Medicine & 1646 & 2.02 & 1.1(1) & -0.2(1) & 1.3 \\
& Engineering & 543 & 1.04 & 0.9(1) & -0.1(1) & 1.6 \\\hline
\multirow{3}{*}{1999}
& Natural Sciences & 1030 & 1.79 & 1.0(1) & -0.4(1) & 1.2 \\
& Medicine & 1770 & 1.62 & 1.1(1) & -0.1(0) & 1.0 \\
& Engineering & 569 & 0.92 & 1.2(2) & 0.1(1) & 1.5 \\\hline
\multirow{3}{*}{2000}
& Natural Sciences & 1062 & 1.66 & 1.0(1) & -0.3(1) & 1.8 \\
& Medicine & 1939 & 1.57 & 1.1(1) & -0.1(1) & 1.3 \\
& Engineering & 637 & 0.92 & 1.1(1) & 0.0(1) & 1.0 \\\hline
\multirow{3}{*}{2001}
& Natural Sciences & 1244 & 1.56 & 1.0(1) & -0.3(1) & 1.6 \\
& Medicine & 2103 & 1.44 & 1.1(1) & -0.1(0) & 1.3 \\
& Engineering & 760 & 0.92 & 1.2(1) & 0.1(0) & 0.8 \\\hline
\multirow{3}{*}{2002}
& Natural Sciences & 1321 & 1.41 & 1.0(1) & -0.4(1) & 1.5 \\
& Medicine & 2262 & 1.36 & 1.1(1) & -0.2(0) & 1.6 \\
& Engineering & 858 & 0.78 & 1.1(1) & 0.1(0) & 1.2 \\\hline
\multirow{3}{*}{2003}
& Natural Sciences & 1319 & 1.26 & 1.2(1) & -0.2(1) & 0.9 \\
& Medicine & 2272 & 1.33 & 1.0(1) & -0.2(1) & 2.1 \\
& Engineering & 921 & 0.69 & 1.0(1) & 0.1(0) & 1.4 \\\hline
\multirow{3}{*}{2004}
& Natural Sciences & 1497 & 0.99 & 1.1(1) & -0.1(0) & 0.9 \\
& Medicine & 2484 & 1.05 & 1.0(1) & -0.1(1) & 2.5 \\
& Engineering & 1045 & 0.66 & 1.1(1) & 0.0(0) & 1.2 \\\hline
\multirow{3}{*}{2005}
& Natural Sciences & 1573 & 0.85 & 1.1(1) & 0.0(0) & 0.9 \\
& Medicine & 2409 & 1.01 & 1.0(1) & -0.2(0) & 1.9 \\
& Engineering & 1018 & 0.55 & 1.1(1) & 0.1(1) & 2.0 \\\hline
\multirow{3}{*}{2006}
& Natural Sciences & 1426 & 0.74 & 0.9(1) & -0.2(0) & 1.0 \\
& Medicine & 2581 & 0.71 & 1.0(1) & -0.1(1) & 2.2 \\
& Engineering & 1100 & 0.45 & 0.9(1) & 0.0(1) & 3.1 \\\hline
\multirow{3}{*}{2007}
& Natural Sciences & 1359 & 0.49 & 1.0(1) & -0.1(0) & 0.8 \\
& Medicine & 2463 & 0.59 & 1.0(1) & -0.2(0) & 1.4 \\
& Engineering & 929 & 0.33 & 0.9(0) & -0.1(0) & 1.1 \\\hline
\end{tabular}
\tsecaption{Faculty data from graphs generated using the $c_{\mathrm{r}}$
measure using 3 parameter fit, $A \cdot F(\cfindex;\mu,\sigma^2)$.} \label{tOurtab3para}
\end{center}
\end{table}

\clearpage
\begin{table}[htbp]\small
  \centering
        \begin{tabular}{|r|r|r|r|r|r|}
    \hline
    Measure & $c_{\mathrm{f,r}}^{*}$ & Bins  & $\chi^2/\mathrm{d.o.f}$ Min & $\chi^2/\mathrm{d.o.f}$ Max & $\chi^2/\mathrm{d.o.f}$ Mean \bigstrut\\
    \hline
    \multicolumn{1}{|c|}{\multirow{15}[30]{*}{$c_{\mathrm{f}}$}} & \multicolumn{1}{c|}{\multirow{7}[14]{*}{0.0}} & 8     & 11.5  & 11.5  & 11.5 \bigstrut\\
\cline{3-6}    \multicolumn{1}{|c|}{} & \multicolumn{1}{c|}{} & 9     & 2.1   & 4.5   & 3.6 \bigstrut\\
\cline{3-6}    \multicolumn{1}{|c|}{} & \multicolumn{1}{c|}{} & 14    & 12.3  & 12.3  & 12.3 \bigstrut\\
\cline{3-6}    \multicolumn{1}{|c|}{} & \multicolumn{1}{c|}{} & 15    & 26.7  & 26.7  & 26.7 \bigstrut\\
\cline{3-6}    \multicolumn{1}{|c|}{} & \multicolumn{1}{c|}{} & 16    & 6.5   & 49.5  & 20.9 \bigstrut\\
\cline{3-6}    \multicolumn{1}{|c|}{} & \multicolumn{1}{c|}{} & 17    & 4.4   & 63.8  & 23.2 \bigstrut\\
\cline{3-6}    \multicolumn{1}{|c|}{} & \multicolumn{1}{c|}{} & 18    & 4.4   & 37.8  & 18.9 \bigstrut\\
\cline{2-6}    \multicolumn{1}{|c|}{} & \multicolumn{1}{c|}{\multirow{8}[16]{*}{0.1}} & 8     & 11.5  & 11.5  & 11.5 \bigstrut\\
\cline{3-6}    \multicolumn{1}{|c|}{} & \multicolumn{1}{c|}{} & 9     & 1.2   & 4.5   & 2.4 \bigstrut\\
\cline{3-6}    \multicolumn{1}{|c|}{} & \multicolumn{1}{c|}{} & 14    & 12.3  & 12.3  & 12.3 \bigstrut\\
\cline{3-6}    \multicolumn{1}{|c|}{} & \multicolumn{1}{c|}{} & 15    & 55.8  & 55.8  & 55.8 \bigstrut\\
\cline{3-6}    \multicolumn{1}{|c|}{} & \multicolumn{1}{c|}{} & 16    & 3.0   & 24.6  & 10.6 \bigstrut\\
\cline{3-6}    \multicolumn{1}{|c|}{} & \multicolumn{1}{c|}{} & 17    & 2.6   & 4.3   & 3.4 \bigstrut\\
\cline{3-6}    \multicolumn{1}{|c|}{} & \multicolumn{1}{c|}{} & 18    & 1.7   & 9.5   & 4.5 \bigstrut\\
\cline{3-6}    \multicolumn{1}{|c|}{} & \multicolumn{1}{c|}{} & 19    & 1.1   & 5.1   & 2.7 \bigstrut\\
    \hline
    \multicolumn{1}{|c|}{\multirow{11}[22]{*}{$c_{\mathrm{r}}$}} & \multicolumn{1}{c|}{\multirow{5}[10]{*}{0.0}} & 9     & 0.5   & 1.0   & 0.7 \bigstrut\\
\cline{3-6}    \multicolumn{1}{|c|}{} & \multicolumn{1}{c|}{} & 15    & 9914.6 & 9914.6 & 9914.6 \bigstrut\\
\cline{3-6}    \multicolumn{1}{|c|}{} & \multicolumn{1}{c|}{} & 17    & 3.1   & 10.1  & 6.2 \bigstrut\\
\cline{3-6}    \multicolumn{1}{|c|}{} & \multicolumn{1}{c|}{} & 18    & 1.1   & 5.3   & 3.0 \bigstrut\\
\cline{3-6}    \multicolumn{1}{|c|}{} & \multicolumn{1}{c|}{} & 19    & 0.6   & 5.7   & 1.8 \bigstrut\\
\cline{2-6}    \multicolumn{1}{|c|}{} & \multicolumn{1}{c|}{\multirow{6}[12]{*}{0.1}} & 8     & 3.0   & 3.0   & 3.0 \bigstrut\\
\cline{3-6}    \multicolumn{1}{|c|}{} & \multicolumn{1}{c|}{} & 9     & 0.8   & 1.3   & 1.0 \bigstrut\\
\cline{3-6}    \multicolumn{1}{|c|}{} & \multicolumn{1}{c|}{} & 16    & 13.0  & 20.1  & 16.5 \bigstrut\\
\cline{3-6}    \multicolumn{1}{|c|}{} & \multicolumn{1}{c|}{} & 17    & 1.8   & 1.9   & 1.8 \bigstrut\\
\cline{3-6}    \multicolumn{1}{|c|}{} & \multicolumn{1}{c|}{} & 18    & 1.0   & 6.3   & 3.4 \bigstrut\\
\cline{3-6}    \multicolumn{1}{|c|}{} & \multicolumn{1}{c|}{} & 19    & 1.1   & 7.8   & 3.0 \bigstrut\\
    \hline
    \end{tabular}
      \tsecaption{Table of $\chi^2$ values for the one parameter lognormal goodness of fit to faculty data. c* denotes the threshold below which publications were not included in the fitting.}
  \label{tabChi2fac}%
\end{table}%

\clearpage
\begin{figure}[htbp]\small
\begin{center}
\includegraphics[width=7cm]{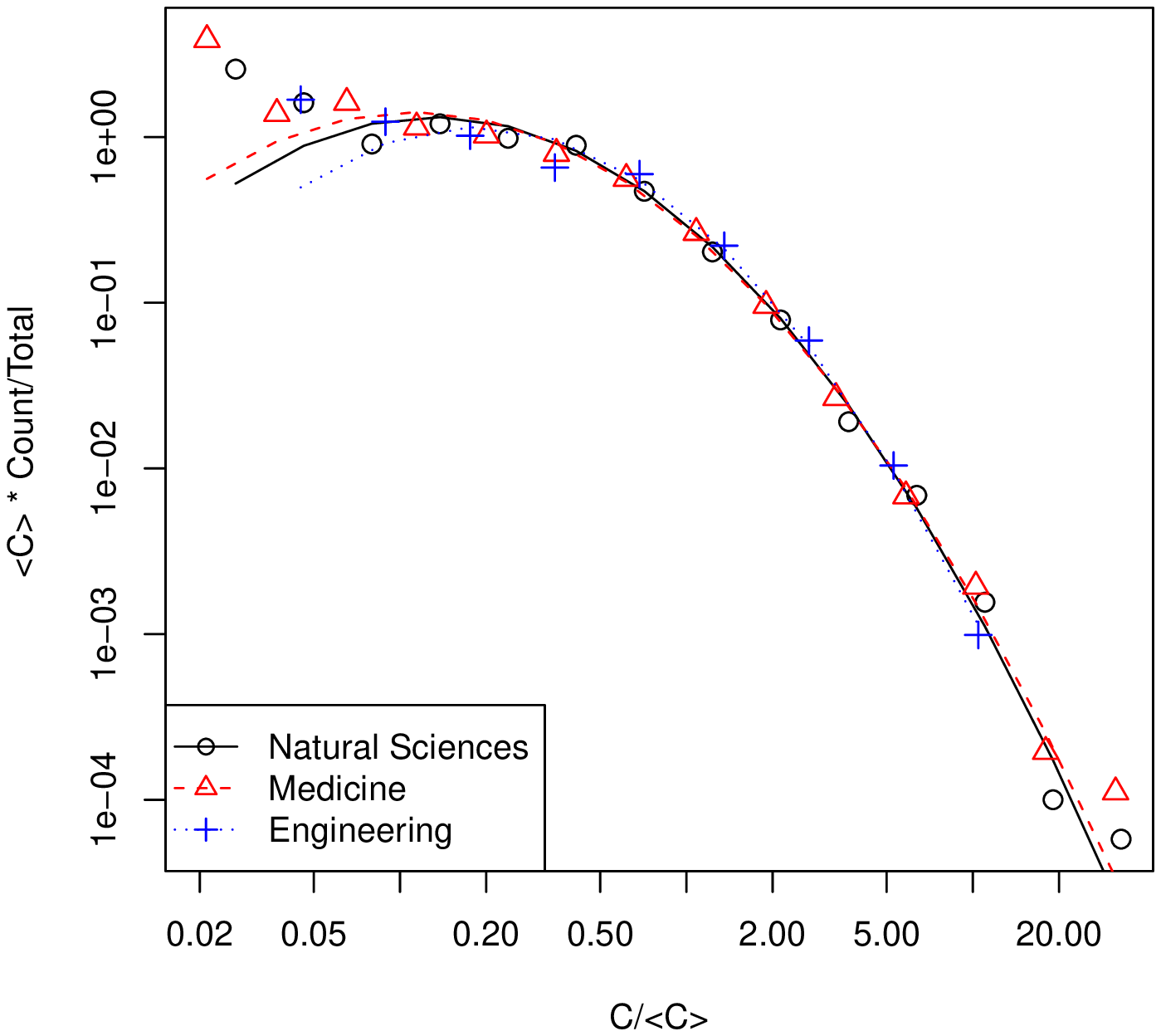}
\includegraphics[width=7cm]{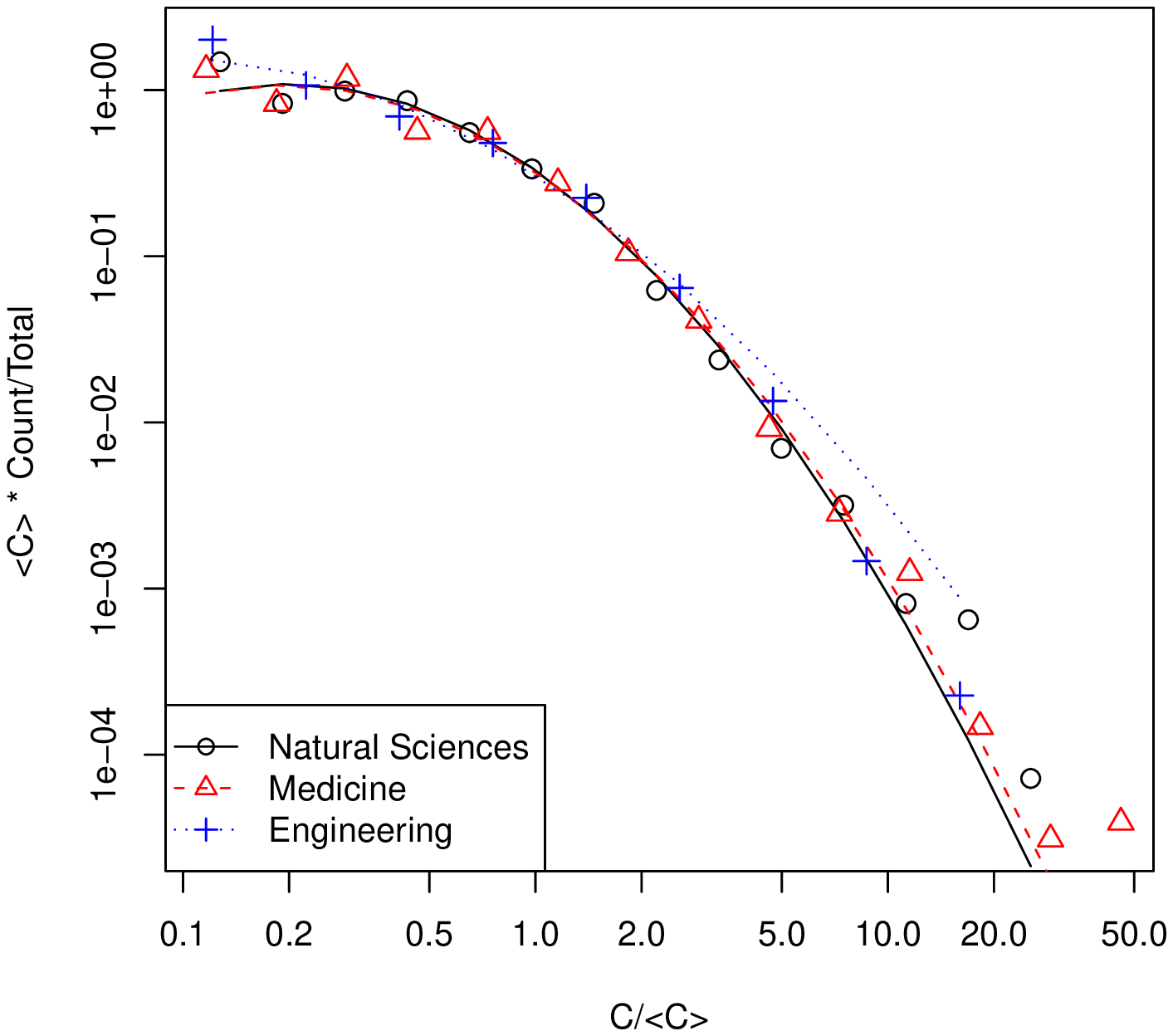}
\end{center}
\tsecaption{The distribution of $c_{\mathrm{f}}$ for
faculty data for all papers published in the year 2007 (left) or only those with
in $c_{\mathrm{f}}>0.1$ (right). The lines are the best fits to lognormal with one
free parameter. Publications with very low citation counts $c_{\mathrm{f}}<0.1$ can be seen to be poorly
described by lognormal distribution.}
\label{frad2007cutoff}
\end{figure}

\begin{figure}[htbp]\small
\begin{center}
\includegraphics[width=7cm]{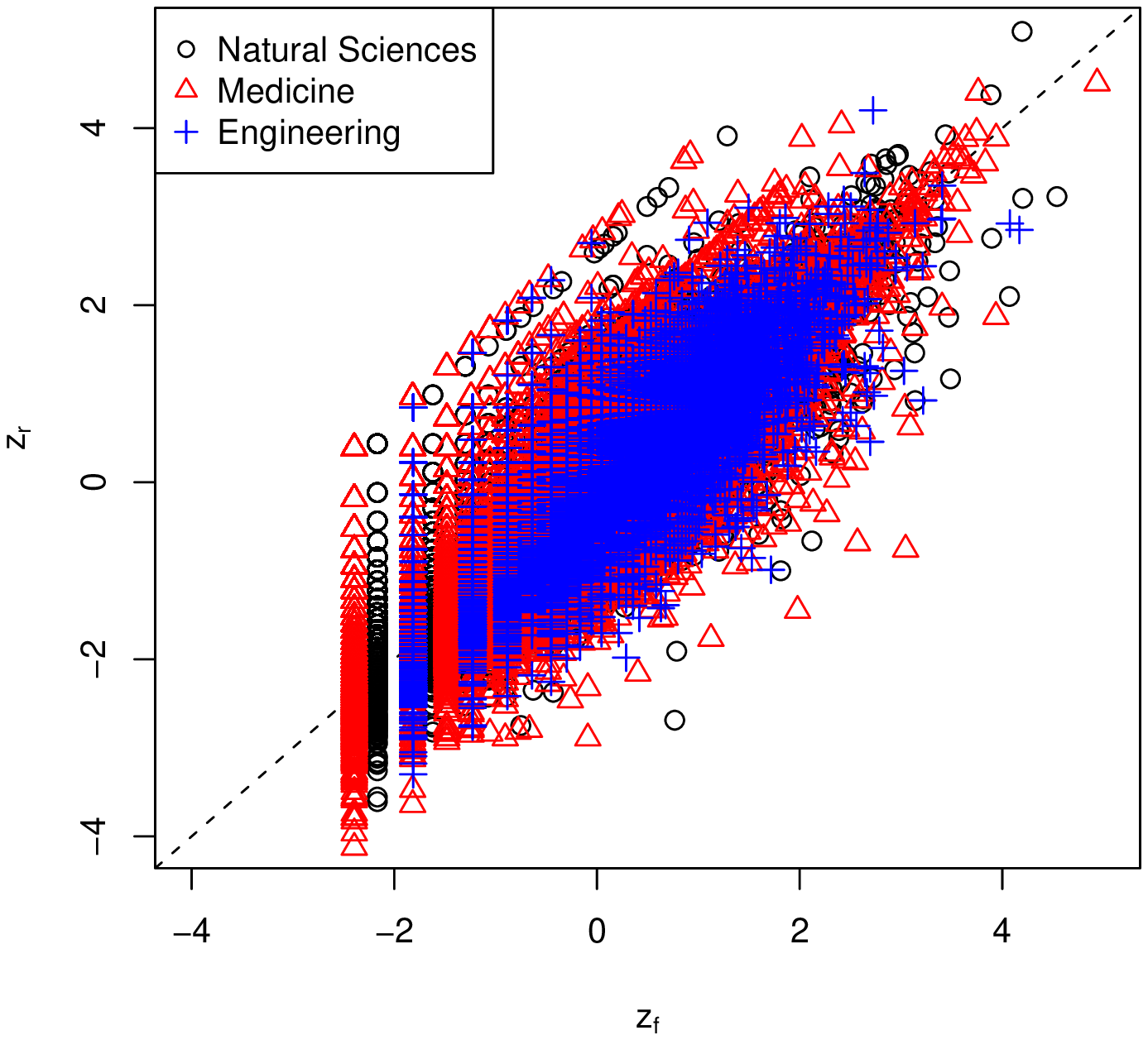}
\includegraphics[width=7cm]{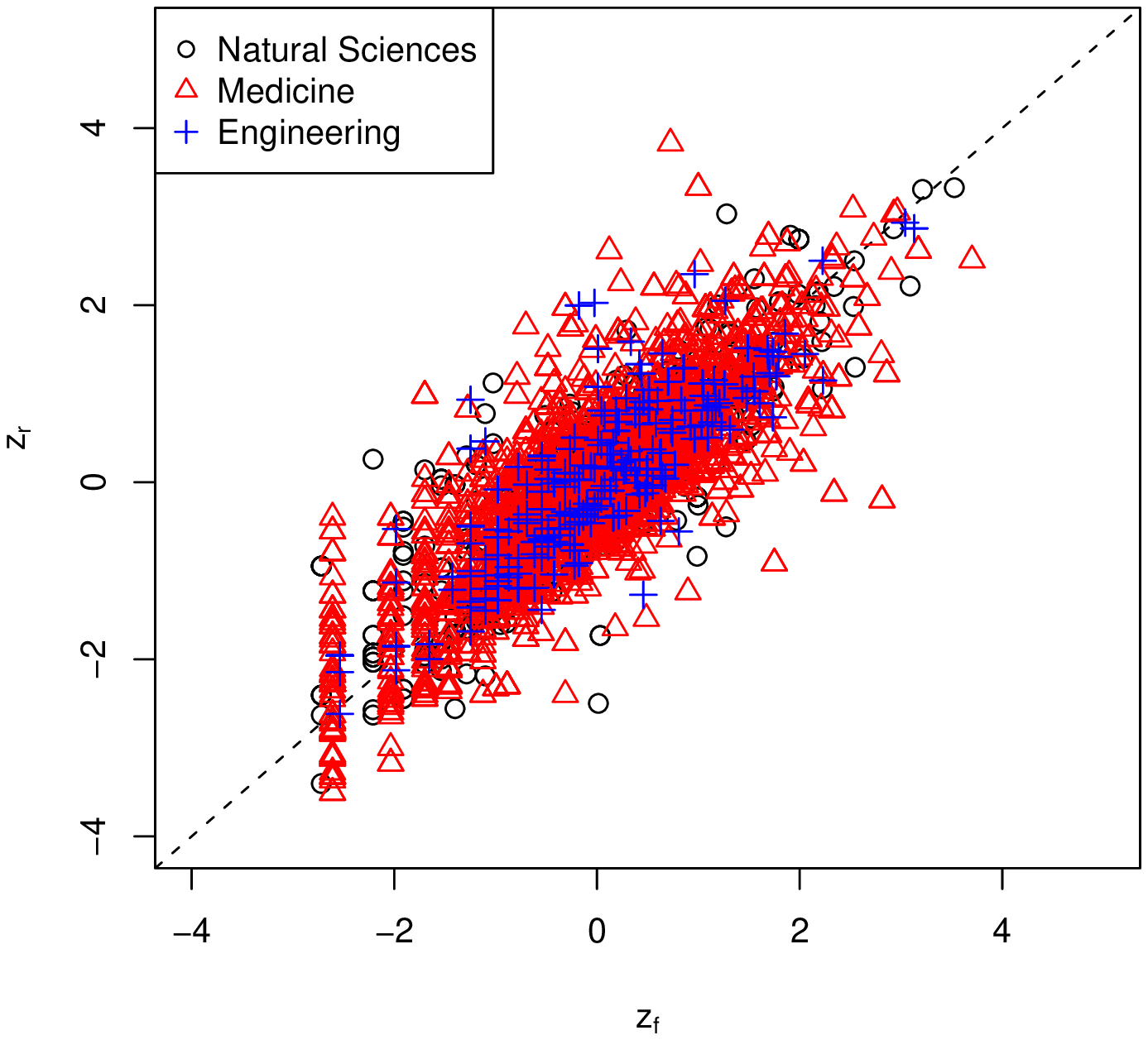}
\end{center}
\tsecaption{Scatter plot of $\zfindex$ vs.\ $\zrindex$ of \tref{zdef} for all items (left)
and review articles only (right).}
\label{fZCorrel}
\end{figure}

\newpage
\subsection*{Departments}

\begin{figure}[htbp]\small
\begin{center}
\includegraphics[width=7cm]{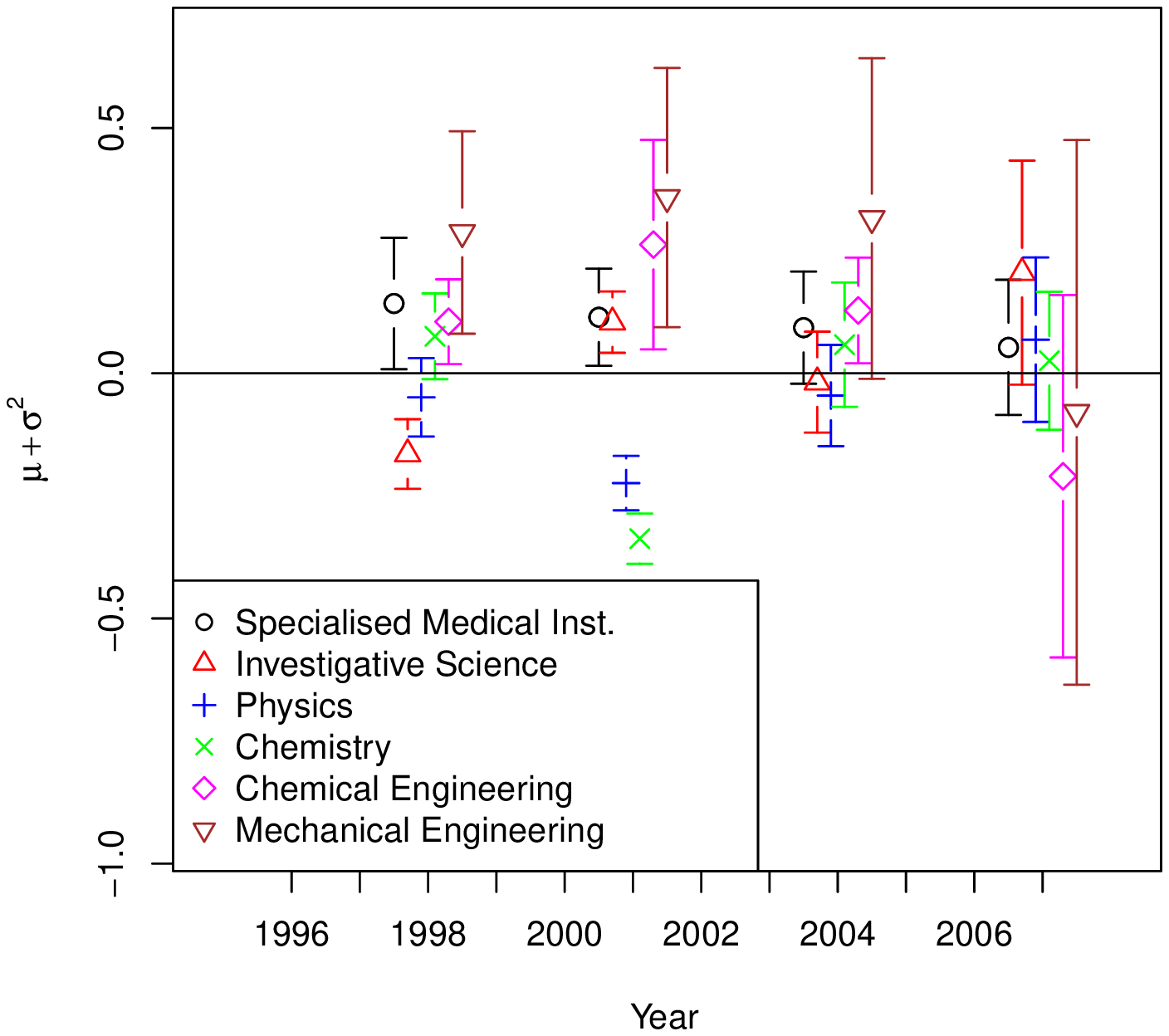}
\includegraphics[width=7cm]{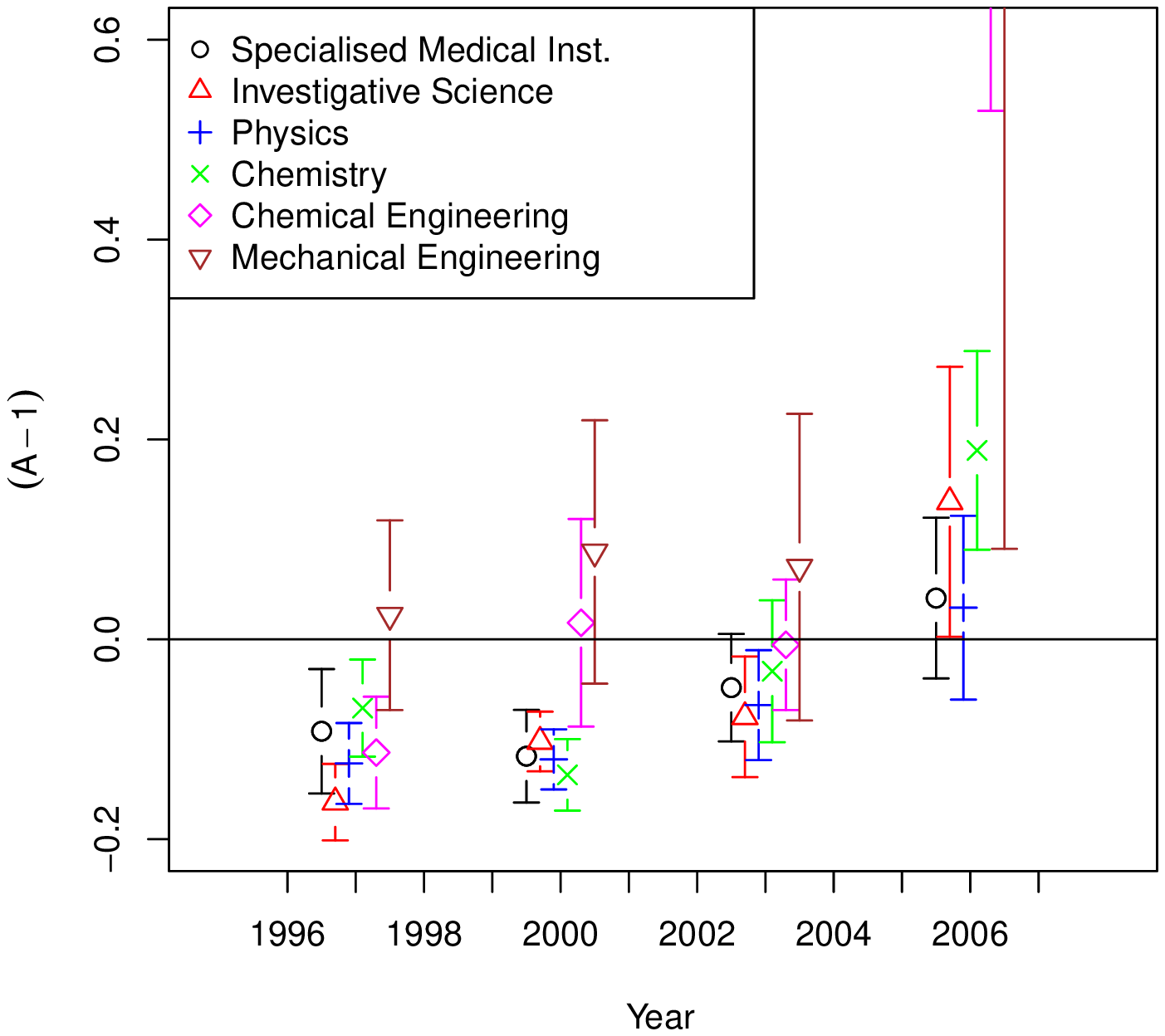}
\end{center}
\tsecaption{A plot of $(\mu+\sigma^2/2)$ (left) and $(A-1)$ (right)
against year obtained by fitting a lognormal to the
$\cfindex$ measure for which zero is expected for both quantities.
Not shown on the right Figure are data points corresponding to Chemical Engineering and
Mechanical Engineering for 2005--2007 corresponding to 1.4(9) and 0.9(7) respectively.}
\label{cdeptmusig}
\end{figure}

\begin{figure}[htbp]\small
\begin{center}
\includegraphics[width=7cm]{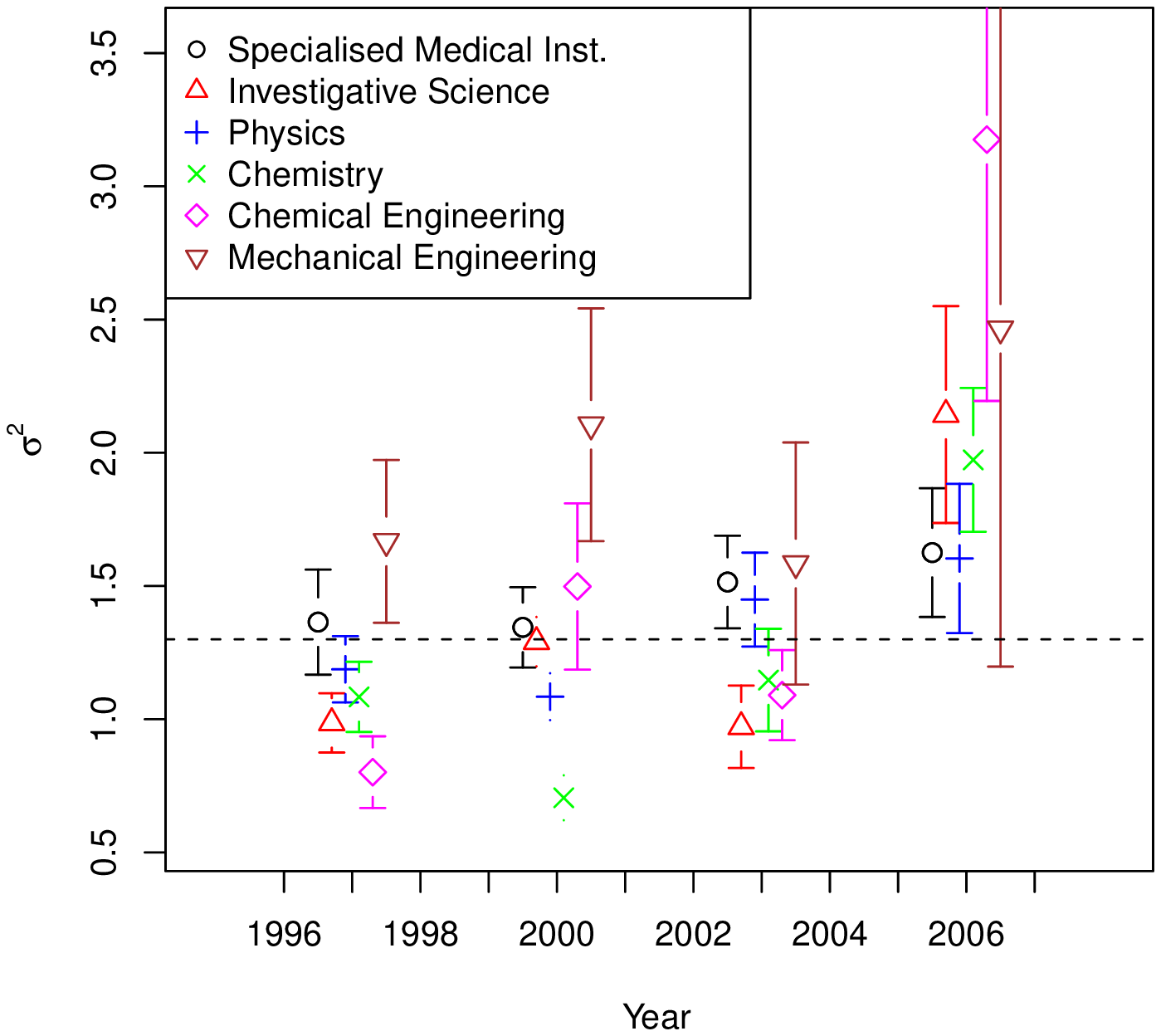}
\includegraphics[width=7cm]{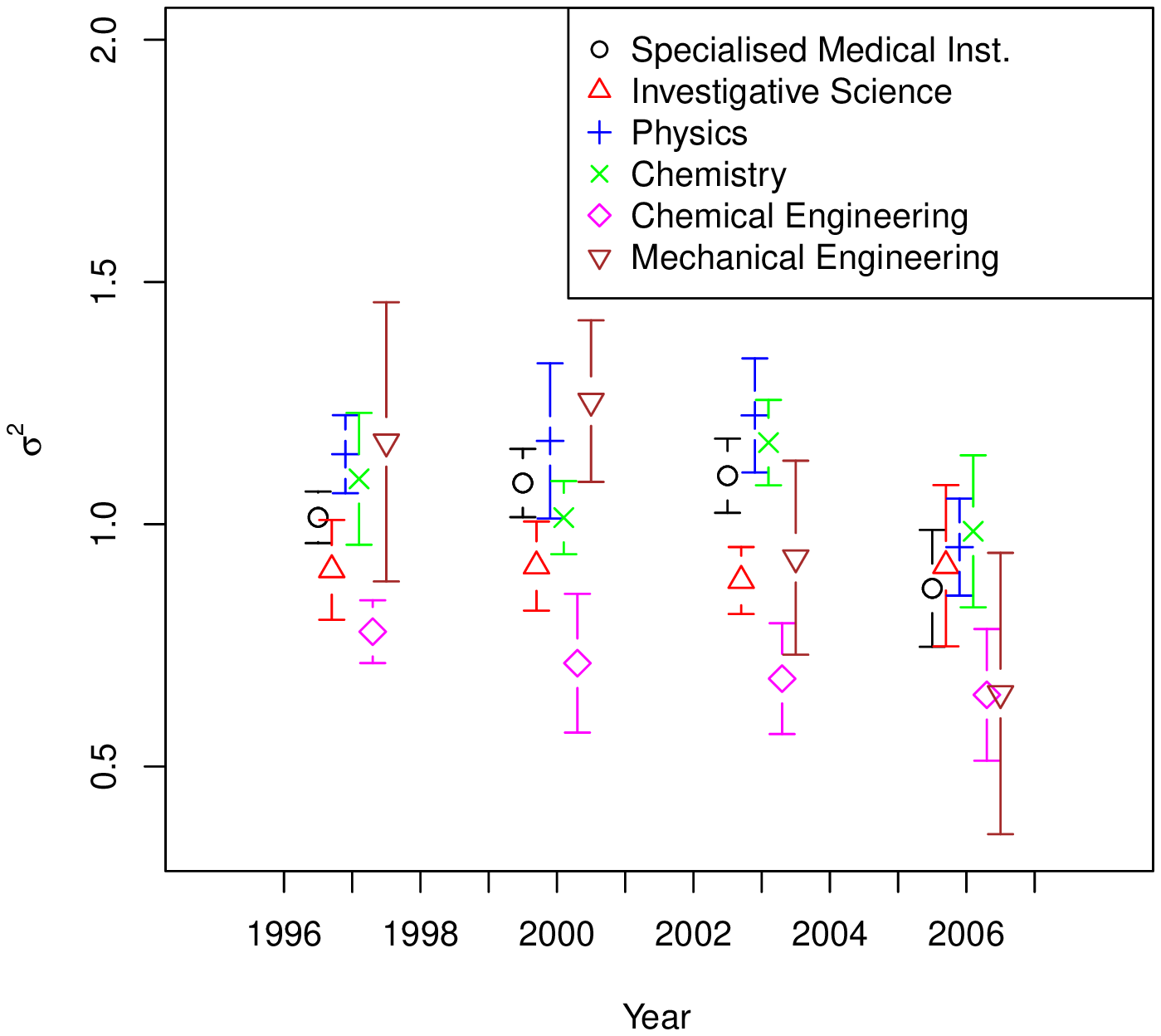}
\end{center}
\tsecaption{A plot of $\sigma^2$ against year resulting from a three
 parameter fit of a lognormal to the $c_{\mathrm{f}}$ (left) and
$c_{\mathrm{r}}$ (right) measure. Error bars mark one standard deviation.
The papers used for each point correspond to publications with $\crindex > 0.1$ binned into three year intervals
for the two most prolific departments in each faculty.}
\label{crdeptsig2b}
\end{figure}

\begin{figure}[htbp]\small
\begin{center}
\includegraphics[width=7cm]{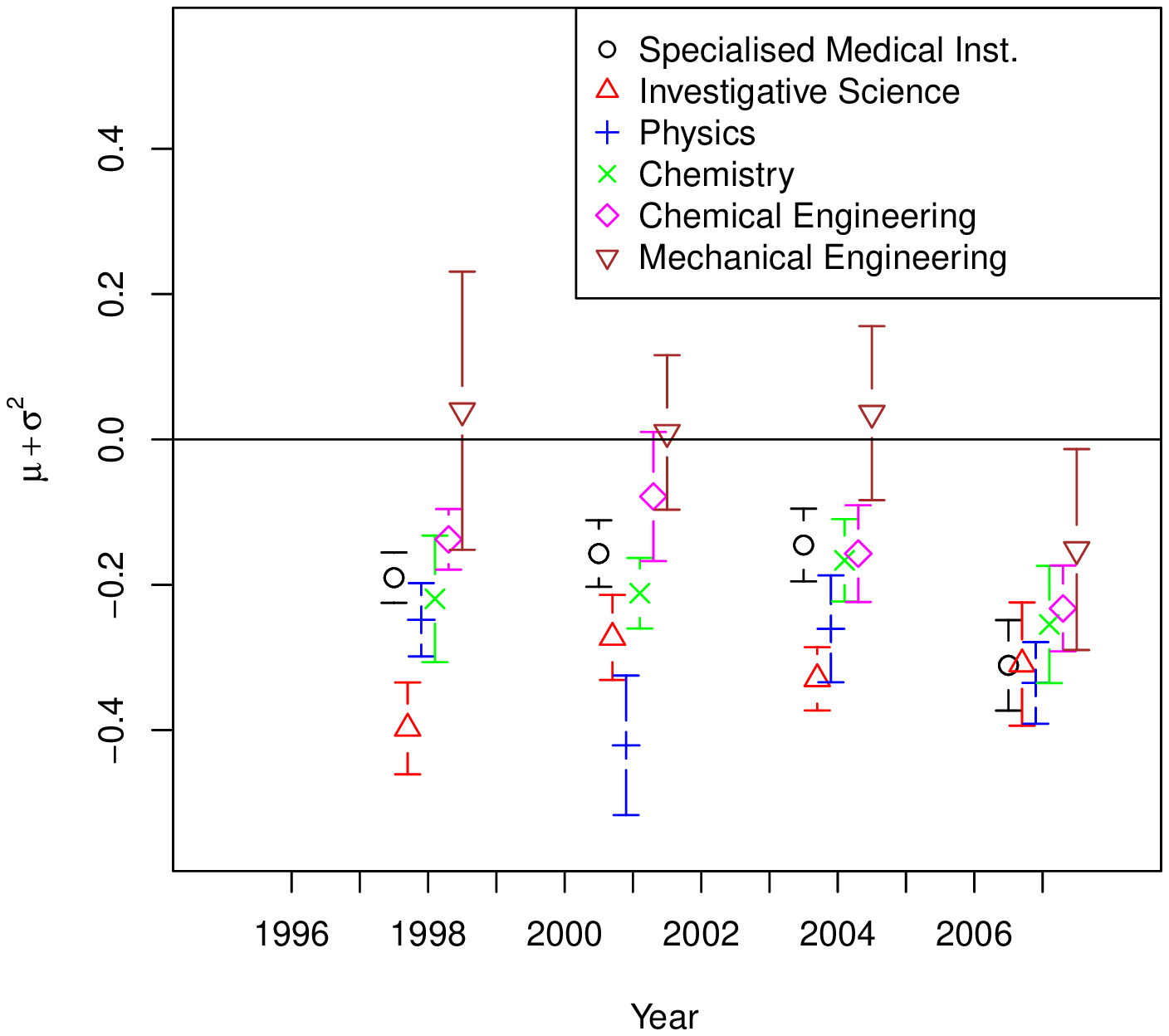}
\includegraphics[width=7cm]{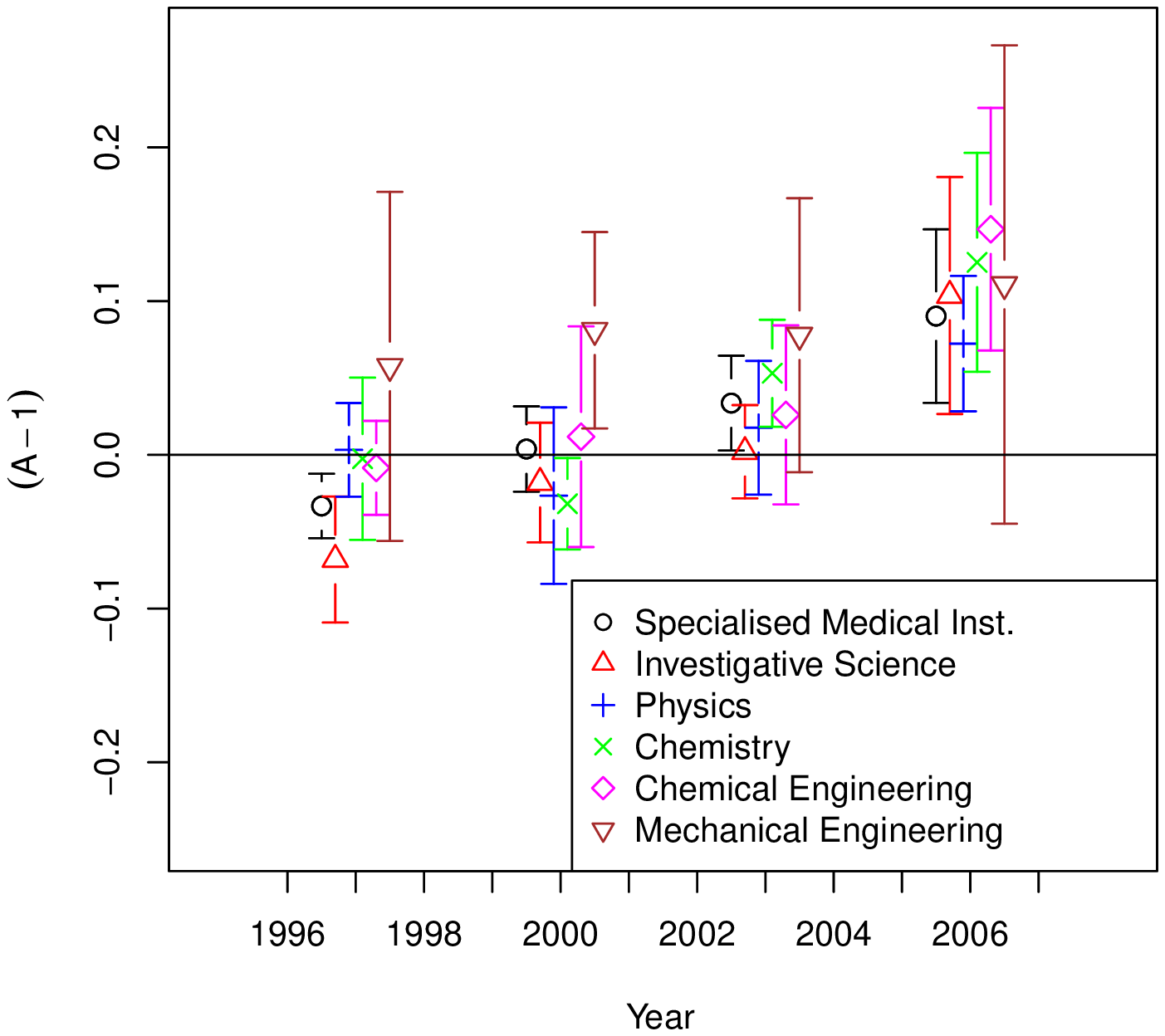}
\end{center}
\tsecaption{A plot of $(\mu+\sigma^2/2)$ (left) and $(A-1)$ (right)
against year obtained by fitting a lognormal to the
$\crindex$ measure for which zero is expected for both quantities.
Error bars are for one standard deviation.
The papers used for each point correspond to publications with $\crindex > 0.1$ binned into three year intervals
for the two most prolific departments in each faculty.}
\label{crdeptsig2}
\end{figure}

\clearpage
\begin{table}[htbp]\small
\begin{center}
\begin{tabular}{|l|l|l|l|l|l|l|l|}
\hline
Year & Department & $N_p$ & $c_0$ & $\sigma^{2}$ &  res.err./d.o.f.  & Bins & $\chi^2/\mathrm{d.o.f.}$ \\ \hline
\multirow{6}{*}{1996-1998}
& Sp.Medical Inst. & 1178 & 38.60 & 1.39(14) & 1.0 & 19 & 3.40 \\
& Invst.Sciences & 585 & 53.08 & 1.41(12) & 0.6 & 16 & 24.57 \\
& Physics & 1001 & 35.48 & 1.45(10) & 0.7 & 19 & 2.01 \\
& Chemistry & 458 & 35.20 & 1.11(9) & 0.3 & 19 & 1.15 \\
& Chem.Eng. & 318 & 21.75 & 0.85(9) & 0.3 & 18 & 1.68 \\
& Mech.Eng. & 209 & 15.76 & 1.40(19) & 0.7 & 9 & 1.22 \\\hline
\multirow{6}{*}{1999-2001}
& Sp.Medical Inst. & 1454 & 34.60 & 1.48(13) & 1.1 & 19 & 3.57 \\
& Invst.Sciences & 724 & 37.67 & 1.37(9) & 0.4 & 19 & 1.06 \\
& Physics & 1087 & 37.40 & 1.49(10) & 0.7 & 18 & 9.46 \\
& Chemistry & 492 & 40.93 & 1.19(13) & 0.7 & 15 & 55.83 \\
& Chem.Eng. & 358 & 17.86 & 1.29(18) & 0.5 & 17 & 2.57 \\
& Mech.Eng. & 232 & 12.58 & 1.61(23) & 0.7 & 9 & 1.50 \\\hline
\multirow{6}{*}{2002-2004}
& Sp.Medical Inst. & 1643 & 27.22 & 1.52(12) & 1.3 & 17 & 4.31 \\
& Invst.Sciences & 950 & 28.93 & 1.07(10) & 0.8 & 19 & 3.68 \\
& Physics & 1301 & 30.13 & 1.64(12) & 0.8 & 19 & 2.37 \\
& Chemistry & 616 & 25.71 & 1.14(11) & 0.7 & 16 & 4.06 \\
& Chem.Eng. & 420 & 14.04 & 1.01(10) & 0.4 & 16 & 3.02 \\
& Mech.Eng. & 307 & 9.48 & 1.26(26) & 1.5 & 9 & 4.49 \\\hline
\multirow{6}{*}{2005-2007}
& Sp.Medical Inst. & 1552 & 13.89 & 1.49(13) & 1.2 & 18 & 3.36 \\
& Invst.Sciences & 925 & 14.03 & 1.61(19) & 0.9 & 18 & 3.35 \\
& Physics & 1435 & 15.21 & 1.47(16) & 1.3 & 19 & 5.09 \\
& Chemistry & 614 & 13.81 & 1.53(15) & 0.4 & 19 & 2.02 \\
& Chem.Eng. & 452 & 6.86 & 1.19(37) & 1.4 & 14 & 12.25 \\
& Mech.Eng. & 289 & 5.90 & 1.22(51) & 2.8 & 8 & 11.49 \\\hline
\end{tabular}
\tsecaption{Department data from graphs generated using the $c_{\mathrm{f}}$
measure using 1 parameter fit, see Equation (\ref{eLognormal}).}
\label{tRadDepttab1para}
\end{center}
\end{table}

\clearpage\begin{table}[htbp]\small
\begin{center}
\begin{tabular}{|l|l|l|l|l|l|l|l|l|}
\hline
Year & Department & $N_p$ & $c_0$ & $\sigma^{2}$ &  res.err./d.o.f.  & Bins & $\chi^2/\mathrm{d.o.f.}$ \\ \hline
\multirow{7}{*}{1996-1998}
& Sp.Medical Inst. & 1178 & 38.60 & 1.39(14) & 0.99 & 19 & 3.4 \\
& Invst.Sciences & 585 & 53.08 & 1.41(12) & 0.63 & 16 & 24.6 \\
& Physics & 1001 & 35.48 & 1.45(10) & 0.66 & 19 & 2.0 \\
& Chemistry & 458 & 35.20 & 1.11(9) & 0.31 & 19 & 1.1 \\
& Chem.Eng. & 318 & 21.75 & 0.85(9) & 0.32 & 18 & 1.7 \\
& Mech.Eng. & 209 & 15.76 & 1.45(32) & 0.33 & 18 & 2.6 \\
& All & 3749 & NA & 1.07(5) & 1.55 & 19 & 38.2 \\\hline
\multirow{7}{*}{1999-2001}
& Sp.Medical Inst. & 1454 & 34.60 & 1.48(13) & 1.10 & 19 & 3.6 \\
& Invst.Sciences & 724 & 37.67 & 1.37(9) & 0.43 & 19 & 1.1 \\
& Physics & 1087 & 37.40 & 1.49(10) & 0.73 & 18 &  9.5 \\
& Chemistry & 492 & 40.93 & 1.19(13) & 0.68 & 15 & 55.8 \\
& Chem.Eng. & 358 & 17.86 & 1.29(18) & 0.48 & 17 & 2.6 \\
& Mech.Eng. & 232 & 12.58 & 1.74(40) & 0.37 & 18 & 3.4 \\
& All & 4347 & NA & 1.14(5) & 1.67 & 19 & 17.5 \\\hline
\multirow{7}{*}{2002-2004}
& Sp.Medical Inst. & 1643 & 27.22 & 1.52(12) & 1.31 & 17 & 4.3 \\
& Invst.Sciences & 950 & 28.93 & 1.07(10) & 0.83 & 19 & 3.7 \\
& Physics & 1301 & 30.13 & 1.64(12) & 0.82 & 19 & 2.4 \\
& Chemistry & 616 & 25.71 & 1.14(11) & 0.67 & 16 & 4.1 \\
& Chem.Eng. & 420 & 14.04 & 1.01(10) & 0.42 & 16 & 3.0 \\
& Mech.Eng. & 307 & 9.48 & 1.41(39) & 0.65 & 16 & 7.5 \\
& All & 5237 & NA & 1.10(2) & 1.01 & 19 & 11.7 \\\hline
\multirow{7}{*}{2005}
& Sp.Medical Inst. & 1552 & 13.89 & 1.49(13) & 1.20 & 18 & 3.4 \\
& Invst.Sciences & 925 & 14.03 & 1.61(19) & 0.90 & 18 & 3.4 \\
& Physics & 1435 & 15.21 & 1.47(16) & 1.25 & 19 & 5.1 \\
& Chemistry & 614 & 13.81 & 1.53(15) & 0.44 & 19 & 2.0 \\
& Chem.Eng. & 452 & 6.86 & 1.19(37) & 1.36 & 14 & 12.3 \\
& Mech.Eng. & 289 & 5.90 & 1.14(56) & 1.03 & 15 & 16.9 \\
& All & 5267 & NA & 1.20(15) & 5.42 & 19 & 28.7 \\\hline
\end{tabular}
\tsecaption{Department data from graphs generated using the $c_{\mathrm{f}}$
measure using 3 parameter fit,  $A \cdot F(\cfindex;\mu,\sigma^2)$.} \label{tRadDepttab3para}
\end{center}
\end{table}

\clearpage\begin{table}[htbp]\small
\begin{center}
\begin{tabular}{|l|l|l|l|l|l|l|l|l|}
\hline
Year & Department & $N_p$ & $\langle c_r \rangle$ & $\langle r \rangle$ & $\sigma^{2}$ & res.err./d.o.f. & Bins & $\chi^2/\mathrm{d.o.f.}$ \\ \hline
\multirow{7}{*}{1996-1998}
& Sp.Medical Inst. & 1160 & 1.73 & 34.71516 & 1.23(6) & 0.68 & 16 & 20.1 \\
& Invst.Sciences & 588 & 2.09 & 33.9882 & 1.43(15) & 0.58 & 18 & 4.9 \\
& Physics & 1016 & 1.61 & 29.06846 & 1.40(10) & 0.63 & 18 & 2.2 \\
& Chemistry & 452 & 1.72 & 33.00604 & 1.32(12) & 0.37 & 18 & 2.6 \\
& Chem.Eng. & 330 & 1.19 & 25.00847 & 0.90(5) & 0.19 & 17 & 1.9 \\
& Mech.Eng. & 194 & 0.88 & 24.52174 & 1.12(12) & 0.14 & 19 & 0.6 \\
& All & 3740 & NA & 0 & 1.22(8) & 2.16 & 19 & 12.8 \\\hline
\multirow{7}{*}{1999-2001}
& Sp.Medical Inst. & 1447 & 1.39 & 38.27095 & 1.23(6) & 0.64 & 19 & 2.4 \\
& Invst.Sciences & 723 & 1.52 & 34.19949 & 1.20(10) & 0.55 & 18 & 2.4 \\
& Physics & 1078 & 1.90 & 28.48872 & 1.69(19) & 1.36 & 16 & 13.0 \\
& Chemistry & 499 & 1.47 & 33.94849 & 1.26(7) & 0.26 & 19 & 1.1 \\
& Chem.Eng. & 340 & 0.76 & 31.96345 & 0.76(8) & 0.30 & 19 & 1.7 \\
& Mech.Eng. & 230 & 0.79 & 22.90421 & 1.15(12) & 0.18 & 18 & 1.0 \\
& All & 4317 & NA & 0 & 1.24(8) & 2.73 & 19 & 28.1 \\\hline
\multirow{7}{*}{2002-2004}
& Sp.Medical Inst. & 1605 & 1.05 & 38.30216 & 1.21(7) & 0.83 & 19 & 3.3 \\
& Invst.Sciences & 934 & 1.06 & 38.889 & 1.22(11) & 0.69 & 19 & 2.6 \\
& Physics & 1361 & 1.41 & 28.85268 & 1.48(12) & 1.02 & 18 & 4.6 \\
& Chemistry & 613 & 1.03 & 36.20420 & 1.28(9) & 0.36 & 18 & 1.0 \\
& Chem.Eng. & 386 & 0.62 & 32.1796 & 0.79(8) & 0.35 & 17 & 1.8 \\
& Mech.Eng. & 252 & 0.59 & 24.49508 & 0.87(10) & 0.19 & 18 & 0.9 \\
& All & 5151 & NA & 0 & 1.20(8) & 2.84 & 19 &  9.6 \\\hline
\multirow{7}{*}{2005-2007}
& Sp.Medical Inst. & 1309 & 0.59 & 39.91351 & 1.07(14) & 1.39 & 18 & 6.3 \\
& Invst.Sciences & 775 & 0.59 & 41.52976 & 1.10(16) & 0.82 & 19 & 3.7 \\
& Physics & 1295 & 0.75 & 32.43258 & 1.21(14) & 1.18 & 19 & 7.8 \\
& Chemistry & 509 & 0.60 & 38.56109 & 1.08(15) & 0.51 & 18 & 2.9 \\
& Chem.Eng. & 298 & 0.38 & 33.19599 & 0.69(11) & 0.29 & 19 & 1.7 \\
& Mech.Eng. & 216 & 0.36 & 25.71429 & 0.71(14) & 0.33 & 16 & 2.3 \\
& All & 4402 & NA & 0 & 1.01(11) & 4.06 & 19 & 21.4 \\\hline
\end{tabular}
\tsecaption{Department data from graphs generated using the $c_{\mathrm{r}}$
measure using 1 parameter fit, see Equation (\ref{eLognormal}).} \label{tOurDepttab1para}
\end{center}
\end{table}

\clearpage\begin{table}[htbp]\small
\begin{center}
\begin{tabular}{|l|l|l|l|l|l|l|}
\hline
Year & Department & $N_p$ & $\langle c_{\mathrm{r}} \rangle$ & $\sigma^{2}$ & $\mu + \frac{\sigma^{2}}{2}$ & res.err./d.o.f. \\ \hline
\multirow{6}{*}{1996-1998}
& Sp.Medical Inst. & 1376 & 38.60 & 1.4(2) & 0.1(1) & 1.0 \\
& Invst.Sciences & 717 & 53.08 & 1.0(1) & -0.1(1) & 0.5 \\
& Physics & 1195 & 35.48 & 1.2(1) & -0.1(1) & 0.6 \\
& Chemistry & 510 & 35.20 & 1.1(1) & 0.1(1) & 0.3 \\
& Chem.Eng. & 363 & 21.75 & 0.8(1) & 0.1(1) & 0.3 \\
& Mech.Eng. & 228 & 15.76 & 1.7(4) & 0.3(2) & 0.8 \\\hline
\multirow{6}{*}{1999-2001}
& Sp.Medical Inst. & 1761 & 34.60 & 1.3(2) & 0.1(1) & 1.0 \\
& Invst.Sciences & 849 & 37.67 & 1.3(1) & 0.1(1) & 0.3 \\
& Physics & 1286 & 37.40 & 1.1(1) & -0.2(1) & 0.6 \\
& Chemistry & 566 & 40.93 & 0.7(1) & -0.3(1) & 0.5 \\
& Chem.Eng. & 385 & 17.86 & 1.5(4) & 0.3(2) & 0.5 \\
& Mech.Eng. & 266 & 12.58 & 2.1(6) & 0.4(3) & 0.9 \\\hline
\multirow{6}{*}{2002-2004}
& Sp.Medical Inst. & 1896 & 27.22 & 1.5(2) & 0.1(1) & 1.5 \\
& Invst.Sciences & 1050 & 28.93 & 1.0(2) & -0.0(1) & 0.9 \\
& Physics & 1576 & 30.13 & 1.4(2) & -0.1(1) & 0.9 \\
& Chemistry & 669 & 25.71 & 1.1(2) & 0.1(1) & 0.8 \\
& Chem.Eng. & 451 & 14.04 & 1.1(2) & 0.1(1) & 0.5 \\
& Mech.Eng. & 307 & 9.48 & 1.6(6) & 0.3(3) & 2 \\\hline
\multirow{6}{*}{2005-2007}
& Sp.Medical Inst. & 1763 & 13.89 & 1.6(3) & 0.1(1) & 1.4 \\
& Invst.Sciences & 1043 & 14.03 & 2.1(6) & 0.2(2) & 1.0 \\
& Physics & 1605 & 15.21 & 1.6(4) & 0.1(2) & 1.5 \\
& Chemistry & 668 & 13.81 & 2.0(4) & 0.0(1) & 0.4 \\
& Chem.Eng. & 452 & 6.86 & 3.2(17) & -0.2(4) & 1.1 \\
& Mech.Eng. & 289 & 5.90 & 2.5(20) & -0.1(6) & 3.2 \\\hline
\end{tabular}
\tsecaption{Department data from graphs generated using the $c_{\mathrm{f}}$
measure using 3 parameter fit,  $A \cdot F(\cfindex;\mu,\sigma^2)$.} \label{tOurDepttab3para}
\end{center}
\end{table}

\clearpage
\begin{table}[htbp]\small
  \centering
    \begin{tabular}{|r|r|r|r|r|r|}
    \hline
    Measure & $c_{\mathrm{f,r}}^{*}$ & Bins  & $\chi^2/\mathrm{d.o.f}$ Min & $\chi^2/\mathrm{d.o.f}$ Max & $\chi^2/\mathrm{d.o.f}$ Mean \bigstrut\\
    \hline
    \multicolumn{1}{|c|}{\multirow{10}[20]{*}{$c_{\mathrm{f}}$}} & \multicolumn{1}{c|}{\multirow{5}[10]{*}{0.0}} & 8     & 2.9   & 6.4   & 4.6 \bigstrut\\
\cline{3-6}    \multicolumn{1}{|c|}{} & \multicolumn{1}{c|}{} & 9     & 4.5   & 10.5  & 6.9 \bigstrut\\
\cline{3-6}    \multicolumn{1}{|c|}{} & \multicolumn{1}{c|}{} & 12    & 33.8  & 33.8  & 33.8 \bigstrut\\
\cline{3-6}    \multicolumn{1}{|c|}{} & \multicolumn{1}{c|}{} & 13    & 8.8   & 12.7  & 10.8 \bigstrut\\
\cline{3-6}    \multicolumn{1}{|c|}{} & \multicolumn{1}{c|}{} & 14    & 4.6   & 38.4  & 19.6 \bigstrut\\
\cline{2-6}    \multicolumn{1}{|c|}{} & \multicolumn{1}{c|}{\multirow{5}[10]{*}{0.1}} & 8     & 3.7   & 3.7   & 3.7 \bigstrut\\
\cline{3-6}    \multicolumn{1}{|c|}{} & \multicolumn{1}{c|}{} & 9     & 1.6   & 5.7   & 2.7 \bigstrut\\
\cline{3-6}    \multicolumn{1}{|c|}{} & \multicolumn{1}{c|}{} & 12    & 24.4  & 24.4  & 24.4 \bigstrut\\
\cline{3-6}    \multicolumn{1}{|c|}{} & \multicolumn{1}{c|}{} & 13    & 3.7   & 5.6   & 4.5 \bigstrut\\
\cline{3-6}    \multicolumn{1}{|c|}{} & \multicolumn{1}{c|}{} & 14    & 1.7   & 9.0   & 3.4 \bigstrut\\
    \hline
    \multicolumn{1}{|c|}{\multirow{10}[20]{*}{$c_{\mathrm{r}}$}} & \multicolumn{1}{c|}{\multirow{5}[10]{*}{0.0}} & 8     & 27.2  & 27.2  & 27.2 \bigstrut\\
\cline{3-6}    \multicolumn{1}{|c|}{} & \multicolumn{1}{c|}{} & 9     & 0.8   & 5.5   & 2.0 \bigstrut\\
\cline{3-6}    \multicolumn{1}{|c|}{} & \multicolumn{1}{c|}{} & 12    & 35.4  & 35.4  & 35.4 \bigstrut\\
\cline{3-6}    \multicolumn{1}{|c|}{} & \multicolumn{1}{c|}{} & 13    & 2.4   & 3.9   & 3.1 \bigstrut\\
\cline{3-6}    \multicolumn{1}{|c|}{} & \multicolumn{1}{c|}{} & 14    & 0.8   & 17.0  & 4.5 \bigstrut\\
\cline{2-6}    \multicolumn{1}{|c|}{} & \multicolumn{1}{c|}{\multirow{5}[10]{*}{0.1}} & 8     & 4.9   & 7.7   & 6.3 \bigstrut\\
\cline{3-6}    \multicolumn{1}{|c|}{} & \multicolumn{1}{c|}{} & 9     & 0.1   & 4.3   & 1.9 \bigstrut\\
\cline{3-6}    \multicolumn{1}{|c|}{} & \multicolumn{1}{c|}{} & 11    & 120.5 & 120.5 & 120.5 \bigstrut\\
\cline{3-6}    \multicolumn{1}{|c|}{} & \multicolumn{1}{c|}{} & 13    & 5.4   & 23.6  & 12.9 \bigstrut\\
\cline{3-6}    \multicolumn{1}{|c|}{} & \multicolumn{1}{c|}{} & 14    & 2.1   & 17.2  & 5.3 \bigstrut\\
    \hline
    \end{tabular}
      \tsecaption{Table of $\chi^2$ values for the one parameter lognormal goodness of fit to departmental data. c* denotes the threshold
  below which publications were not included in the fitting.}
  \label{tabChi2dept}%
\end{table}%

\newpage
\section*{arXiv}

\begin{table}[htbp]\small
\begin{center}
\begin{tabular}{|r|c|c|c|c|}
\hline
 & \multicolumn{2}{|c|}{All papers}  & \multicolumn{2}{|c|}{Papers with $c,r >0$}  \\ \hline
\textbf{Sub-archives} & \textbf{Number} & \textbf{\%} & \textbf{Number} & \textbf{\% total} \\ \hline
astro-ph & 69934 & 34.1\% & 53032 & 30.2\% \\ \hline
cond-mat & 1 & 0.0\% & 0 & 0.0\% \\ \hline
gr-qc & 15675 & 7.6\% & 13843 & 7.9\% \\ \hline
hep-ex & 7193 & 3.5\% & 6373 & 3.6\% \\ \hline
hep-lat & 7597 & 3.7\% & 6905 & 3.9\% \\ \hline
hep-ph & 49632 & 24.2\% & 46555 & 26.5\% \\ \hline
hep-th & 40891 & 19.9\% & 36871 & 21.0\% \\ \hline
nucl-ex & 2643 & 1.3\% & 2287 & 1.3\% \\ \hline
nucl-th & 11514 & 5.6\% & 9748 & 5.6\% \\ \hline
quant-ph & 1 & 0.0\% & 0 & 0.0\% \\ \hline
\textbf{TOTAL} & \textbf{205081} & \multicolumn{1}{l|}{\textbf{}} & \textbf{175614} & \multicolumn{1}{l|}{\textbf{}} \\ \hline
\end{tabular}
\end{center}
\tsecaption{Different sub-archives in arXiv data.}
\label{tarXivNumbers}
\end{table}

\begin{figure}[htbp]\small
\begin{center}
\includegraphics[width=7cm]{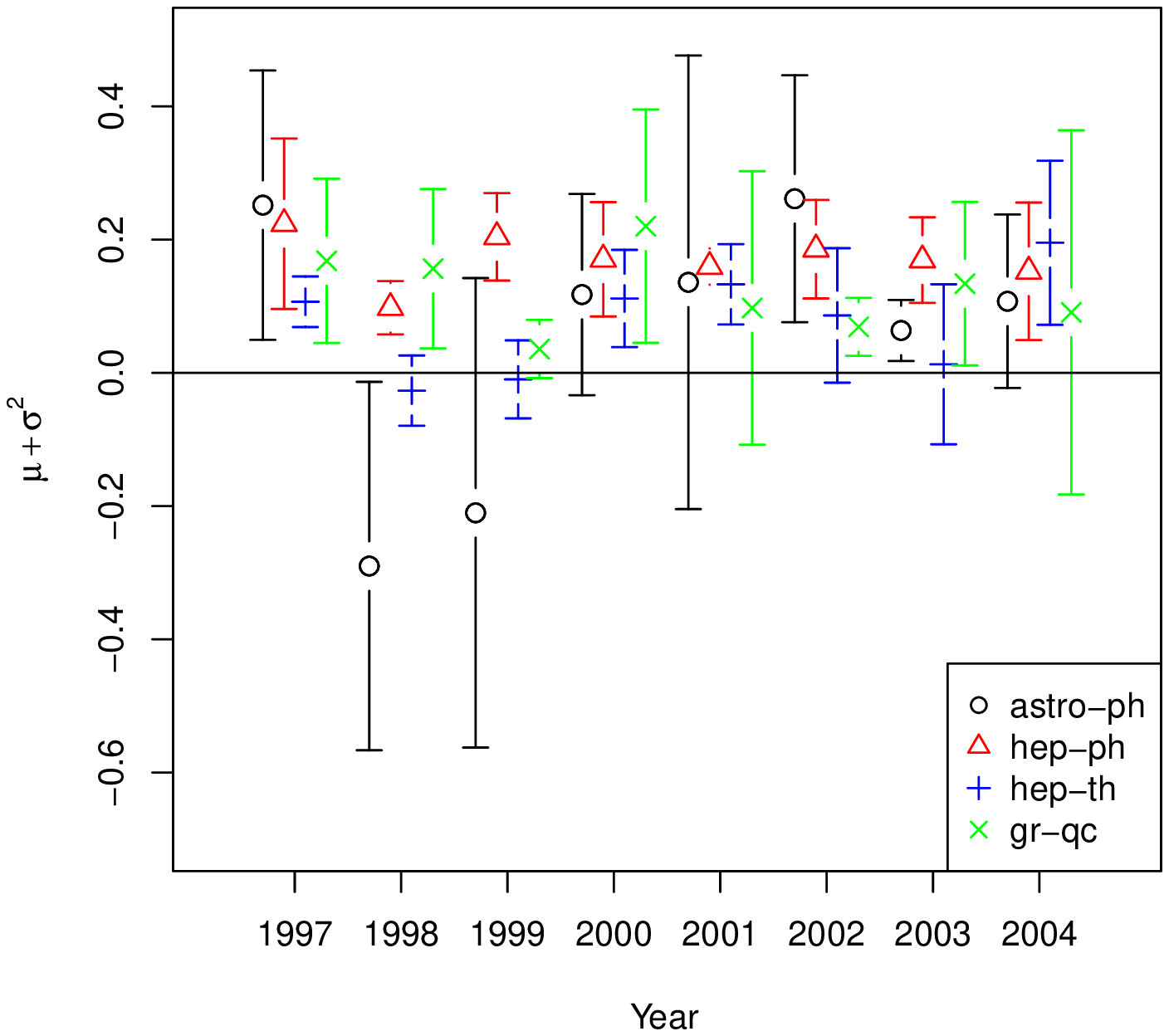}
\includegraphics[width=7cm]{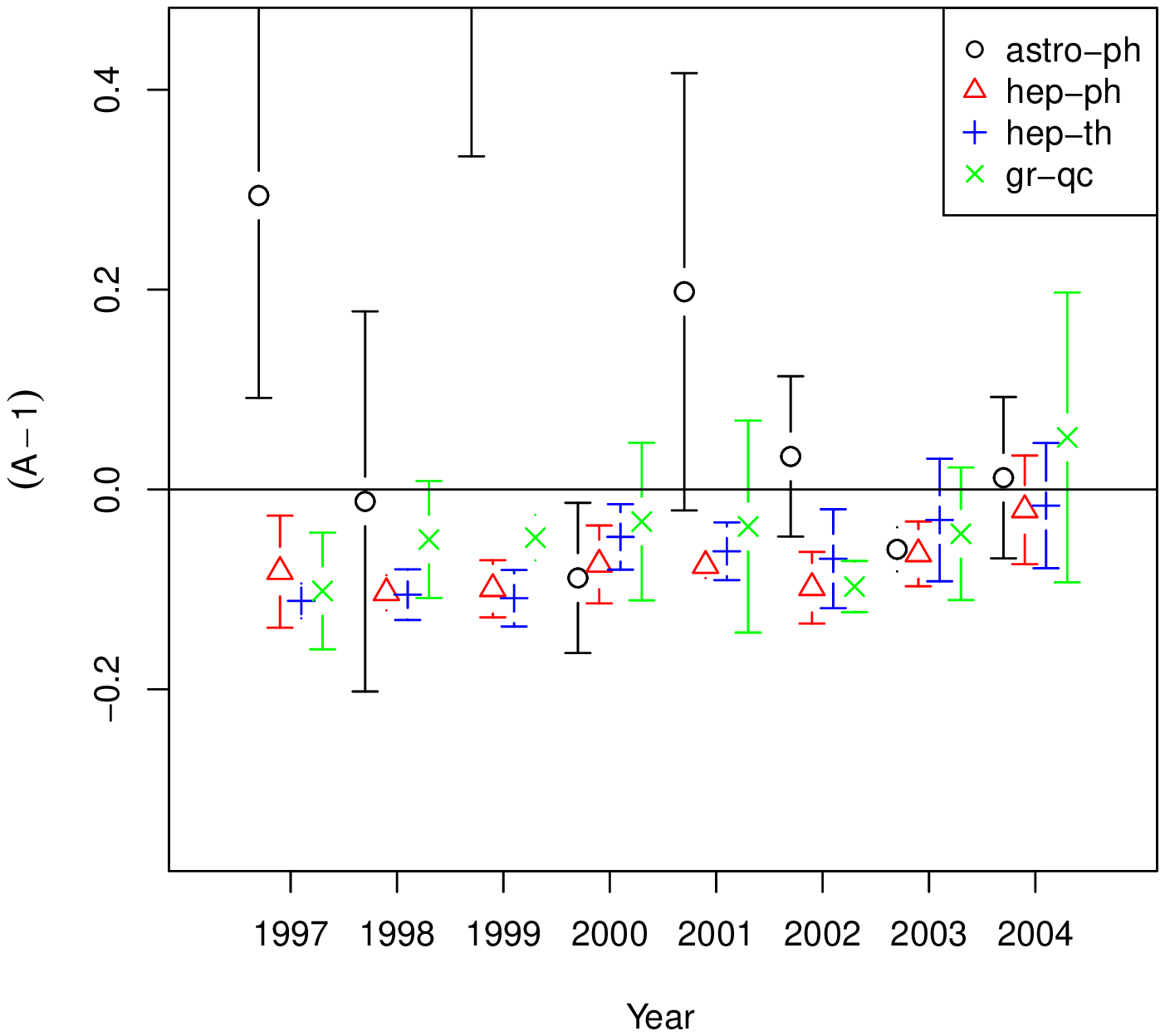}
\end{center}
\tsecaption{A plot of $(\mu+\sigma^2/2)$ (left) and $(A-1)$ (right)
against year obtained by fitting a lognormal to the
$\cfindex$ measure for which zero is expected for both quantities.}
\label{carXivmusigb}
\end{figure}

\begin{figure}[htbp]\small
\begin{center}
\includegraphics[width=7cm]{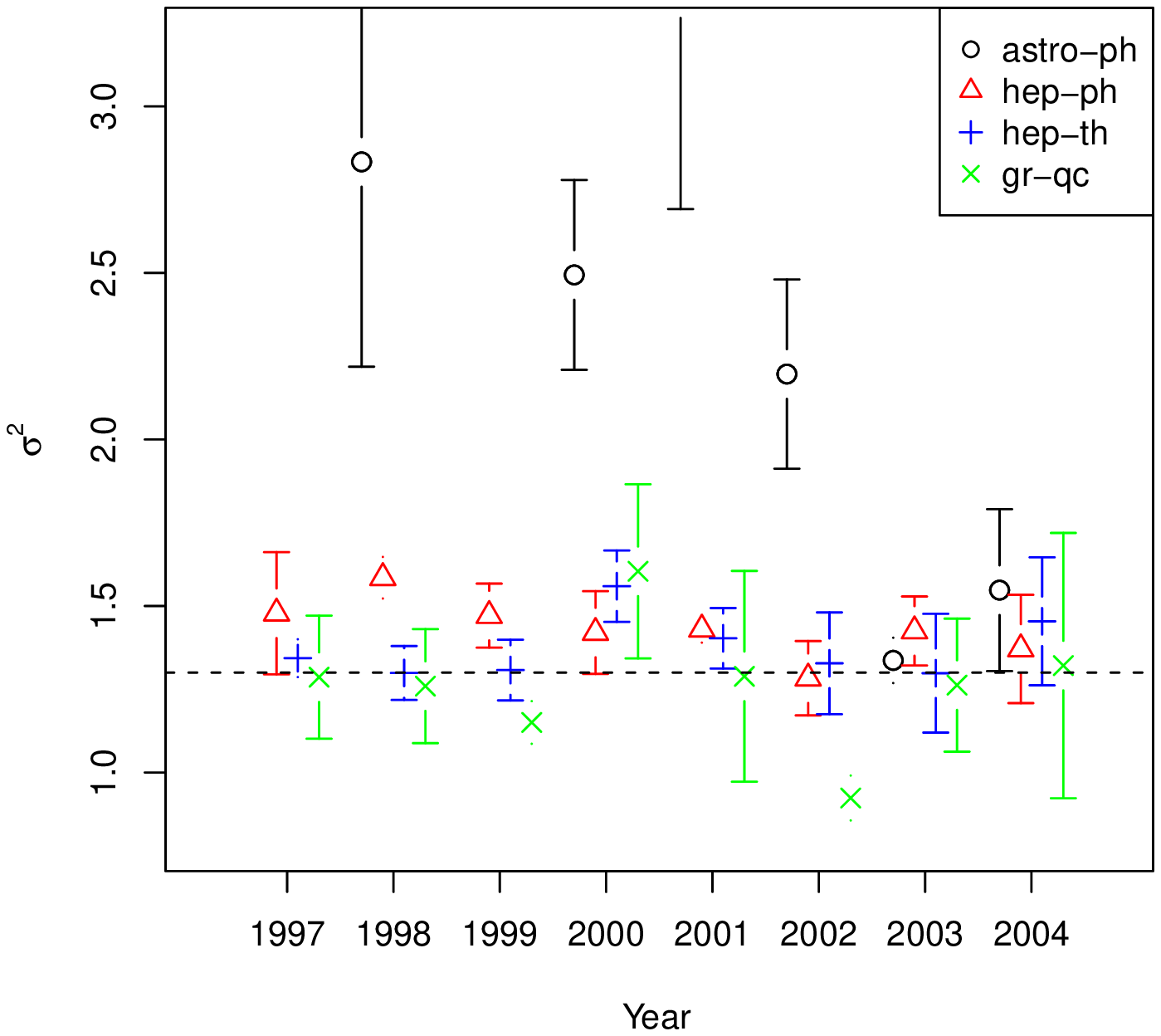}
\includegraphics[width=7cm]{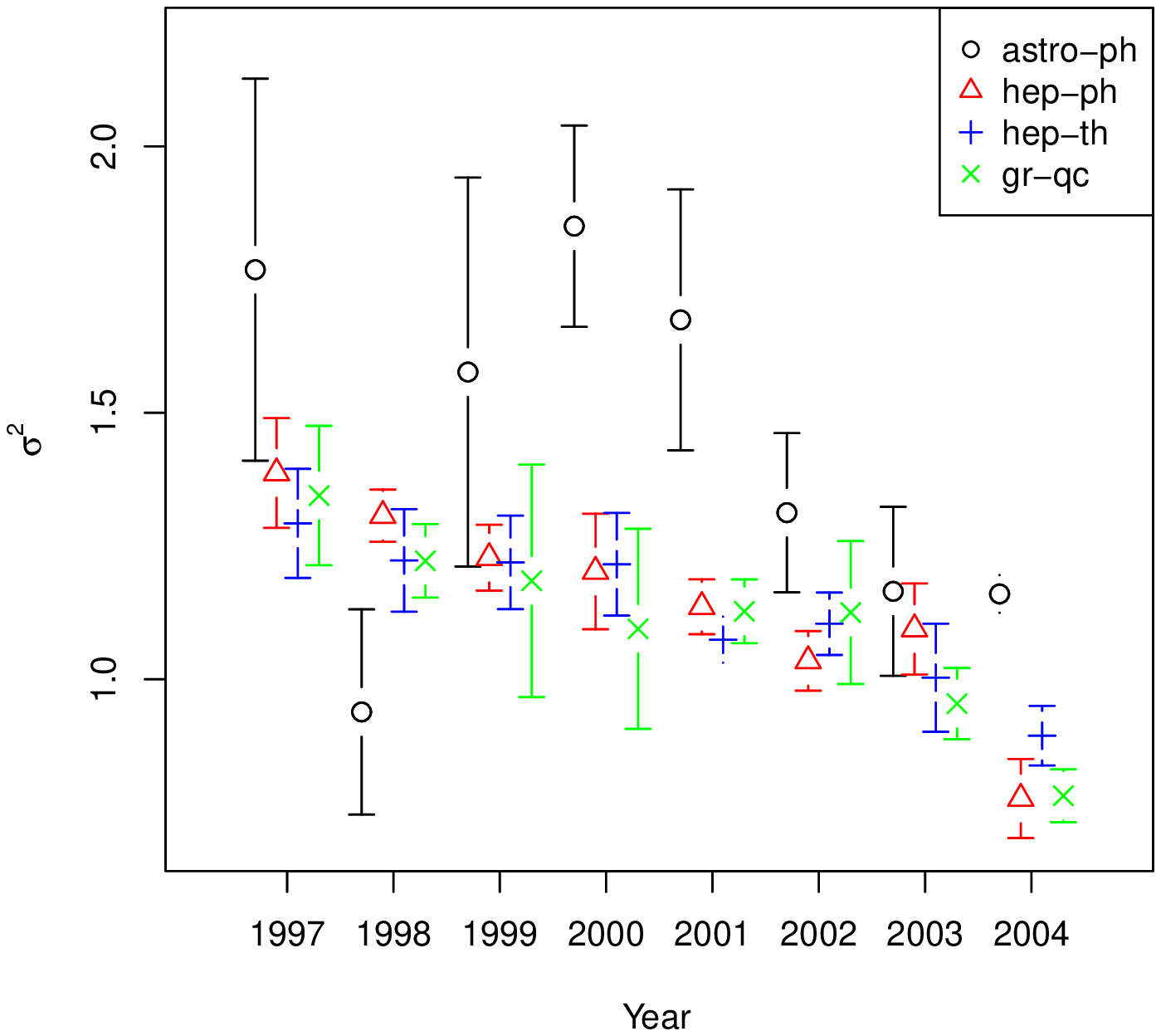}
\end{center}
\tsecaption{A plot of $\sigma^2$ against year resulting from a three
parameter fit of a lognormal to the $c_{\mathrm{f}}$ (left)
and $c_{\mathrm{r}}$ (right) measure.
Error bars correspond to one standard deviation.
Omitted from the left plot are markers corresponding to astro-ph 1997, astro-ph 1999 and astro-ph 2001 with
values $5.57\pm1.11$, $6.19\pm2.52$ and $3.34\pm1.18$.
}
\label{crarXivsig2}
\end{figure}
\newpage

\begin{figure}[htbp]\small
\begin{center}
\includegraphics[width=7cm]{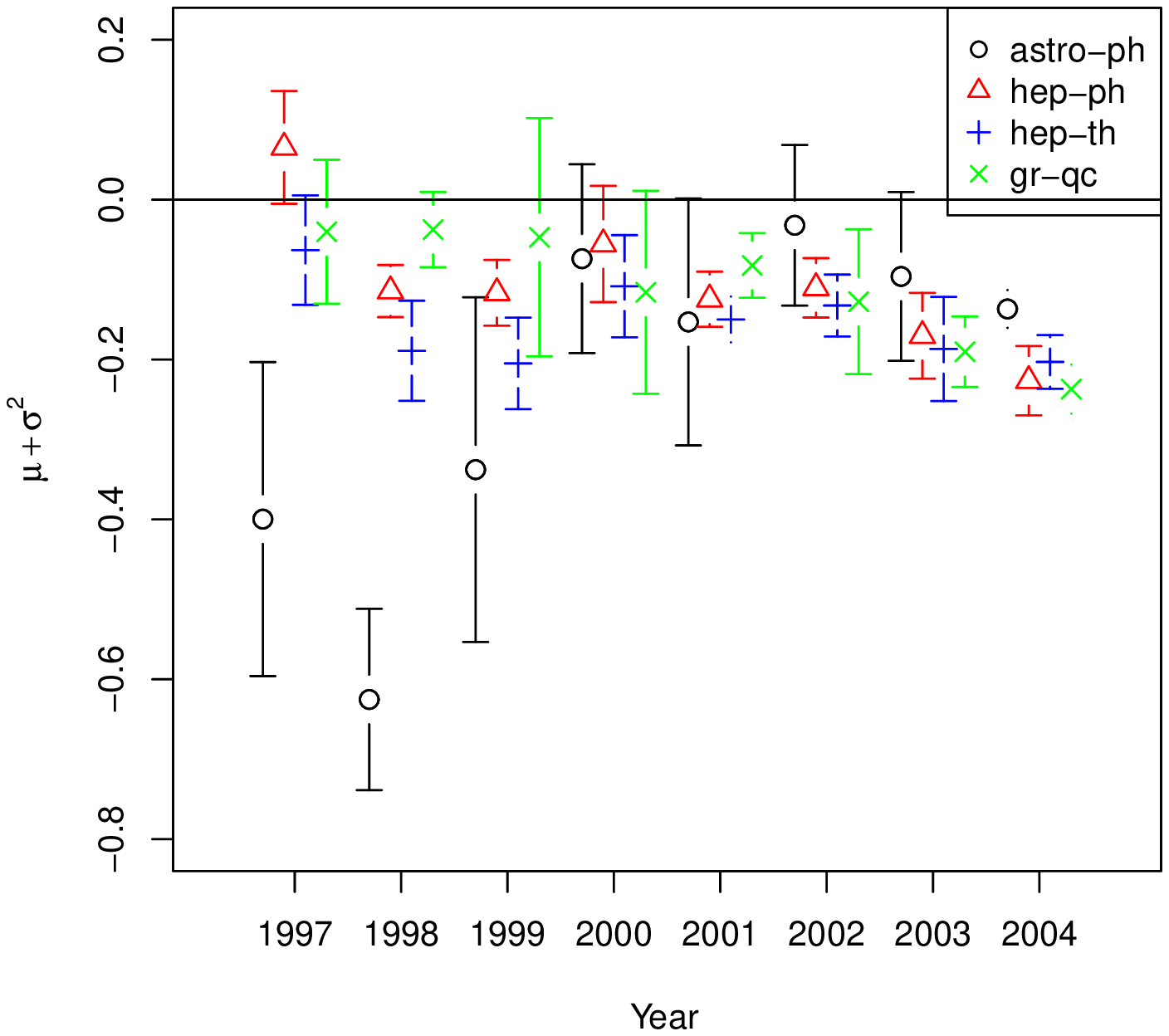}
\includegraphics[width=7cm]{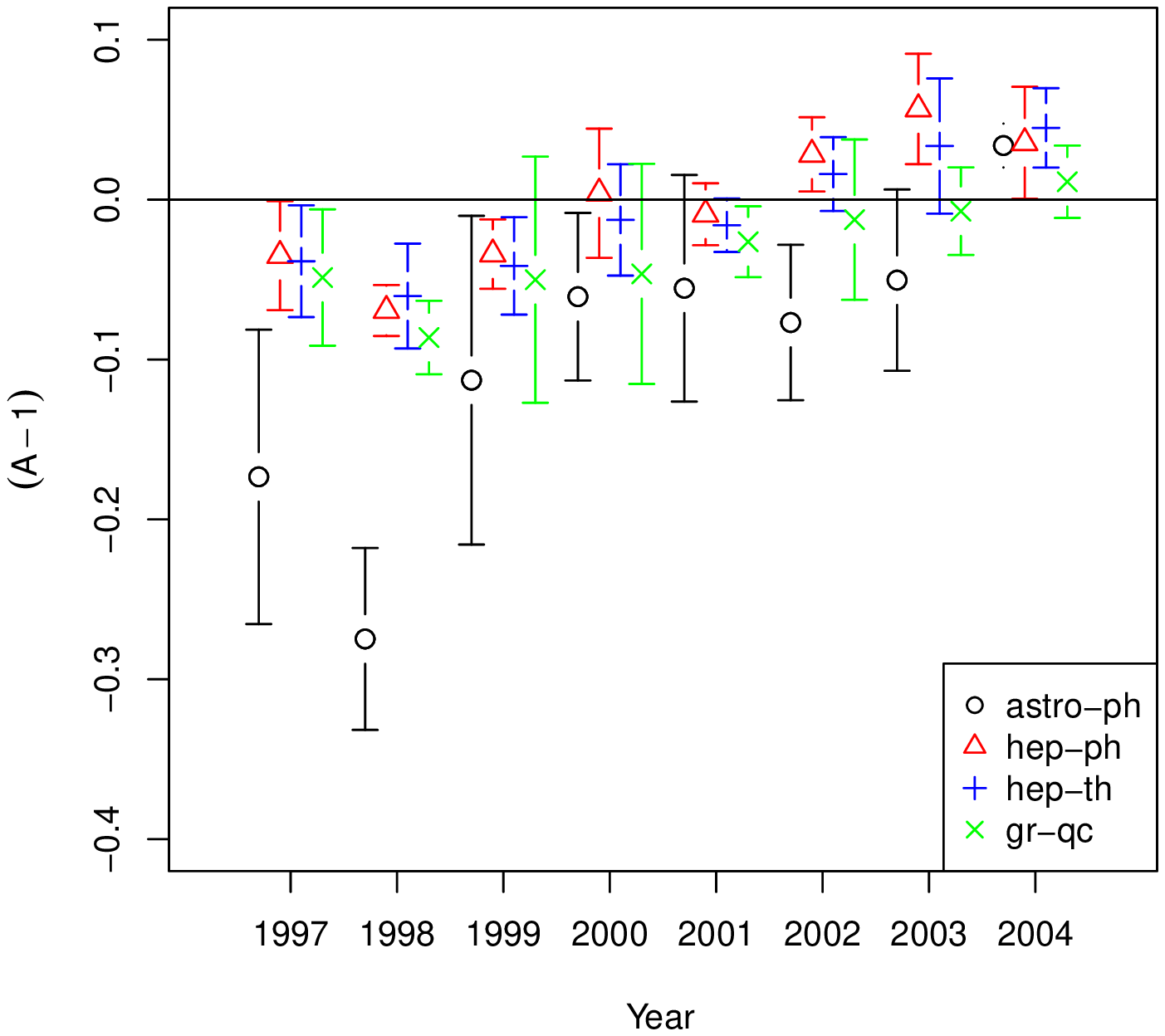}
\end{center}
\tsecaption{A plot of $(\mu+\sigma^2/2)$ (left) and $(A-1)$ (right)
against year obtained by fitting a lognormal to the
$\crindex$ measure for which zero is expected for both quantities.}
\label{crarXivmusig}
\end{figure}

\clearpage
\begin{table}[htbp]\small
  \centering
    \begin{tabular}{|r|r|r|r|r|r|}
    \hline
    Measure & $c_{\mathrm{f,r}}^{*}$ & Bins  & $\chi^2/\mathrm{d.o.f}$ Min & $\chi^2/\mathrm{d.o.f}$ Max & $\chi^2/\mathrm{d.o.f}$ Mean \bigstrut\\
    \hline
    \multicolumn{1}{|c|}{\multirow{8}[16]{*}{$c_{\mathrm{f}}$}} & \multicolumn{1}{c|}{\multirow{3}[6]{*}{0.0}} & 9     & 3.9   & 14.7  & 10.6 \bigstrut\\
\cline{3-6}    \multicolumn{1}{|c|}{} & \multicolumn{1}{c|}{} & 13    & 21.6  & 59.6  & 34.3 \bigstrut\\
\cline{3-6}    \multicolumn{1}{|c|}{} & \multicolumn{1}{c|}{} & 14    & 23.0  & 57.2  & 40.1 \bigstrut\\
\cline{2-6}    \multicolumn{1}{|c|}{} & \multicolumn{1}{c|}{\multirow{5}[10]{*}{0.1}} & 8     & 2.8   & 2.8   & 2.8 \bigstrut\\
\cline{3-6}    \multicolumn{1}{|c|}{} & \multicolumn{1}{c|}{} & 9     & 1.5   & 14.7  & 4.8 \bigstrut\\
\cline{3-6}    \multicolumn{1}{|c|}{} & \multicolumn{1}{c|}{} & 12    & 6.0   & 16.2  & 11.1 \bigstrut\\
\cline{3-6}    \multicolumn{1}{|c|}{} & \multicolumn{1}{c|}{} & 13    & 2.0   & 87.0  & 18.8 \bigstrut\\
\cline{3-6}    \multicolumn{1}{|c|}{} & \multicolumn{1}{c|}{} & 14    & 3.0   & 13.5  & 6.5 \bigstrut\\
    \hline
    \multicolumn{1}{|c|}{\multirow{7}[14]{*}{$c_{\mathrm{r}}$}} & \multicolumn{1}{c|}{\multirow{3}[6]{*}{0.0}} & 9     & 0.5   & 3.7   & 1.6 \bigstrut\\
\cline{3-6}    \multicolumn{1}{|c|}{} & \multicolumn{1}{c|}{} & 13    & 8.0   & 10.0  & 9.0 \bigstrut\\
\cline{3-6}    \multicolumn{1}{|c|}{} & \multicolumn{1}{c|}{} & 14    & 2.0   & 10.3  & 5.0 \bigstrut\\
\cline{2-6}    \multicolumn{1}{|c|}{} & \multicolumn{1}{c|}{\multirow{4}[8]{*}{0.1}} & 9     & 0.9   & 4.0   & 2.5 \bigstrut\\
\cline{3-6}    \multicolumn{1}{|c|}{} & \multicolumn{1}{c|}{} & 12    & 29.6  & 29.6  & 29.6 \bigstrut\\
\cline{3-6}    \multicolumn{1}{|c|}{} & \multicolumn{1}{c|}{} & 13    & 2.8   & 45.4  & 15.4 \bigstrut\\
\cline{3-6}    \multicolumn{1}{|c|}{} & \multicolumn{1}{c|}{} & 14    & 2.6   & 11.5  & 6.2 \bigstrut\\
    \hline
    \end{tabular}
      \tsecaption{Table of $\chi^2$ values for the one parameter lognormal goodness of fit to arXiv data. c* denotes the threshold below which publications were not included in the fitting.}
  \label{tabChi2arXiv}
\end{table}

\clearpage
\begin{table}[htbp]\small
\begin{center}
\begin{tabular}{|l|l|l|l|l|l|l|l|l|}
\hline
Year & sub-archive & $N_p$ & $c_0$ & $\sigma^{2}$ &  res.err./d.o.f.  & Bins & $\chi^2/\mathrm{d.o.f.}$ \\ \hline
\multirow{5}{*}{1997}
& astro-ph & 1116 & 30.32 & 3.86(13) & 0.88 & 13 & 2.0 \\
& hep-ph & 2753 & 37.91 & 1.43(15) & 4.15 & 14 & 10.1 \\
& hep-th & 1969 & 33.00 & 1.45(11) & 2.91 & 12 & 16.2 \\
& gr-qc & 718 & 22.27 & 1.32(16) & 2.48 & 9 & 3.0 \\
& All & 6556 & NA & 1.26(8) & 3.98 & 19 & 67.5 \\\hline
\multirow{5}{*}{1998}
& astro-ph & 1829 & 25.47 & 3.26(28) & 3.53 & 12 & 6.0 \\
& hep-ph & 2735 & 40.47 & 1.74(11) & 2.56 & 14 & 4.2 \\
& hep-th & 2072 & 34.53 & 1.51(9) & 2.08 & 13 & 11.1 \\
& gr-qc & 893 & 17.99 & 1.20(12) & 2.68 & 9 & 2.6 \\
& All & 7529 & NA & 1.54(9) & 3.60 & 19 & 18.1 \\\hline
\multirow{5}{*}{1999}
& astro-ph & 2690 & 18.89 & 2.92(37) & 5.03 & 14 & 13.5 \\
& hep-ph & 3050 & 34.12 & 1.49(13) & 4.69 & 13 & 8.6 \\
& hep-th & 2118 & 32.38 & 1.53(10) & 2.29 & 13 & 6.6 \\
& gr-qc & 978 & 17.60 & 1.17(5) & 1.39 & 9 & 1.5 \\
& All & 8836 & NA & 1.57(9) & 4.33 & 19 & 14.9 \\\hline
\multirow{5}{*}{2000}
& astro-ph & 2605 & 21.22 & 2.75(16) & 2.56 & 14 & 4.6 \\
& hep-ph & 3130 & 28.12 & 1.39(11) & 3.80 & 14 & 6.2 \\
& hep-th & 2483 & 29.50 & 1.53(9) & 1.96 & 14 & 3.0 \\
& gr-qc & 914 & 15.51 & 1.47(17) & 2.90 & 9 & 4.7 \\
& All & 9132 & NA & 1.39(8) & 4.59 & 19 & 16.1 \\\hline
\multirow{5}{*}{2001}
& astro-ph & 3254 & 16.90 & 2.47(26) & 4.99 & 14 & 10.4 \\
& hep-ph & 3205 & 26.61 & 1.42(9) & 2.88 & 14 & 3.8 \\
& hep-th & 2460 & 25.29 & 1.39(9) & 2.21 & 14 & 3.4 \\
& gr-qc & 947 & 14.23 & 1.26(18) & 3.71 & 9 & 5.0 \\
& All & 9866 & NA & 1.49(11) & 5.94 & 19 & 20.4 \\\hline
\multirow{5}{*}{2002}
& astro-ph & 4176 & 17.64 & 1.87(17) & 5.36 & 14 &  9.4 \\
& hep-ph & 3114 & 22.84 & 1.31(12) & 4.69 & 13 & 7.4 \\
& hep-th & 2560 & 24.93 & 1.36(11) & 3.09 & 14 & 5.6 \\
& gr-qc & 945 & 13.35 & 0.98(8) & 2.25 & 9 & 2.0 \\
& All & 10795 & NA & 1.26(8) & 5.64 & 19 & 18.3 \\\hline
\multirow{5}{*}{2003}
& astro-ph & 5478 & 17.42 & 1.37(7) & 4.81 & 13 & 87.0 \\
& hep-ph & 2887 & 20.16 & 1.42(10) & 2.82 & 14 & 5.0 \\
& hep-th & 2639 & 19.14 & 1.33(11) & 3.18 & 14 & 5.9 \\
& gr-qc & 997 & 11.58 & 1.23(13) & 3.26 & 8 & 2.8 \\
& All & 12001 & NA & 1.22(4) & 4.31 & 19 & 156.8 \\\hline
\multirow{5}{*}{2004}
& astro-ph & 4929 & 10.59 & 1.44(13) & 6.70 & 13 & 8.9 \\
& hep-ph & 3042 & 14.25 & 1.29(11) & 3.37 & 14 & 6.0 \\
& hep-th & 2534 & 14.65 & 1.33(13) & 3.10 & 14 & 6.1 \\
& gr-qc & 1171 & 8.65 & 1.17(21) & 5.97 & 9 & 14.7 \\
& All & 11676 & NA & 1.09(9) & 8.37 & 19 & 23.4 \\\hline
\end{tabular}
\tsecaption{arXiv data from graphs generated using the $c_{\mathrm{f}}$
measure using 1 parameter fit, see Equation (\ref{eLognormal}).} \label{tRadarXivtab1para}
\end{center}
\end{table}

\clearpage
\begin{table}[htbp]\small
\begin{center}
\begin{tabular}{|l|l|l|l|l|l|l|l|}
\hline
Year & sub-archive & $N_p$ & $c_0$ & $\sigma^{2}$ & $\mu + \frac{\sigma^{2}}{2}$ & res.err./d.o.f.  \\ \hline
\multirow{4}{*}{1997}
& astro-ph & 1116 & 30.32 & 5.57(111) & 0.3(2) & 0.9 \\
& hep-ph & 2753 & 37.91 & 1.48(22) & 0.2(1) & 3.8 \\
& hep-th & 1969 & 33.00 & 1.34(7) & 0.1(0) & 1.2 \\
& gr-qc & 718 & 22.27 & 1.29(21) & 0.2(1) & 2.4 \\\hline
\multirow{4}{*}{1998}
& astro-ph & 1829 & 25.47 & 2.83(104) & -0.3(3) & 4.5 \\
& hep-ph & 2735 & 40.47 & 1.58(8) & 0.1(0) & 1.2 \\
& hep-th & 2072 & 34.53 & 1.30(9) & 0.0(1) & 1.6 \\
& gr-qc & 893 & 17.99 & 1.26(19) & 0.2(1) & 3.1 \\\hline
\multirow{4}{*}{1999}
& astro-ph & 2690 & 18.89 & 6.19(252) & -0.2(4) & 3.3 \\
& hep-ph & 3050 & 34.12 & 1.47(12) & 0.2(1) & 2.6 \\
& hep-th & 2118 & 32.38 & 1.31(10) & 0.0(1) & 1.8 \\
& gr-qc & 978 & 17.60 & 1.15(7) & 0.0(0) & 1.4 \\\hline
\multirow{4}{*}{2000}
& astro-ph & 2605 & 21.22 & 2.49(45) & 0.1(2) & 2.9 \\
& hep-ph & 3130 & 28.12 & 1.42(15) & 0.2(1) & 3.1 \\
& hep-th & 2483 & 29.50 & 1.56(13) & 0.1(1) & 1.9 \\
& gr-qc & 914 & 15.51 & 1.6(33) & 0.2(2) & 3.6 \\\hline
\multirow{4}{*}{2001}
& astro-ph & 3254 & 16.90 & 3.34(118) & 0.1(3) & 6.0 \\
& hep-ph & 3205 & 26.61 & 1.43(5) & 0.2(0) & 1.0 \\
& hep-th & 2460 & 25.29 & 1.40(11) & 0.1(1) & 1.7 \\
& gr-qc & 947 & 14.23 & 1.29(36) & 0.1(2) & 5.4 \\\hline
\multirow{4}{*}{2002}
& astro-ph & 4176 & 17.64 & 2.2(42) & 0.3(2) & 6.1 \\
& hep-ph & 3114 & 22.84 & 1.28(13) & 0.2(1) & 3.2 \\
& hep-th & 2560 & 24.93 & 1.33(18) & 0.1(1) & 3.3 \\
& gr-qc & 945 & 13.35 & 0.92(6) & 0.1(0) & 1.5 \\\hline
\multirow{4}{*}{2003}
& astro-ph & 5478 & 17.42 & 1.34(8) & 0.1(0) & 3.9 \\
& hep-ph & 2887 & 20.16 & 1.43(12) & 0.2(1) & 2.1 \\
& hep-th & 2639 & 19.14 & 1.30(20) & 0.0(1) & 4.0 \\
& gr-qc & 997 & 11.58 & 1.26(22) & 0.1(1) & 4.2 \\\hline
\multirow{4}{*}{2004}
& astro-ph & 4929 & 10.59 & 1.55(30) & 0.1(1) & 8.4 \\
& hep-ph & 3042 & 14.25 & 1.37(19) & 0.2(1) & 3.7 \\
& hep-th & 2534 & 14.65 & 1.45(23) & 0.2(1) & 3.3 \\
& gr-qc & 1171 & 8.65 & 1.32(46) & 0.1(3) & 9.0 \\\hline
\end{tabular}
\tsecaption{ArXiv data from graphs generated using the $c_{\mathrm{f}}$
measure using 3 parameter fit,  $A \cdot F(\cfindex;\mu,\sigma^2)$.} \label{tRadarXivtab3para}
\end{center}
\end{table}

\clearpage
\begin{table}[htbp]\small
\begin{center}
\begin{tabular}{|l|l|l|l|l|l|l|l|l|}
\hline
Year & sub-archive & $N_p$ & $\langle c_r \rangle$ & $\langle r \rangle$ & $\sigma^{2}$ & res.err./d.o.f. & Bins & $\chi^2/\mathrm{d.o.f.}$ \\ \hline
\multirow{5}{*}{1997}
& astro-ph & 1401 & 4.50 & 12.59243 & 3.04(19) & 1.80 & 14 & 5.1 \\
& hep-ph & 3143 & 2.01 & 27.99745 & 1.66(9) & 3.08 & 14 & 8.9 \\
& hep-th & 2255 & 2.03 & 23.82306 & 1.61(6) & 1.63 & 14 & 4.4 \\
& gr-qc & 762 & 2.62 & 14.46325 & 1.67(9) & 1.58 & 9 & 1.9 \\
& All & 7561 & NA & 0 & 1.92(6) & 2.85 & 19 & 5.2 \\\hline
\multirow{5}{*}{1998}
& astro-ph & 2107 & 3.24 & 12.44186 & 2.66(17) & 2.95 & 13 & 8.0 \\
& hep-ph & 3308 & 2.10 & 27.8815 & 1.83(5) & 1.75 & 14 & 4.8 \\
& hep-th & 2357 & 2.02 & 24.81375 & 1.71(4) & 1.01 & 14 & 2.0 \\
& gr-qc & 898 & 2.15 & 15.88419 & 1.72(9) & 1.66 & 9 & 1.6 \\
& All & 8670 & NA & 0 & 2.00(6) & 3.18 & 19 & 5.4 \\\hline
\multirow{5}{*}{1999}
& astro-ph & 2714 & 2.67 & 12.47973 & 2.48(10) & 2.22 & 14 & 5.9 \\
& hep-ph & 3575 & 1.61 & 30.89538 & 1.64(4) & 1.85 & 14 & 6.8 \\
& hep-th & 2477 & 1.77 & 26.16795 & 1.72(4) & 1.12 & 14 & 2.5 \\
& gr-qc & 981 & 1.91 & 15.80224 & 1.54(5) & 1.02 & 9 & 0.5 \\
& All & 9747 & NA & 0 & 1.88(3) & 1.99 & 19 & 6.3 \\\hline
\multirow{5}{*}{2000}
& astro-ph & 3126 & 2.61 & 13.3151 & 2.33(9) & 2.30 & 14 & 4.9 \\
& hep-ph & 3545 & 1.19 & 32.50945 & 1.45(6) & 2.67 & 14 & 4.0 \\
& hep-th & 2773 & 1.48 & 27 & 1.52(4) & 1.45 & 14 & 2.1 \\
& gr-qc & 968 & 1.59 & 16.12293 & 1.38(5) & 1.21 & 9 & 3.7 \\
& All & 10412 & NA & 0 & 1.73(3) & 2.32 & 19 & 2.4 \\\hline
\multirow{5}{*}{2001}
& astro-ph & 3443 & 2.14 & 14.14551 & 2.27(8) & 2.37 & 14 & 5.0 \\
& hep-ph & 3686 & 1.12 & 33.89935 & 1.57(6) & 2.60 & 14 & 7.0 \\
& hep-th & 2786 & 1.21 & 28.22505 & 1.40(3) & 1.19 & 14 & 4.4 \\
& gr-qc & 997 & 1.41 & 17.08325 & 1.45(5) & 1.17 & 9 & 0.5 \\
& All & 10912 & NA & 0 & 1.72(4) & 3.15 & 19 & 6.6 \\\hline
\multirow{5}{*}{2002}
& astro-ph & 4137 & 2.41 & 13.94199 & 1.73(7) & 3.02 & 14 & 3.3 \\
& hep-ph & 3656 & 0.86 & 36.68545 & 1.36(7) & 3.52 & 14 & 7.3 \\
& hep-th & 2900 & 1.07 & 30.37 & 1.36(5) & 2.06 & 14 & 3.7 \\
& gr-qc & 1025 & 1.28 & 18.14927 & 1.42(5) & 1.11 & 9 & 1.1 \\
& All & 11718 & NA & 0 & 1.49(5) & 4.37 & 19 &  9.3 \\\hline
\multirow{5}{*}{2003}
& astro-ph & 5512 & 1.90 & 15.28538 & 1.45(4) & 3.04 & 14 & 6.7 \\
& hep-ph & 3406 & 0.72 & 39.10423 & 1.40(3) & 1.23 & 14 & 2.8 \\
& hep-th & 2826 & 0.80 & 31.52937 & 1.27(3) & 1.29 & 13 & 10.0 \\
& gr-qc & 1110 & 1.00 & 20.16667 & 1.45(3) & 0.76 & 9 & 0.8 \\
& All & 12854 & NA & 0 & 1.40(4) & 4.06 & 19 & 10.4 \\\hline
\multirow{5}{*}{2004}
& astro-ph & 5414 & 0.98 & 18.73735 & 1.44(2) & 1.27 & 14 & 10.3 \\
& hep-ph & 3395 & 0.52 & 40.162 & 1.25(4) & 2.02 & 14 & 3.9 \\
& hep-th & 2823 & 0.61 & 34.2136 & 1.26(4) & 1.47 & 14 & 4.0 \\
& gr-qc & 1135 & 0.63 & 22.19031 & 1.24(6) & 1.69 & 9 & 2.3 \\
& All & 12767 & NA & 0 & 1.29(3) & 3.74 & 19 & 19.2 \\\hline
\end{tabular}
\tsecaption{ArXiv Data from graphs generated using the $c_{\mathrm{r}}$
measure using 1 parameter fit, see Equation (\ref{eLognormal}).} \label{tOurarXivtab1para}
\end{center}
\end{table}

\clearpage
\begin{table}[htbp]\small
\begin{center}
\begin{tabular}{|l|l|l|l|l|l|l|l|}
\hline
Year & sub-archive & $N_p$ & $\langle c_{\mathrm{r}} \rangle$ & $\sigma^{2}$ & $\mu + \frac{\sigma^{2}}{2}$ & res.err./d.o.f.  \\ \hline
\multirow{4}{*}{1997}
& astro-ph & 945 & 4.75 & 1.77(48) & -0.4(2) & 1.8 \\
& hep-ph & 2620 & 2.18 & 1.39(12) & 0.1(1) & 2.2 \\
& hep-th & 1909 & 2.17 & 1.29(12) & -0.1(1) & 1.7 \\
& gr-qc & 667 & 2.73 & 1.34(15) &0.0(1) & 1.7 \\\hline
\multirow{4}{*}{1998}
& astro-ph & 1463 & 3.42 & 0.94(19) & -0.6(1) & 3.3 \\
& hep-ph & 2700 & 2.27 & 1.31(6) & -0.1(0) & 1.4 \\
& hep-th & 1984 & 2.15 & 1.22(11) & -0.2(1) & 1.7 \\
& gr-qc & 751 & 2.28 & 1.22(8) &0.0(0) & 1.0 \\\hline
\multirow{4}{*}{1999}
& astro-ph & 2008 & 2.87 & 1.58(46) & -0.3(2) & 4.9 \\
& hep-ph & 3018 & 1.75 & 1.23(7) & -0.1(0) & 1.8 \\
& hep-th & 2101 & 1.9 & 1.22(10) & -0.2(1) & 1.8 \\
& gr-qc & 867 & 1.99 & 1.18(24) &0.0(1) & 3.8 \\\hline
\multirow{4}{*}{2000}
& astro-ph & 2421 & 2.77 & 1.85(26) & -0.1(1) & 2.6 \\
& hep-ph & 3055 & 1.31 & 1.2(12) & -0.1(1) & 3.0 \\
& hep-th & 2422 & 1.6 & 1.22(11) & -0.1(1) & 2.4 \\
& gr-qc & 846 & 1.69 & 1.09(20) & -0.1(1) & 3.6 \\\hline
\multirow{4}{*}{2001}
& astro-ph & 2692 & 2.32 & 1.67(32) & -0.2(2) & 4.6 \\
& hep-ph & 3098 & 1.27 & 1.14(5) & -0.1(0) & 1.9 \\
& hep-th & 2452 & 1.31 & 1.07(4) & -0.1(0) & 1.2 \\
& gr-qc & 889 & 1.5 & 1.13(6) & -0.1(0) & 1.1 \\\hline
\multirow{4}{*}{2002}
& astro-ph & 3417 & 2.55 & 1.31(17) &0.0(1) & 4.3 \\
& hep-ph & 3147 & 0.98 & 1.03(6) & -0.1(0) & 1.8 \\
& hep-th & 2568 & 1.18 & 1.1(6) & -0.1(0) & 1.4 \\
& gr-qc & 905 & 1.39 & 1.13(14) & -0.1(1) & 2.6 \\\hline
\multirow{4}{*}{2003}
& astro-ph & 4764 & 2.01 & 1.16(17) & -0.1(1) & 7.0 \\
& hep-ph & 2889 & 0.84 & 1.09(9) & -0.2(1) & 2.7 \\
& hep-th & 2507 & 0.89 & 1.00(10) & -0.2(1) & 3.5 \\
& gr-qc & 975 & 1.12 & 0.95(7) & -0.2(0) & 1.6 \\\hline
\multirow{4}{*}{2004}
& astro-ph & 4828 & 1.09 & 1.16(4) & -0.1(0) & 2.0 \\
& hep-ph & 2716 & 0.64 & 0.78(7) & -0.2(0) & 2.4 \\
& hep-th & 2366 & 0.72 & 0.89(5) & -0.2(0) & 1.6 \\
& gr-qc & 959 & 0.74 & 0.78(4) & -0.2(0) & 1.3 \\\hline
\end{tabular}
\tsecaption{ArXiv data from graphs generated using the $c_{\mathrm{f}}$
measure using 3 parameter fit,  $A \cdot F(\cfindex;\mu,\sigma^2)$.} \label{tOurXivtab3para}
\end{center}
\end{table}

\end{document}